\documentclass[times, twocolumn]{aastex631}
\usepackage{amsfonts,amsmath,amssymb,amsthm,epsfig,graphicx,float,tabularx,multirow,booktabs,gensymb}
\usepackage{booktabs,color,textcomp}
\usepackage{natbib}
\usepackage{rotating}
\usepackage{makecell}
\usepackage{multirow}
\usepackage{appendix}
\providecommand{\e}[1]{\ensuremath{\times 10^{#1}}} 

\setcellgapes{0pt}

\newcolumntype{S}{>{\centering\arraybackslash}m{0.05\linewidth}}
\newcolumntype{E}{>{\centering\arraybackslash}m{0.05\linewidth}}
\newcolumntype{M}{>{\centering\arraybackslash}m{0.1\linewidth}}

\newcolumntype{N}{>{\centering\arraybackslash}m{0.0005\linewidth}}

\received{November 22, 2024}
\revised{April 21, 2025}
\accepted{May 1, 2025, The Astronomical Journal}

\usepackage{xcolor}
\usepackage{amsmath}

\usepackage{graphicx}
\usepackage{natbib}

\usepackage{multirow}
\usepackage{booktabs}
\usepackage{tabularx}

\usepackage{savesym}
\savesymbol{tablenum}
\usepackage{siunitx}
\restoresymbol{SIX}{tablenum}

\usepackage{xspace} 

\newcommand{\I}{$\mathcal{I}$\xspace}
\newcommand{\Q}{$\mathcal{Q}$\xspace}
\newcommand{\U}{$\mathcal{U}$\xspace}
\newcommand{\V}{$\mathcal{V}$\xspace}
\newcommand{\Qr}{$\mathcal{Q}_\phi$\xspace}
\newcommand{\Ur}{$\mathcal{U}_\phi$\xspace}
\newcommand{\Lir}{$L_{\mathrm{IR}}/L_\star$\xspace}

\begin{document}

\title{The Disks In Scorpius-Centaurus Survey (DISCS) I: Four Newly-Resolved Debris Disks in Polarized Intensity Light}

\correspondingauthor{Justin Hom}
\email{jrhom@arizona.edu}

\author[0000-0001-9994-2142]{Justin Hom}
\affiliation{Steward Observatory and Department of Astronomy, University of Arizona, Tucson AZ 85721}

\author[0000-0002-0792-3719]{Thomas M. Esposito}
\affiliation{Astronomy Department, University of California, Berkeley, CA 94720, USA}
\affiliation{SETI Institute, Carl Sagan Center, 189 Bernardo Ave., Mountain View CA 94043, USA}

\author[0000-0003-4909-256X]{Katie A. Crotts}
\affiliation{Space Telescope Science Institute, Baltimore, MD 21218, USA}

\author[0000-0002-5092-6464]{Gaspard Duch\^ene}
\affiliation{Universit\'e Grenoble Alpes / CNRS, Institut de Plan\'etologie et d'Astrophysique de Grenoble, 38000 Grenoble, France}
\affiliation{Astronomy Department, University of California, Berkeley, CA 94720, USA}

\author{Jennifer Patience}
\affiliation{School of Earth and Space Exploration, Arizona State University, Tempe, AZ 85281, USA}

\author[0000-0002-9133-3091]{Johan Mazoyer}
\affiliation{LESIA, Observatoire de Paris, Universit{\'e} PSL, CNRS, Universit{\'e} Paris Cit{\'e}, Sorbonne Universit{\'e}, 5 place Jules Janssen, F-92195 Meudon, France}

\author[0000-0002-4918-0247]{Robert J. De Rosa}
\affiliation{European Southern Observatory, Alonso de C\'{o}rdova 3107, Vitacura, Santiago, Chile}

\author[0000-0002-8382-0447]{Christine H. Chen}
\affiliation{Space Telescope Science Institute, Baltimore, MD 21218, USA}

\author[0000-0002-6221-5360]{Paul Kalas}
\affiliation{Astronomy Department, University of California, Berkeley, CA 94720, USA}
\affiliation{SETI Institute, Carl Sagan Center, 189 Bernardo Ave., Mountain View CA 94043, USA}
\affiliation{Institute of Astrophysics, FORTH, GR-71110 Heraklion, Greece}

\author[0000-0003-1212-7538]{Bruce Macintosh}
\affiliation{Kavli Institute for Particle Astrophysics and Cosmology, Stanford University, Stanford, CA 94305, USA}
\affiliation{University of California Observatories, 1156 High Street, Santa Cruz, CA 95064, USA}
\affiliation{Department of Astronomy and Astrophysics, University of California, Santa Cruz, Santa Cruz, CA 95064, USA}

\author[0000-0003-3017-9577]{Brenda C. Matthews}
\affiliation{National Research Council of Canada Herzberg, 5071 West Saanich Road, Victoria, BC V9E 2E7, Canada}

\begin{abstract}
The presence of infrared excesses around stars directly correlates to spatially-resolved imaging detections of circumstellar disks at both mm and optical/near-infrared wavelengths. High contrast imagers have resolved dozens of circumstellar disks with scattered light polarimetric imaging. Many of these detections are members of the Scorpius-Centaurus OB association, demonstrating it to be a rich sample for investigating planetary system architectures and planet-disk interactions. With the goal of expanding the sample of directly imaged debris disks in Scorpius-Centaurus, we conducted the Disks In Scorpius-Centaurus Survey (DISCS) leveraging knowledge of high-IR excesses and the power of high contrast polarimetric differential imaging. In combination with the GPIES polarimetric disk survey, we observe seven new Scorpius-Centaurus targets to achieve a 60\% complete survey of debris disks with IR excesses exceeding $2.5\times10^{-4}$, resolving four new debris disks. HD 98363, HD 109832, and HD 146181 are resolved for the first time, and HD 112810 is resolved for the first time in polarized intensity. We identify morphological structures that may be indications of planet-disk interactions. We place the systems in the greater context of resolved debris disks, identifying factors of ten differences in scattered light contrast for a given IR excess and implying gaps in our understanding of the smallest and largest dust grains of a system. We conclude that while thermal emission measurements are correlated with scattered light detection, they poorly predict the magnitude of scattered light brightness. We also establish Scorpius-Centaurus debris disks as critical benchmarks in understanding the properties of disks in the scattering regime.
\end{abstract}


\section{Introduction}
Circumstellar debris disks are expansive collections of dust grains and planetesimals, representing a stage of planet formation and evolution that occurs after the gas content of protoplanetary disks dissipate and evaporate \citep{wyatt2008,hughes2018}. These dust grains, absorbing energy from the stars they orbit, re-emit that energy in the mid- to far-infrared, causing their host stars to display excesses of IR emission in their spectral energy distributions \citep[SED; e.g.][]{aumann1984}. Direct correlations have been identified between levels of IR excess and resolved detection of debris disks \citep[e.g.,][]{smith1984,holland1998}, and the evolution of IR excess appears to peak around 10-20 Myr of stellar age before declining over time \citep{wyatt2008}. The advancement of spatially-resolved, high contrast imaging with instruments such as the Gemini Planet Imager \citep[GPI;][]{macintosh2014} and the Spectro-Polarimetric High-contrast Exoplanet REsearch (SPHERE; \citealt{beuzit2019}) has further expanded our understanding of debris disk evolution and structure, with the presence of asymmetric structural features such as warps \citep[e.g., $\beta$ Pic;][]{lagrange2009} and eccentricity \citep[e.g. HD 106906,][]{kalas2015,crotts2021} potentially originating from perturbations of substellar companions \citep[e.g.,][]{lee2016}. This is further supported by the presence of debris disks in older systems, as processes such as ongoing planetesimal collisions \citep{backman1993} or catastrophic planet collisions \citep{cameron1997} must continually replenish dust grains in a debris disk, as the dust grains themselves exist through relatively short timescales.

As the closest OB association ($\sim$110 -- 140 pc), Scorpius-Centaurus \citep[Sco-Cen;][]{blaauw1946,preibisch2008,dezeeuw1999} has been shown to contain a rich sample of young planetary systems. The moving group is also at an age where IR excesses around stars are expected to be at their highest, and several studies have identified debris disk targets in the mid-IR with \textit{Spitzer} \citep{chen2014} and at mm wavelengths with the Atacama Large Millimeter Array \citep[ALMA;][]{lieman-sifry2016}. The ages of stars (11--16 Myr) in the Upper Centaurus Lupus (UCL) and Lower Centaurus Crux (LCC) subregions of Sco-Cen \citep{pecaut2016} roughly correspond to the later stages of planet formation, when planets may directly perturb the normally axisymmetrical dust grain orbits into asymmetric behavior. Of the total number of unique debris disk detections from the Gemini Planet Imager Exoplanet Survey (GPIES) presented in \cite{esposito2020} (26 in total), 13 are members of Sco-Cen. Additional Sco-Cen scattered light-resolved debris disks have also been reported with SPHERE \citep[e.g.,][]{bonnefoy2017,bonnefoy2021,perrot2023,pawellek2024}, along with other systems such as protoplanetary and transition disks \citep[e.g.][]{janson2016,currie2015a,benisty2017,bohn2019} and substellar companions \citep[e.g.][]{chauvin2017,rameau2013,bailey2014}. \textit{Spitzer} IR and ALMA measurements have detected even more debris disks in the Sco-Cen association \citep{chen2014,lieman-sifry2016}, demonstrating that the moving group is a rich environment from which to investigate the properties of young planetary systems in a common age and formation environment. Several asymmetric debris disks have been identified in Sco-Cen, including the eccentric HD 106906 \citep{kalas2015,lagrange2016}, the warped HD 110058 \citep{stasevic2023,lopez2023}, and the asymmetrically bright HD 111520 \citep{draper2016,crotts2022}. The external substellar companion HD 106906 b \citep{bailey2014} is a likely candidate for inducing the asymmetric structure seen in its corresponding debris disk, as demonstrated in dynamical modeling of the system in \cite{nesvold2017}.

In this work, we present the first results of the Disks in Scorpius-Centaurus Survey (DISCS), a polarimetric imaging campaign of high IR-excess debris disk targets in Sco-Cen. This survey expands upon debris disk imaging campaigns from the Gemini Planet Imager Exoplanet Survey (GPIES) presented in \cite{esposito2020}, completing GPI observations of high-IR excess debris disks down to \Lir $=2.5\times 10^{-4}$ in Sco-Cen by leveraging the efficacy of polarimetric differential imaging in unbiased detection of circumstellar disk structure. In \S \ref{sec:target_sample}-\ref{sec:datareduction}, we describe the DISCS sample, observations, and data reduction. In \S \ref{sec:results}, we show the polarized and total intensity images for all systems. In \S \ref{sec:diskchar}, we perform ellipse-fitting disk modeling analyses to characterize the four resolved debris disk detections. In \S \ref{sec:discussion}, we place our resolved debris disk properties in the greater context of Sco-Cen resolved debris disks and directly-imaged scattered light debris disks as a whole, identifying trends between surface brightnesses, IR excesses, disk morphology, and host star properties. In \S \ref{sec:conclusion} we give the summary and implications of our findings.

\section{DISCS Sample} \label{sec:target_sample}
The DISCS program seeks to complete a survey of Sco-Cen debris disks down to \Lir$=2.5\times 10^{-4}$ already initiated by the GPIES polarimetric debris disk campaign \citep{esposito2020}. \cite{esposito2020} found \Lir$=2.5\times10^{-4}$ to be a loose lower limit on what GPI could reasonably detect in scattered light. Sco-Cen was selected as the number of GPIES-detected debris disks were heavily biased towards members of the OB association, with 13 debris disks and 2 protoplanetary disks out of 29 total circumstellar disks detected. Completing the sample would therefore provide a sufficiently large number of systems in which to understand planetary system architectures in a common age and formation environment. DISCS sample criteria included: 1) having IR excess $\geq 2.5 \times 10^{-4}$, 2) being a debris disk member of Sco-Cen, 3) having a host star bright enough for GPI observations ($R<$9), and 4) containing no known stellar companion within the GPI field-of-view, since the coronagraph can only block the light of the target star. The $R$ magnitude requirement effectively limits the spectral type coverage of the sample to stars earlier than mid-K type (37 systems). Only seven debris disks around G- and K-type stars were excluded from DISCS sample due to having $R>9$. \cite{jang-condell2015} determined from Spitzer IRS excess measurements, however, that the majority of late spectral type US stars had disks consistent with being T Tauri/protoplanetary systems. While there are other late spectral type stars in UCL and LCC with IR excesses consistent with debris disks, they are too faint to be observed with current high contrast imaging instrumentation. IR excess values were calculated using the same method described in \cite{esposito2020}, leveraging IR measurements from \textit{IRAS}, \textit{Spitzer}, \textit{Herschel}, and \textit{WISE} and utilizing a similar methodology described in \cite{cotten2016}. \cite{esposito2020} observed 16 of 44 (36\%) Sco-Cen debris disks with $L_{IR}/L_* \geq 2.5\times 10^{-4}$ and spectral type earlier than mid-K. Four additional debris disks in Sco-Cen have also been detected with SPHERE (HD 120326; \citealt{bonnefoy2017}; HD 141011; \citealt{bonnefoy2021}; HD 121617; \citealt{perrot2023}; HD 131488; \citealt{pawellek2024}), with three satisfying DISCS sample minimum $L_{IR}/L_*$. In combination with the Sco-Cen sample observed in \cite{esposito2020} and SPHERE detections, the new observations in this program facilitate a $\sim$60\% complete survey of Sco-Cen debris disks with $L_{IR}/L_* \geq 2.5\times 10^{-4}$ and spectral type earlier than mid-K. The combined observing samples also complete a 100\% complete survey of Sco-Cen debris disks with $L_{IR}/L_* \gtrsim 8 \times 10^{-4}$. With 20 detections, this makes the detection rate of the observed sample 77\%. In addition to high IR excess values, two of our newly-observed systems also had unresolved ALMA detections and one system had a resolved ALMA detection \citep{lieman-sifry2016,moor2017}. The properties of the DISCS sample are summarized in Table \ref{tab:obs_sample} and distinguishes between the GPIES subset of systems presented in \cite{esposito2020} and supplemental observations presented in \cite{hom2020} and this work.

\begin{deluxetable*}{ccccccccc}




\tablecaption{Stellar Properties of the DISCS Sample.}\label{tab:obs_sample}


\tablehead{\colhead{Name} & \colhead{Subgroup} & \colhead{SpTy} & \colhead{$H$} & \colhead{D} & \colhead{$T_{eff}$} & \colhead{$M_*$} & \colhead{\Lir} & \colhead{ALMA Detected?} \\ 
\colhead{} & \colhead{} & \colhead{} & \colhead{} & \colhead{(pc)} & \colhead{(K)} & \colhead{($M_{\odot}$)} & \colhead{($10^{-4}$)} & \colhead{} } 

\startdata
\multicolumn{9}{c}{DISCS--GPIES Subset Detections} \\ \hline	
HD 106906 & LCC & F5V & 6.8 & 103.3 & 6500 & 2.70 & 50.4 & \citet{fehr2022} \\
HD 110058 & LCC & A0V & 7.5 & 130.0 & 8000 & 1.70 & 26.2 & \citet{lieman-sifry2016} \\
HD 111161 & LCC & A3III/IV & 7.2 & 109.4 & 7800 & 1.72 & 42.3 & \citet{lieman-sifry2016} \\
HD 111520 & LCC & F5/6V & 7.7 & 108.9 & 6500 & 1.26 & 10.3 & \citet{lieman-sifry2016} \\
HD 114082 & LCC & F3V & 7.2 & 95.7 & 7000 & 1.42 & 36.3 & \citet{lieman-sifry2016} \\
HD 115600 & LCC & F2IV/V & 7.3 & 109.6 & 7000 & 1.54 & 22.6 & \citet{lieman-sifry2016} \\
HD 117214 & LCC & F6V & 6.9 & 107.6 & 6500 & 1.47 & 26.7 & \citet{lieman-sifry2016} \\
HD 129590 & UCL & G3V & 7.8 & 136.0 & 5910 & 1.40 & 69.6 & \citet{lieman-sifry2016} \\
HD 131835 & UCL & A2IV & 7.5 & 133.7 & 8100 & 1.77 & 30.9 & \citet{lieman-sifry2016} \\
HD 143675 & UCL & A5IV/V & 7.6 & 139.2 & 7900 & 1.78 & 5.6 & N$^a$ \\
HD 145560 & UCL & F5V & 7.8 & 120.4 & 6500 & 1.29 & 12.7 & \citet{lieman-sifry2016} \\
HD 146897 & US & F2/3V & 7.8 & 131.5 & 6200 & 1.28 & 101.9 & \citet{lieman-sifry2016} \\
HD 156623 & UCL & A0V & 7.0 & 111.8 & 8350 & 1.90 & 43.3 & \citet{lieman-sifry2016} \\ \hline
\multicolumn{9}{c}{DISCS--Supplemental Subset Detections} \\ \hline
HD 98363 & LCC & A2V & 7.5 & 138.6 & 8830 & 1.92 & 6.4 & \citet{moor2017} \\
HD 109832 & LCC & A9V & 8.1 & 107.3 & 7400 & 1.75 & 3.5 & N$^a$ \\
HD 112810 & LCC & F3/5IV/V & 8.1 & 133.7 & 6637 & 1.40 & 5.9 & \citet{lieman-sifry2016} \\
HD 146181 & UCL & F6V & 8.1 & 127.5 & 6350 & 1.25 & 19.0 & \citet{lieman-sifry2016} \\ \hline
\multicolumn{9}{c}{DISCS--GPIES Subset Nondetections} \\ \hline
HD 95086 & LCC & A8III & 6.9 & 86.4 & 7600 & 1.61 & 8.4 & \citet{su2017} \\
HD 108857 & LCC & F7V & 7.2 & 104.5 & 6000 & 1.39 & 6.9 & N \\
HD 138813 & US & A0V & 7.2 & 137.4 & 8640 & 2.15 & 13.4 & \citet{lieman-sifry2016} \\
HD 142315 & US & B9V & 6.7 & 145.3 & 8800 & 4.17 & 6.4 & \citet{lieman-sifry2016} \\ \hline
\multicolumn{9}{c}{DISCS--Supplemental Subset Nondetections} \\ \hline
HD 108904 & LCC & F6V & 6.8 & 107.1 & 6350 & 1.25 & 3.3 & N \\
HD 113556 & LCC & F2V & 7.3 & 100.5 & 6820 & 1.46 & 4.4 & N$^b$ \\
HD 119718 & LCC & F5V & 6.9 & 115.5 & 6550 & 1.33 & 2.6 & N \\
\enddata


\tablecomments{$T_{eff}$ and $M_*$ for HD 112810 was found in \cite{matthews2023}, all other effective temperatures and masses were estimated by querying \cite{pecaut2013}. Distances were determined from parallaxes presented in Gaia EDR3 \citep{gaia2020}. \textit{Other Notes:} (a) upper limits obtained in \citet{moor2017}, (b) upper limit obtained in \citet{lieman-sifry2016}.}

\tablerefs{1. \cite{dezeeuw1999}, 2. \cite{houk1975}, 3. \cite{goldman2018}, 4. \cite{houk1978}.}

\end{deluxetable*}

\section{Observations} \label{observations}
Prior to the new observations presented in this work, previous GPIES observations of Sco-Cen systems were reported in \cite{esposito2020}, \cite{hung2015}, \cite{kalas2015}, \cite{draper2016}, and \cite{hom2020}. The new observations for this study were taken in the 2019A semesters with GPI at Gemini-South (GS-2019A-Q-109; PI: Patience) and summarized in Table \ref{tab:GPI_Obs}. Our targets were observed in the $H$-band using GPI's polarimetric mode \citep{perrin2015} and consisted of 36---64 frames of 60 s or 90 s each, depending on the brightness of the star and observing conditions. \textit{H}-band was selected for consistency with polarization observations in \cite{esposito2020}, which allows for uniform comparisons between the GPIES and DISCS samples. In contrast to a majority of polarimetric observations reported in \cite{esposito2020} which were either or both $\sim$20 minute ``snapshot"  and/or $\lesssim$40 minute ``deep" observations, we conducted even deeper polarimetric observations for a total of at least $\sim$45 minutes integration time per target. This allows for a higher SNR in addition to a higher amount of field rotation, increasing the likelihood of a detection in total intensity light after applying post-processing techniques that utilize angular differential imaging \citep[ADI;][]{marois2006}. As the geometries of these systems are poorly constrained from unresolved \textit{Spitzer} and ALMA data, we determined that the polarimetric mode gave us the best likelihood of disk detection, while still providing a chance of detecting the disk in total intensity light.

\begin{deluxetable}{ccccc}




\tablecaption{Summary of new GPI observations.}
\label{tab:GPI_Obs}


\tablehead{\colhead{HD} & \colhead{Date} & \colhead{$N \times t_{exp}$} & \colhead{$t_{int}$} & \colhead{$\Delta \theta$} \\ 
\colhead{} & \colhead{(UT)} & \colhead{(s)} & \colhead{(min)} & \colhead{(deg)} } 

\startdata
98363 & 2019 Feb 20 & 36$\times$90 & 54 & $28.^{\circ}6$ \\
108904 & 2019 Feb 28 & 40$\times$90 & 60 & $28.^{\circ}8$ \\
109832 & 2019 May 14 & 64$\times$90 & 96 & $44.^{\circ}4$ \\
112810 & 2019 Feb 28 & 36$\times$90 & 54 & $39.^{\circ}7$ \\
113556 & 2019 May 15 & 36$\times$90 & 54 & $28.^{\circ}1$ \\
119718 & 2019 May 15 & 37$\times$90 & 55 & $28.^{\circ}0$ \\
146181 & 2019 Mar 29 & 46$\times$60 & 46 & $62.^{\circ}5$ \\
\enddata


\tablecomments{$N$ is the number of exposures, $t_{exp}$ is the integration time per exposure, $t_{int}$ is the total integration time, and $\Delta \theta$ is the total accumulated field rotation for the observational sequence. Observations of HD 98363 were previously presented in \cite{hom2020}.}


\end{deluxetable}

\section{Data Reduction} \label{sec:datareduction}
We summarize the critical steps of the data reduction process for GPI polarimetry data here, but we direct the reader to \S 4 of \cite{esposito2020} for a detailed description of the data reduction process.

\subsection{Pre-Processing of GPI Polarization Data} \label{sec:pre-processing}
The GPI Data Reduction Pipeline \citep[DRP;][]{perrin2014,wang2018} was used to reduce all raw GPI data and includes dark subtraction, correlated noise cleaning, and bad pixel correction. Flexure correction was also performed for each frame and consequently combined into a polarization datacube, followed by calibration with a polarized flat field and the use of a double differencing algorithm to correct for non-common path errors \citep{perrin2015}. Each datacube was smoothed with a 1 pixel Gaussian kernel. We then measure the intrinsic stellar and instrumental polarization in an annulus with an extent of 13 to 15 pixels from the image center (except in the case of HD 119718 which was 10 to 13 pixels from the image center); this measurement is then subtracted from each pixel (after being scaled by the total intensity of the pixel). The size and location of the annulus were varied to minimize stellar and instrumental polarization noise. Finally, the datacubes were all rotated to align with north pointing up and also corrected for geometric distortion. We refer the reader to \cite{esposito2020} for a detailed description of the GPI polarization data reduction procedure.

\subsection{Generating Stokes and Azimuthal Stokes Images} \label{sec:polReductions}
From the datacubes for each dataset, we generated a single Stokes datacube with each slice corresponding to the Stokes vectors \I, \Q, \U, and \V. This is done by taking the inverse of a measurement matrix that describes the conversion between orthogonal polarization states from each datacube and a Stokes vector as described in \cite{perrin2015} and \cite{esposito2020}. Following \cite{esposito2020}, we convert the \Q and \U images into their azimuthal components \Qr and \Ur following the equations described in \cite{schmid2006}.

As discussed in \cite{esposito2020}, the estimation of the instrumental polarization was not always perfect, sometimes resulting in a quadrupole pattern persisting in both \Qr and \Ur images. To account for this, we apply the same approach discussed in \cite{esposito2020} and fit a function of the form $B=B_0 \mathcal{I}_r \sin{2(\theta + \theta_0)}$ to the \Ur image. $B_0$ is a scalar factor, $\theta_0$ is the offset angle, and $\mathcal{I}_r$ is the azimuthally-averaged total intensity as a function of radius. Both $B_0$ and $\theta_0$ were varied in order to minimize the squared residual sum, and the final best-fit $B$ was subtracted from the \Ur image and the \Qr image after rotating $45\degr$ counterclockwise. Finally, the Stokes \Qr and \Ur images were flux converted from ADU coadd$^{-1}$ to mJy/arcsec$^2$ following the approach described in \cite{hung2016} and \cite{esposito2020}.

\subsection{Total Intensity Post-Processing} \label{sec:totint_postproc}
To generate total intensity images, we considered two different post-processing approaches: the Locally-Optimized Combination of Images \citep[LOCI;][]{lafreniere2007} for HD 98363 only and Karhunen-Lo{\`e}ve Image Projection \citep[KLIP;][]{soummer2012,amara2012} for the rest of the sample. For KLIP specifically, we use the \texttt{pyKLIP} implementation \citep{wang2015}. Both approaches utilize angular differential imaging for creating PSF references \citep{marois2006} for GPI data only after collapsing each datacube along the orthogonal polarization slices. We approach our total intensity post-processing with parameters optimized for two imaging scenarios: disk-focused and planet-focused. In the disk-focused case, we choose conservative parameters (7 KL modes, no image subdivision for KL mode calculation) to mitigate the effect of over/self-subtraction \citep{milli2012}. In the planet-focused case, we chose aggressive parameters (30 KL modes, image subdivision into annuli ranging from 9 to 100) to maximize sensitivity to substellar companions. By choosing conservative parameters, we optimize our ability to detect disks at the cost of lower overall contrast for detecting substellar companions and vice versa.

\section{Results} \label{sec:results}
\subsection{Polarized Intensity Images} \label{sec:polimages}
\begin{figure*}
    \centering
    \includegraphics[width=\textwidth]{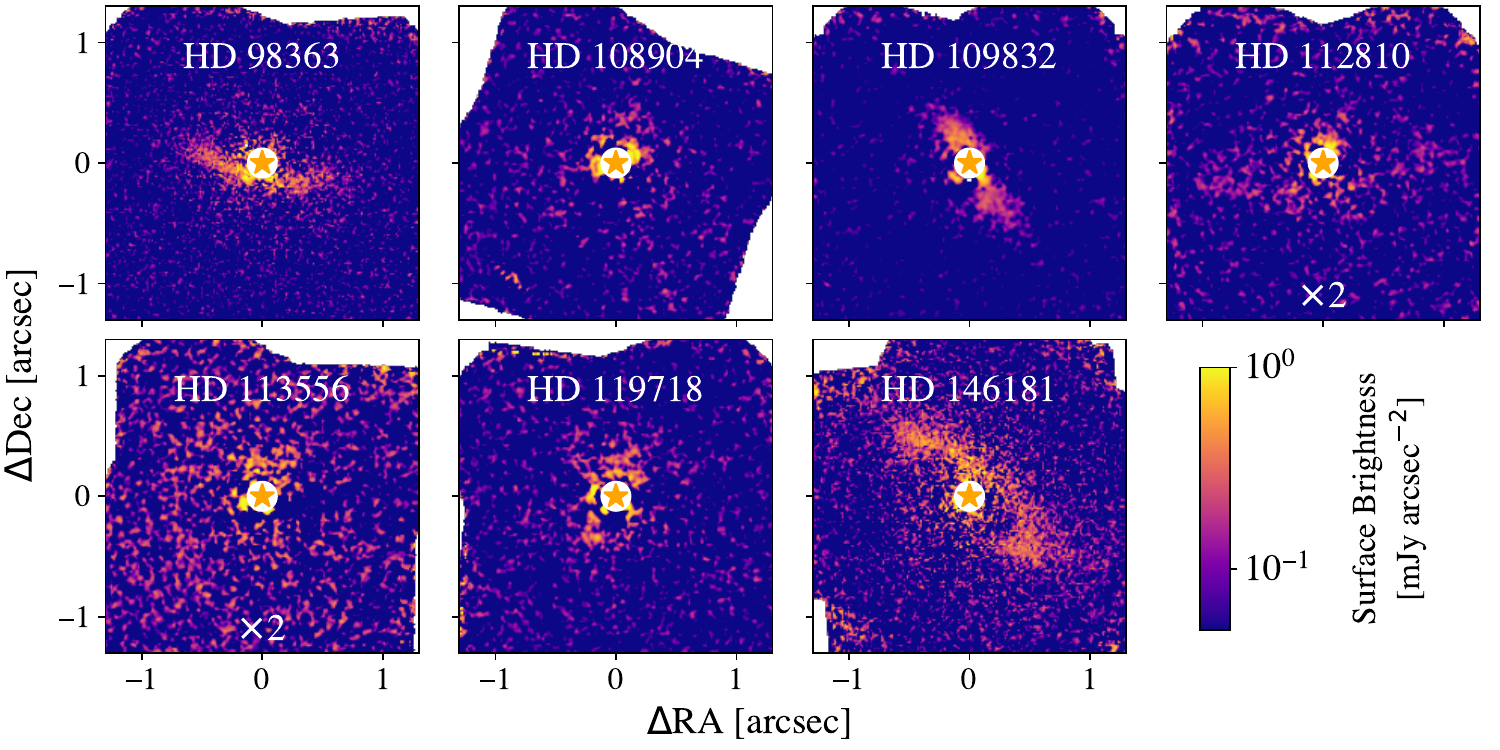}
    \caption{\Qr images of all sample targets, oriented with North pointing up and East pointing left, in flux units of mJy arcsec$^{-2}$. The white circle represents the extent of the FPM of GPI, 0$\farcs$12 in radius. To maintain a consistent colorbar for all images, the \Qr images of HD 112810 and HD 113556 are scaled up by a factor of two. HD 98363, HD 109832, and HD 146181 show the clearest detections. The HD 112810 image shows a lower S/N detection, and a hint of an arc is seen east of the star in the image of HD 113556. The detections in HD 112810 and HD 113556 become more obvious after smoothing with a Gaussian kernel, shown in Figure \ref{fig:gauss_smooth_disk}.}
    \label{fig:Qr_Images}
\end{figure*}

\begin{figure*}
    \centering
    \includegraphics[width=\textwidth]{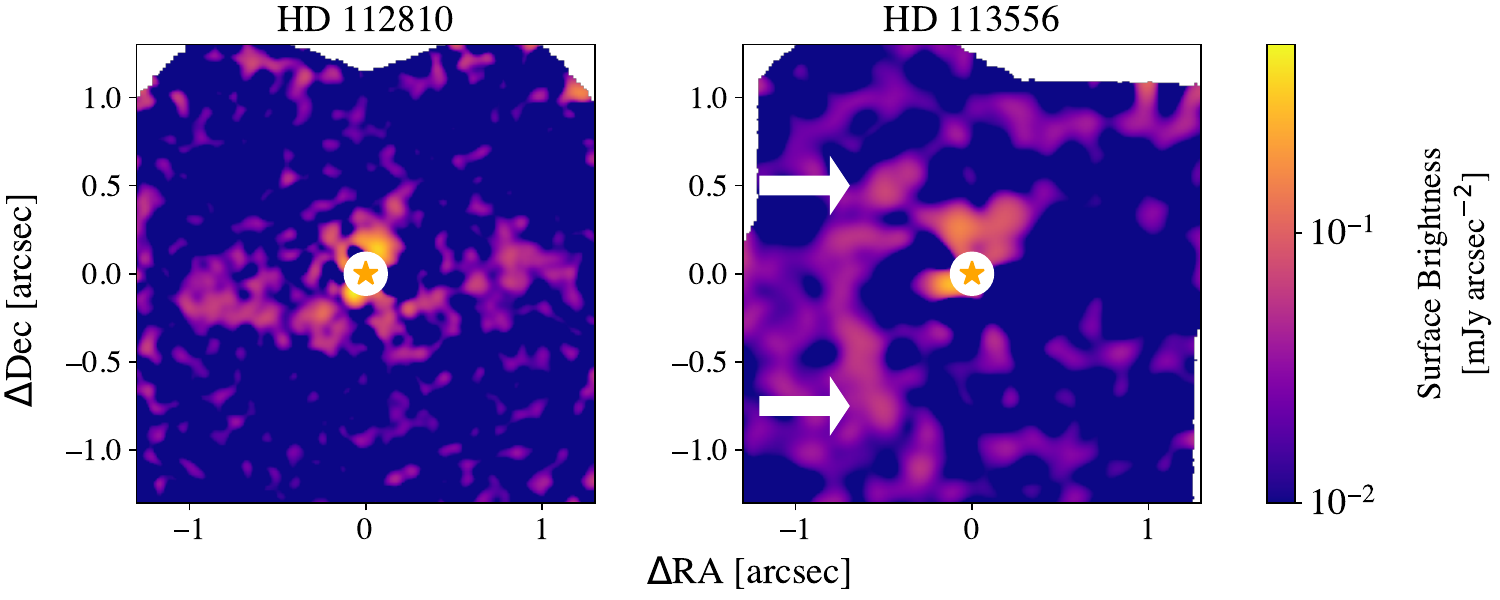}
    \caption{\Qr images of HD 112810 and HD 113556, smoothed with Gaussian kernels of FWHM$\sim2$ and FWHM$\sim4$ respectively. The Eastern arc (indicated by white arrows) seen in HD 113556 is more noticeable, but still faint.}
    \label{fig:gauss_smooth_disk}
\end{figure*}

Our reduced \Qr images are shown in Figure \ref{fig:Qr_Images}. High SNR detections are identified in HD 98363, HD 109832, and HD 146181, of which HD 109832 and HD 146181 are detected for the first time in scattered light. HD 112810, first reported in \cite{matthews2023} and detected in polarized intensity for the first time in this work, contains a much lower SNR detection, while a hint of an arc is seen in HD 113556. To demonstrate that we can detect HD 112810 and marginally detect HD 113556, we smoothed with Gaussian kernels of FWHM$\sim2$ and FWHM$\sim4$ pixels respectively, shown in Figure \ref{fig:gauss_smooth_disk}.

HD 98363, first reported in \cite{hom2020}, is an inclined and asymmetric/eccentric disk, reminiscent of eccentric structures seen in the debris disk around HD 106906 \citep{kalas2015,crotts2021}. HD 109832 appears as a close to edge-on, relatively compact system, with what appears to be a significant brightness asymmetry between the NE and SW sides. HD 146181 is an inclined ring, with what tentatively appears to be a moderate brightness asymmetry. HD 112810, first reported in \cite{matthews2023}, is a large, highly-inclined and faint ring. Finally, the Eastern arc seen in HD 113556 is suggestive of a more face-on geometry, but this cannot be definitively determined given the low SNR.

\subsection{Total Intensity Images} \label{sec:totimages}
\begin{figure*}
    \centering
    \includegraphics[width=\textwidth]{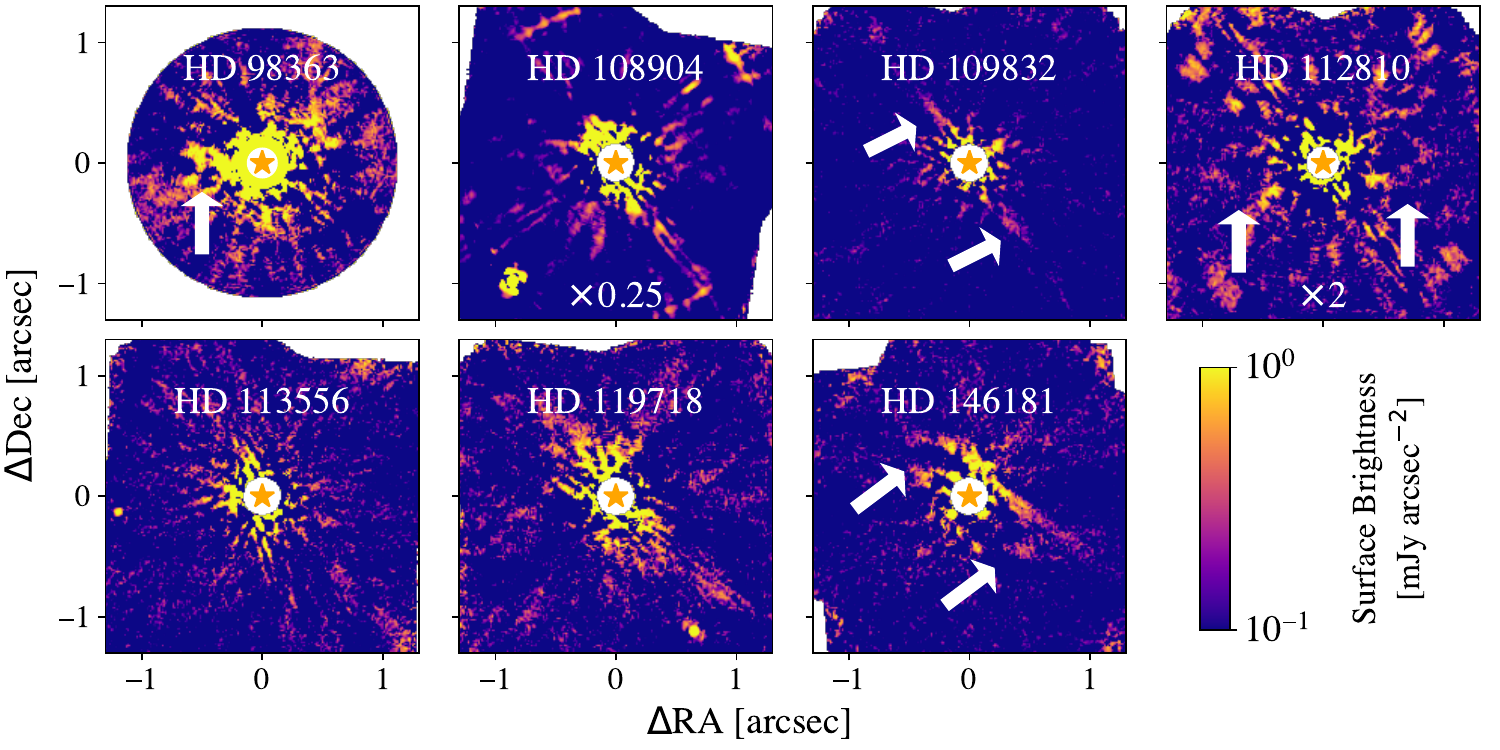}
    \caption{Total intensity images of all sample targets, in surface brightness units of mJy arcsec$^{-2}$. The HD 98363 image was produced using LOCI, while all other target images were produced using KLIP. As marked by the white arrows, all polarized intensity-detected systems have some level of detection in total intensity, with HD 109832 providing the clearest, highest SNR detection. As the images were processed with ADI, surface brightness is underestimated due to self-subtraction \citep{milli2012}. HD 108904, HD 113556, and HD 119718 show bright candidate companions, with the candidates identified in HD 108904 and HD 113556 confirmed to be background sources (see Appendix \ref{sec:vettingcompanions}).}
    \label{fig:totintimage}
\end{figure*}

HD 109832 is the only system with a marginally significant detection in total intensity, as shown in Figure \ref{fig:totintimage}. The HD 98363, HD 112810, and HD 146181 total intensity images contain structure co-located with the polarized intensity detections, but have very low S/N. The lower significance detections are unsurprising, as these systems have moderately-inclined geometries and more broad structures. As a result, they are more subject to the effects of PSF self-subtraction \citep{milli2012}. HD 109832 has the strongest total intensity detection, due to its closer to edge-on inclination compared to the other systems. The ``broken" ring structure shown in HD 112810 appears visually similar to single-epoch SPHERE-IRDIS observations presented in \cite{matthews2023}, despite a contrast degradation from poor FPM centering.

Due to the low SNR achieved in these reductions, they are not considered for the disk characterization analysis presented in \S \ref{sec:diskchar}. The effects of PSF self-subtraction would need to be accounted for with forward modeling \citep[e.g.,][]{mazoyer2020,hom2024}.

\subsection{Total Intensity Contrasts} \label{sec:contrasts_limits}
\begin{figure}
    \centering
    \includegraphics[width=\linewidth]{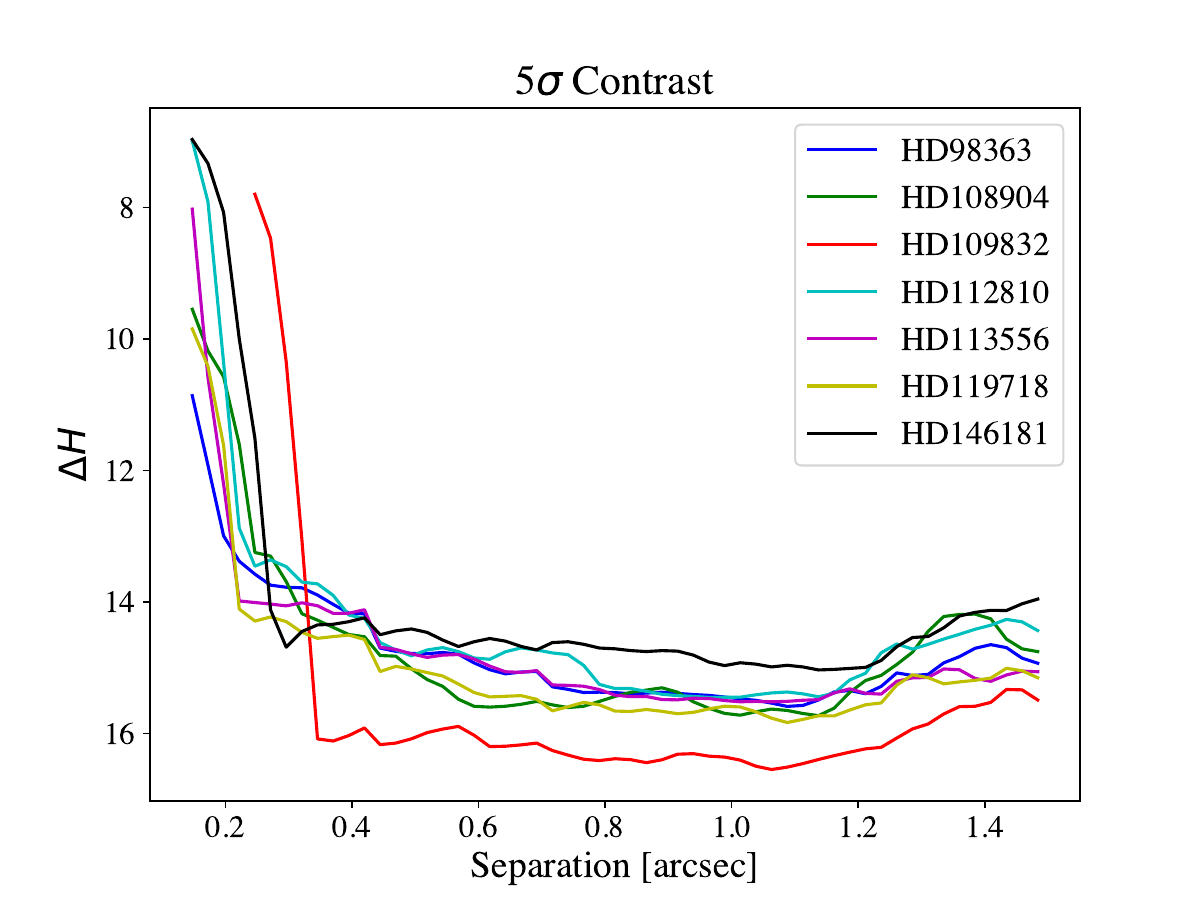}
    \caption{5$\sigma$ contrast measurements in $\Delta H$ magnitudes for all targets, calibrated against self-subtraction.}
    \label{fig:totintcontrasts}
\end{figure}

Per \cite{wang2015}, assuming azimuthally symmetric noise, pyKLIP calculates the 5$\sigma$ noise level at a range of radial separations throughout the image. To assess sensitivity to planets, 12 fake planets of known brightness are injected into the pyKLIP-reduced images. The brightnesses of the planets scale to the detection threshold at different radial separations. The images are passed through pyKLIP again and the flux of each injected planet is retrieved to calculate the final calibrated contrast curves. All contrast curves were calculated using a pyKLIP reduction using 30 KL modes and subdivided into a number of concentric annuli ranging from 9 to 100 pixels. In the case of HD 108904, HD 113556, and HD 119718, we mask bright candidate companions that would otherwise bias the 5$\sigma$ contrast measurement. Our final, throughput-calibrated 5$\sigma$ contrast curves are shown in Figure \ref{fig:totintcontrasts}, with all sequences achieving contrast levels typical of GPI performance \citep[e.g.,][]{macintosh2014,ruffio2017}.

\section{Disk Characterization} \label{sec:diskchar}
\subsection{Spine-Fitting Model Setup} \label{sec:spinemodel_setup}
We characterize the disk geometry through an empirical approach, utilizing the framework described in \cite{crotts2024}, where ellipses are fit along the disk spines. This approach assumes that a disk can be well represented by a narrow and vertically/radially thin ring, and that any vertical/radial dependencies for ``broader" systems will not affect the overall fit to the midplane of the vertical/radial profile. For the disks where the back side is detected/marginally detected (HD 112810 and HD 146181), we fit a Gaussian profile to the \Qr images along radial cuts spaced 2$\degr$ apart from each other, tracing the geometry between ranges of -180$\degr$ to 180$\degr$ relative to the disk PA, depending on how much of the back side of the disk is detected. At each radial cut, the Gaussian mean, FWHM, and offset location with respect to the host star are determined. For the compact systems (HD 98363 and HD 109832), we fit a Gaussian profile to the \Qr images along vertical cuts perpendicular to the midplane of the disk to derive the same properties. Uncertainties are derived from the \Ur images, where the standard deviation is calculated in 1 pixel wide annuli from the center. The final noisemap consists of the concentric annuli with each annulus set equal to its corresponding standard deviation in the \Ur image. In all cases, a simple ring model is fit to the Gaussian mean locations, with five parameters--the radius of the ring ($R_d$), the offset of ring center along the major axis ($dx$), the offset of ring center along the minor axis ($dy$), the inclination ($i$), and the position angle ($PA$)--fit using the affine-invariant Markov Chain Monte Carlo sampler \texttt{emcee} \citep{foreman-mackey2013} for 200 walkers and at least 2000 iterations. The initial parameters and prior ranges are described in Table \ref{tab:ringMCMCpriors}.

\begin{deluxetable*}{cccccc}




\tablecaption{MCMC initial parameters for spine-fitting analysis.}
\label{tab:ringMCMCpriors}


\tablehead{\colhead{Name} & \colhead{$R_d$} & \colhead{$dx$} & \colhead{$dy$} & \colhead{$i$} & \colhead{$PA$} \\ 
\colhead{} & \colhead{(au)} & \colhead{(au)} & \colhead{(au)} & \colhead{(deg)} & \colhead{(deg)} } 

\startdata
HD 98363 & 76 [27...83] & 0.0 [-14...14] & 0.0 [-14...14] & 75 [72...78] & 76 [73...79] \\
HD 109832 & 48 [21...54] & 0.0 [-21...21] & 0.0 [-21...21] & 83 [80...86] & 31 [28...34] \\
HD 112810 & 113 [93...134] & 0.0 [-27...27] & 0.0 [-27...27] & 73 [68...78] & 98 [93...103] \\
HD 146181 & 83 [65...90] & 0.0 [-13...13] & 0.0 [-13...13] & 66.0 [61...71] & 50.0 [45...55] \\
\enddata


\tablecomments{Prior ranges in square brackets. For $R_d$, $dx$, and $dy$, the sample parameter space is defined in arcseconds, with sizes determined from the star distance. We show these values in au for easier comparison to previous literature results. A positive $dx$ value is in the direction of West, while a positive $dy$ value corresponds to North. All priors are uniform.}


\end{deluxetable*}

\subsection{Modeling Results} \label{sec:model_results}
\begin{figure*}
    \centering
    \includegraphics[width=\textwidth]{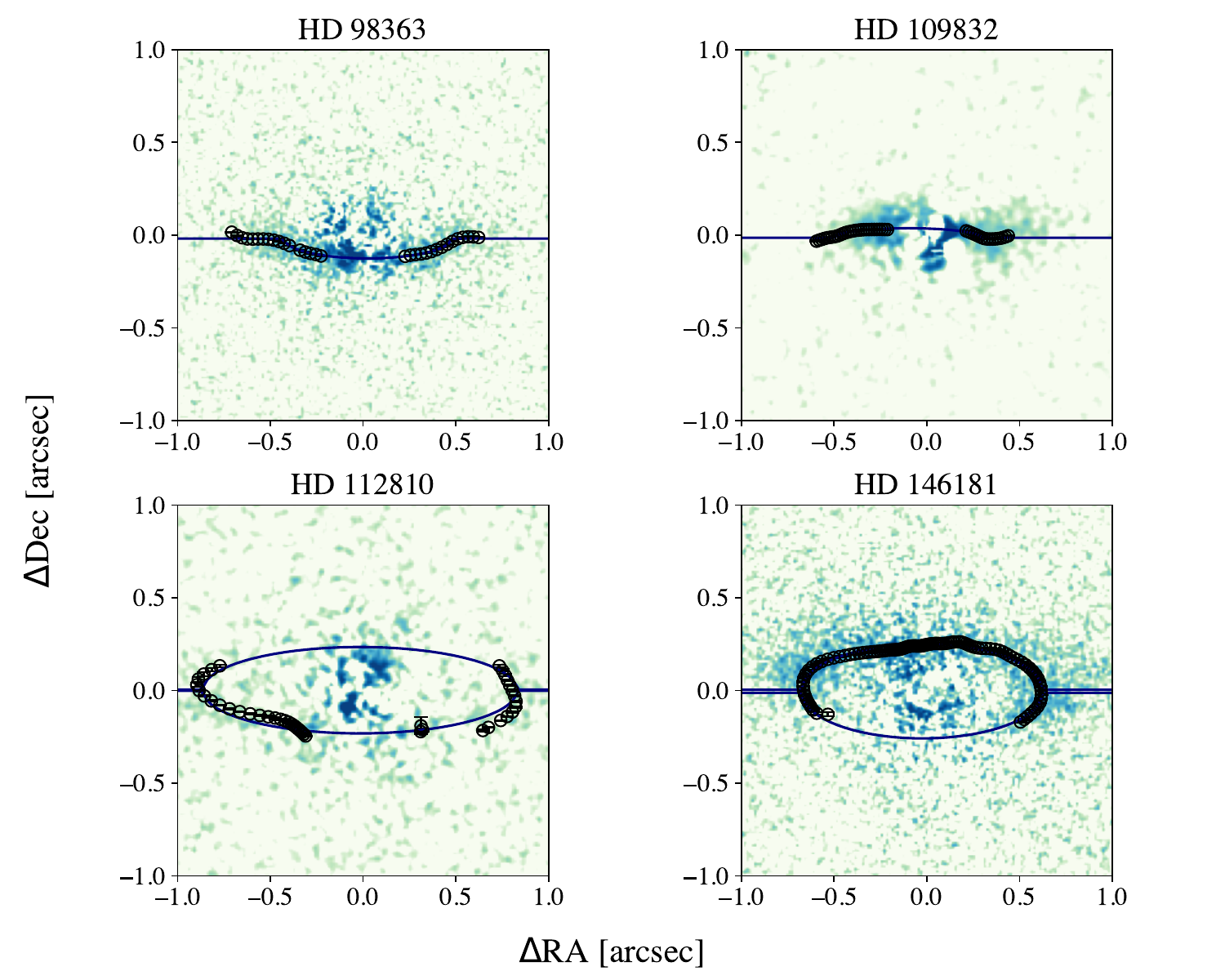}
    \caption{Best-fitting ellipse models for the spine-fitting analysis of all four detected systems overlaid on the disk images. Geometric asymmetries appear present to some degree in all systems, with the strongest asymmetries apparent in HD 98363 and HD 109832.}
    \label{fig:spinemodels}
\end{figure*}

\begin{figure*}
    \centering
    \includegraphics[width=\textwidth]{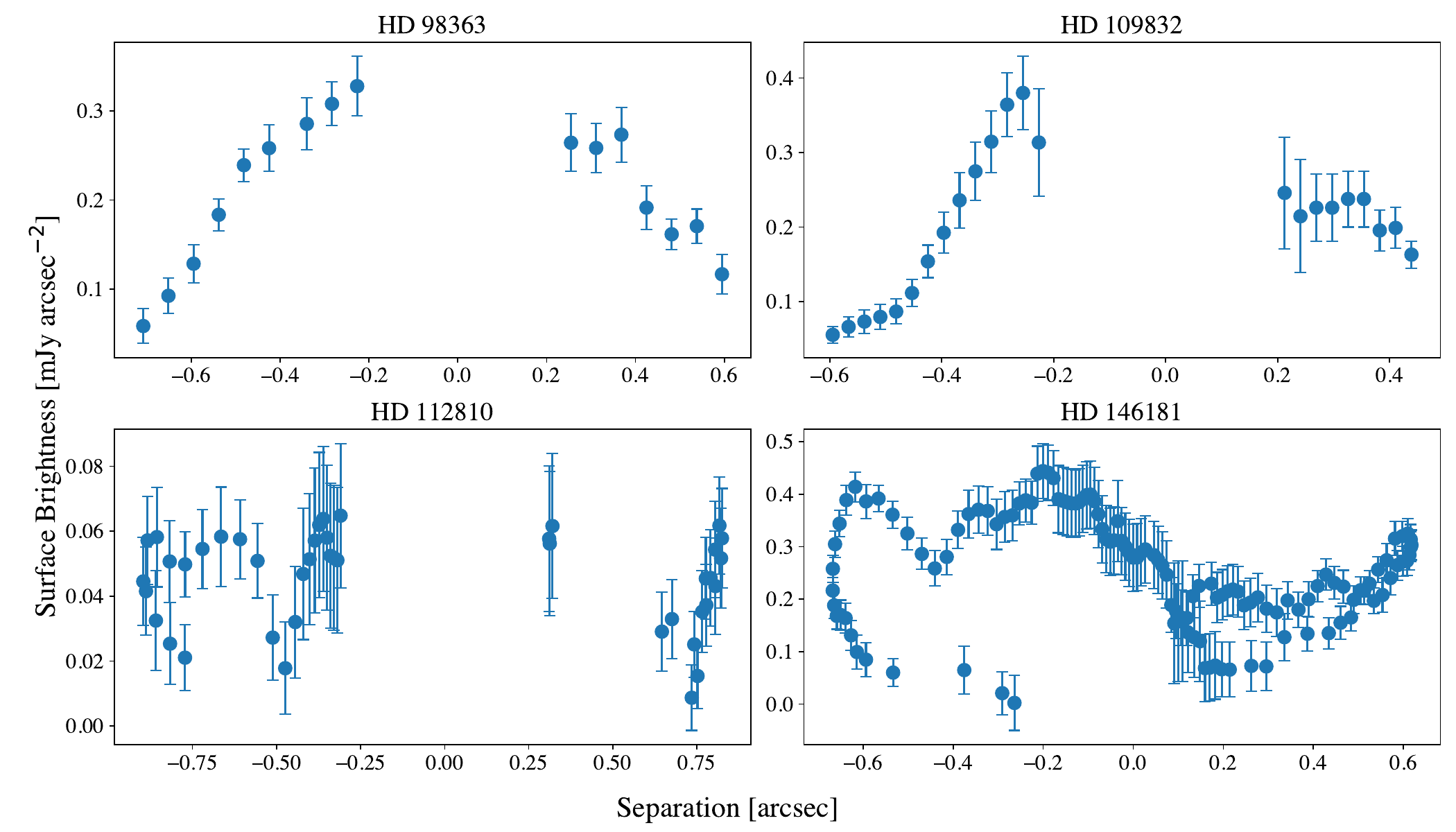}
    \caption{Measured surface brightness profiles for all four detected systems.}
    \label{fig:SBprofiles}
\end{figure*}

The median likelihood constrained parameters after excluding at least 200 iterations as burn-in with 3$\sigma$ uncertainties are presented in Table \ref{tab:ringMCMCposteriors} for the spine-fitting analysis, and the maximum likelihood models are overlaid with the disk images and corresponding FWHM points in Figure \ref{fig:spinemodels}. Given the relatively poor spine fits to the data, our derived uncertainties are expected to be underestimated, with the primary limitations being the low SNR of the data for HD 112810 and the irregular morphologies seen in HD 98363 and HD 109832. While the posterior distribution functions for HD 98363 are bimodal, the difference between families of models is not resolvable from the GPI data. The bimodality in the HD 109832 posteriors is likely due to the irregular morphology of the disk. This bimodality also leads to highly asymmetric uncertainties in the constrained parameters. For HD 112810, we are consistent within 1$\sigma$ for most parameters to the radial profile fit derived in \cite{matthews2023}, and within 2$\sigma$ of the other model fits presented in the same work. From our constrained parameters, we can also measure surface brightness profiles along the disk spine as constrained by the previous spine-fitting analysis. We follow the method explained in \cite{crotts2024}, where the reduced image is rotated by the best-fit $PA$ such that the disk is horizontal. We then bin the image into 2$\times$2 pixel bins to account for pixel-to-pixel correlation, and the derived Gaussian-fit mean from the spine-fitting analysis defines the y-coordinate of a 3$\times$3 (85$\times$85 mas) aperture. The mean of the aperture is then taken as the surface brightness for a given position. For the low-inclination systems HD 112810 and HD 146181, we measure the surface brightness along the disk by rotating between the same angles used in the spine-fitting analysis. For the high-inclination systems HD 98363 and HD 109832, we measure directly along the spine of the disk. The measured surface brightness profiles are shown in Figure \ref{fig:SBprofiles}, with uncertainties derived from placing apertures in the \Ur map.

From the derived Gaussian mean locations and surface brightness profiles, we can also derive a degree of asymmetry between the East and West sides of the disk, again following the approach described in \cite{crotts2024}. For HD 112810 and HD 146181, 3 square apertures are placed along both the East and West sides on the front side of the disk up until the ansae. For HD 98363 and HD 109832, a single rectangular aperture is placed on each side of the disk. The height of the apertures is given by the average of the derived FWHM values from fitting the previously described Gaussian profiles, and the width of the rectangular apertures for HD 98363 and HD 109832 are selected to include the highest disk S/N. We find asymmetry values of 1.03$\pm$0.10, 1.95$\pm$0.38, 1.13$\pm$0.19, and 1.31$\pm$0.11 for HD 98363, HD 109832, HD 112810, and HD 146181 respectively. HD 109832 and HD 146181 have significant brightness asymmetries, which are also evident in their measured surface brightness profiles.

\begin{deluxetable*}{cccccccc}




\tablecaption{Median likelihood constrained parameters for spine-fitting analysis.}
\label{tab:ringMCMCposteriors}


\tablehead{\colhead{Name} & \colhead{$R_d$} & \colhead{$dx$} & \colhead{$dy$} & \colhead{$i$} & \colhead{$PA$} & \colhead{$e$} & \colhead{$\chi^{2}_{\rm red}$} \\ 
\colhead{} & \colhead{(au)} & \colhead{(au)} & \colhead{(au)} & \colhead{(deg)} & \colhead{(deg)} & \colhead{} & \colhead{}} 

\startdata
HD 98363 & $62.43\substack{+1.42\\-0.86}$ & $3.32\substack{+1.10\\-0.49}$  & $-2.59\substack{+0.56\\-0.32}$ & $76.25\substack{+0.32\\-0.47}$ & $77.31\substack{+0.50\\-0.18}$ & $0.07\substack{+0.007\\-0.009}$ & 25.5 \\
HD 109832 & $42.55\substack{+11.10\\-0.57}$ & $-11.01\substack{+11.24\\-1.14}$ & $-1.54\substack{+0.13\\-2.34}$ & $82.28\substack{+0.30\\-0.35}$ & $29.66\substack{+0.33\\-2.60}$ & $0.26\substack{+0.024\\-0.195}$ & 9.6 \\
HD 112810 & $113.04\substack{+0.99\\-1.06}$ & $-2.26\substack{+0.92\\-0.82}$ & $-0.21\substack{+0.55\\-0.52}$ & $74.23\pm0.15$ & $97.50\substack{+0.31\\-0.30}$ & $0.02\pm0.007$ & 65.5 \\
HD 146181 & $82.67\substack{+0.43\\-0.30}$ & $-3.10\pm0.24$ & $0.45\substack{+0.19\\-0.32}$ & $67.56\substack{+0.25\\-0.27}$ & $52.31\substack{+0.22\\-0.34}$ & $0.04\pm0.003$ & 12.2 \\
\enddata


\tablecomments{3$\sigma$ uncertainties. Eccentricity is derived from sampling the posterior distributions of $dx$ and $dy$. The significant asymmetries in the error bars for median likelihood parameters for HD 109832 stem from the bimodality observed in its posterior distribution functions (see Appendix \ref{sec:cornerplots}).}


\end{deluxetable*}

\section{Discussion} \label{sec:discussion}
\subsection{Model Limitations}
Our disk characterization efforts are primarily limited by disk SNR, particularly for HD 112810. The impact of extended radial/vertical structure and inclination of the disks can limit our ability to robustly fit for some derived parameters such as offsets (see \citealt{crotts2024} for a detailed discussion). Similar to \cite{crotts2024}, we also find 1$\sigma$ uncertainties to likely be underestimated, especially given the large values of $\chi^{2}_{\rm red}$. Therefore, we choose to present 3$\sigma$ uncertainties. The simple circular ring morphologies assumed may also not be sufficient in truly capturing the eccentric structure seen in systems such as HD 98363. Regardless of these shortcomings, the constraints on the morphological results are sufficient for the empirical analyses and comparisons we conduct in the rest of this work.

\subsection{Notable Features of Sample Targets}
As noted in \cite{hom2020}, the HD 98363 has a morphological asymmetry, with the Eastern side appearing more extended than the Western side. This feature is reminiscent of the morphological asymmetry observed in HD 106906 \citep{kalas2015}, another debris disk in Scorpius-Centaurus with a companion orbiting exterior to the debris disk at a high mutual inclination \citep{bailey2014}. Dynamical modeling of the system has demonstrated that the interactions between HD 106906 b and the corresponding debris disk could explain the eccentric shape of the system \citep{nesvold2017}. This morphological asymmetry is further demonstrated in the constrained eccentricity from the spine-fitting analysis, suggesting an eccentricity greater than 0 with at least 5$\sigma$ confidence. Visually, the system also appears to have a kink on the Western side, which provides additional evidence that some dynamically perturbing body, e.g., a highly eccentric perturbing companion, may be present in the system \citep{lee2016}.

The spine-fitting analysis of HD 109832 may suggest asymmetry at $>1\sigma$ confidence, with the derived spine profile bearing visual similarities to the spine profile measured for HD 110058 in \cite{crotts2024}, another compact disk in Sco-Cen. The posterior distribution functions give two families of solutions, each with slightly different values of disk sizes, inclinations, and position angles. The profile provides tentative evidence for a warp, and the simple ring model is likely not sufficient for characterizing the system with high precision. Although the system lies within the image region with the highest noise, qualitative inspection of the detected total intensity ansae reveals a slight misalignment and provides tentative evidence that the warp is potentially real.

HD 112810 is the faintest debris disk detected by GPI in polarized intensity, with a measured surface brightness of $\sim$50 $\mu$Jy arcsec$^{-2}$ at the ansae. 
At the ansae, our polarized intensity surface brightness measurement of $\sim$50 $\mu$Jy arcsec$^{-2}$ appears commensurable with the total intensity scattered light image from \cite{matthews2023}. However, the actual polarization fraction remains uncertain, as the true value of the total intensity surface brightness in \cite{matthews2023} may be underestimated due to PSF self-subtraction.

The narrow ring geometry of HD 146181 may be indicative of enhanced collisional activity. The brightness asymmetry between the ansae may also indicate interesting dynamical interactions in the system, and asymmetric dust clumps could suggest influence from one or multiple perturbing bodies in resonance \citep[e.g.,][]{krivov2007,stuber2023}. The system also shares a similar inclination to the HR 4796A debris disk \citep[e.g.,][]{schneider1999} but is much broader radially. The scattering properties of HR 4796A are also considered unusual compared to other debris disks \citep{milli2017}, and HD 146181 may therefore be a valuable comparison against HR 4796A for understanding the dominant grain transport mechanisms in both systems.

The properties of these four systems add to the diversity of planetary systems in Scorpius-Centaurus, further highlighting Sco-Cen as a valuable region in which to investigate young planetary system formation and evolution.

\subsection{Morphological Properties Inferred from Scattered Light and Thermal Emission}
A key finding from previous studies has shown that most resolved debris disk radii are typically much larger than inferred radii from SEDs by a factor of $\sim$3 \citep{rodriguez2012,booth2013,morales2016,esposito2020,hom2020}. We also find that the model-fit scattered light radii are larger than the inferred blackbody radii, consistent with similar findings by \cite{cotten2016}, with the ratio of $R_d$ to $R_{bb}$ being 5.05, 1.49, 2.31, and 1.78 for HD 98363, HD 109832, HD 112810, and HD 146181 respectively, adding further evidence that SED-inferred blackbody radii generally underestimate radii determined from resolved observations. \cite{pawellek2015} defined a blackbody radius correction factor $\Gamma$ that is dependent on the luminosity of the host star and coefficients dependent on the dust grain composition, defined as
\begin{equation}
    \Gamma = A(L/L_{\odot})^{B}.
    \label{eq:pawellekgamma}
\end{equation}

If we apply the $\Gamma$ correction factor for our systems with the various combinations of A and B from Table 4 in \cite{pawellek2015}, we find that the derived radii are only consistent for HD 146181, with all compositions providing consistent radii within 1$\sigma$ of our model-constrained radius of $82.67\substack{+0.43\\-0.30}$. The $\Gamma$-corrected $R_{bb}$ for HD 98363 underestimates our model derived $R_d$ by a factor of $\sim2$, while the $\Gamma$-corrected radii for both HD 109832 and HD 112810 overestimate our model derived radii by a factor of $\sim$2.

All four of our systems have also been observed in mm continuum emission. \cite{lieman-sifry2016} resolved HD 112810 along the major axis of the disk, and from fitting the visibility, derive an upper limit of 28 au to the inner radius of the disk and an outer radius of $130\substack{+80\\-70}$au. Their broad profile is in agreement with our own spine-fitting radius, and a detailed analysis in \cite{matthews2023} also finds agreement between the two-temperature SED fit of the system from \cite{chen2014} after applying the \cite{pawellek2015} correction factor. HD 146181 was also only resolved along the major axis in \cite{lieman-sifry2016}, with a visibility-fit inner radius of $73\substack{+14\\-19}$ au and radial width $<50$ au. Our modeling result is in good agreement with \cite{lieman-sifry2016}. While \cite{moor2017} observed HD 98363 and HD 109832 in 1.33mm continuum emission, HD 98363 was unresolved and only an upper limit on flux was reported for HD 109832.

\subsection{Scattered Light Flux Ratio} \label{sec:scatteredlightfluxratio}
\begin{figure*}
    \centering
    \includegraphics[width=\linewidth]{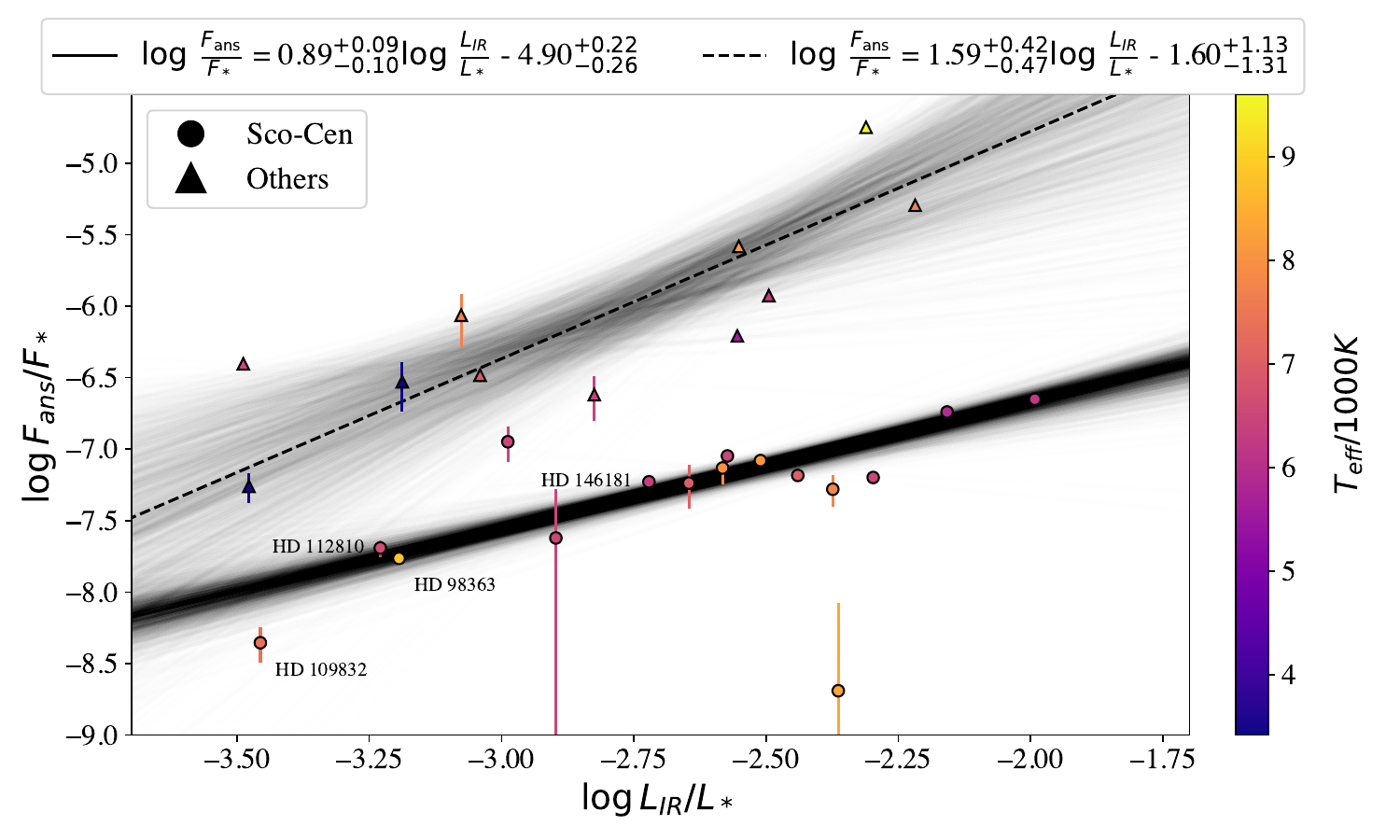}
    \caption{Measured $F_{ans}/F_*$ as a function of system infrared excess $L_{IR}/L_*$ and stellar $T_{eff}$. The black lines represent the median likelihood solutions with 1$\sigma$ uncertainty to separate linear bootstrapping fits (gray lines). The non-Sco-Cen sample (triangles and dashed line) shows considerably higher $F_{ans}/F_*$ overall compared to Sco-Cen systems (circles and solid line) at similar levels of $L_{IR}/L_*$. }
    \label{fig:ansaeVsexcess}
\end{figure*}

To place our newly detected systems in the broader context of GPI-imaged Sco-Cen and non-Sco-Cen disks, we measure a scattered light flux ratio (hereafter $F_{ans}/F_*$) and compare against various other system properties, including $L_{IR}/L_*$ and $T_{eff}$. We specifically measure the scattered light flux ratio at the ansae, as it is the least likely to be influenced by significant noise if the front side of the disk lies close to or within the FPM. Additionally, scattering flux at the ansae should correspond to a scattering angle of $\sim90\degr$, regardless of the system inclination, allowing for a consistent comparison. Ansae surface brightness measurements for the systems not highlighted in this work are collected from \textit{H}-band surface brightness profiles of the GPI polarimetric disk sample presented in \cite{crotts2024}. Utilizing the radii determined from the spine-fitting analyses conducted in this work and in \cite{crotts2024}, we determine the angular separation of the ansae and average the adjacent surface brightness measurement values. To determine $F_{ans}/F_*$, we multiply the ansae surface brightness measurement in mJy arcsec$^{-2}$ by the squared pixel scale of GPI (14.16 mas)$^2$ to get the flux. We then divide that by the stellar $H$-band flux in mJy. As the intensity of light falls off as $\propto \frac{1}{R^2}$, we normalize each ratio by multiplying by $(\frac{R_d}{100 au})^2$. $L_{IR}/L_*$ measurements are collated from \cite{esposito2020}. We set a 10\% floor for all $F_{ans}/F_*$ uncertainties to account for any minor differences in flux calibration and measurement methodologies. Furthermore, we divide the sample into two groups: Sco-Cen objects and non-Sco-Cen objects. Our results are shown in Figure \ref{fig:ansaeVsexcess}. While the range of IR excess appears consistent between the two samples, there is a clear division in the measured $F_{ans}/F_*$, with Sco-Cen systems showing a much lower $F_{ans}/F_*$ overall compared to the other systems. From our measurements, we calculate the Pearson correlation coefficients for the Sco-Cen and non-Sco-Cen systems separately in log-log space. We find $r = 0.51$ and $r = 0.82$ for the Sco-Cen and non-Sco-Cen systems respectively. The lower $r$ for Sco-Cen systems is likely driven by the low $F_{ans}/F_*$ outlier. We can also perform a two-sample Kolmogorov-Smirnov (K-S) test between the Sco-Cen and non-Sco-Cen IR excess and $F_{ans}/F_*$ samples to assess whether they could be derived from the same parent population. From the combined IR excess measurements alone, we determine a p-value of 0.628, meaning that we cannot reject the null hypothesis that the measurements come from the same distribution. A K-S test of the combined ansae flux ratio values, on the other hand, gives a p-value of 4.14$\times 10^{-6}$, meaning that the null hypothesis can be rejected. Finally, we perform linear bootstrap fitting with 2000 iterations in log-log space for the non-Sco-Cen and Sco-Cen samples separately, assuming that the uncertainties in the IR excess values are 10\%. The solid and dashed lines in Figure \ref{fig:ansaeVsexcess} are the median likelihood lines and associated 1$\sigma$ uncertainties from the bootstrap fitting. Most notably, the best-fit intercepts (i.e., the offsets between lines) are significantly different. All of these results imply that the IR excess levels between the Sco-Cen and non-Sco-Cen systems span similar ranges while $F_{ans}/F_*$ does not.

\begin{figure*}
    \centering
    \includegraphics[width=\linewidth]{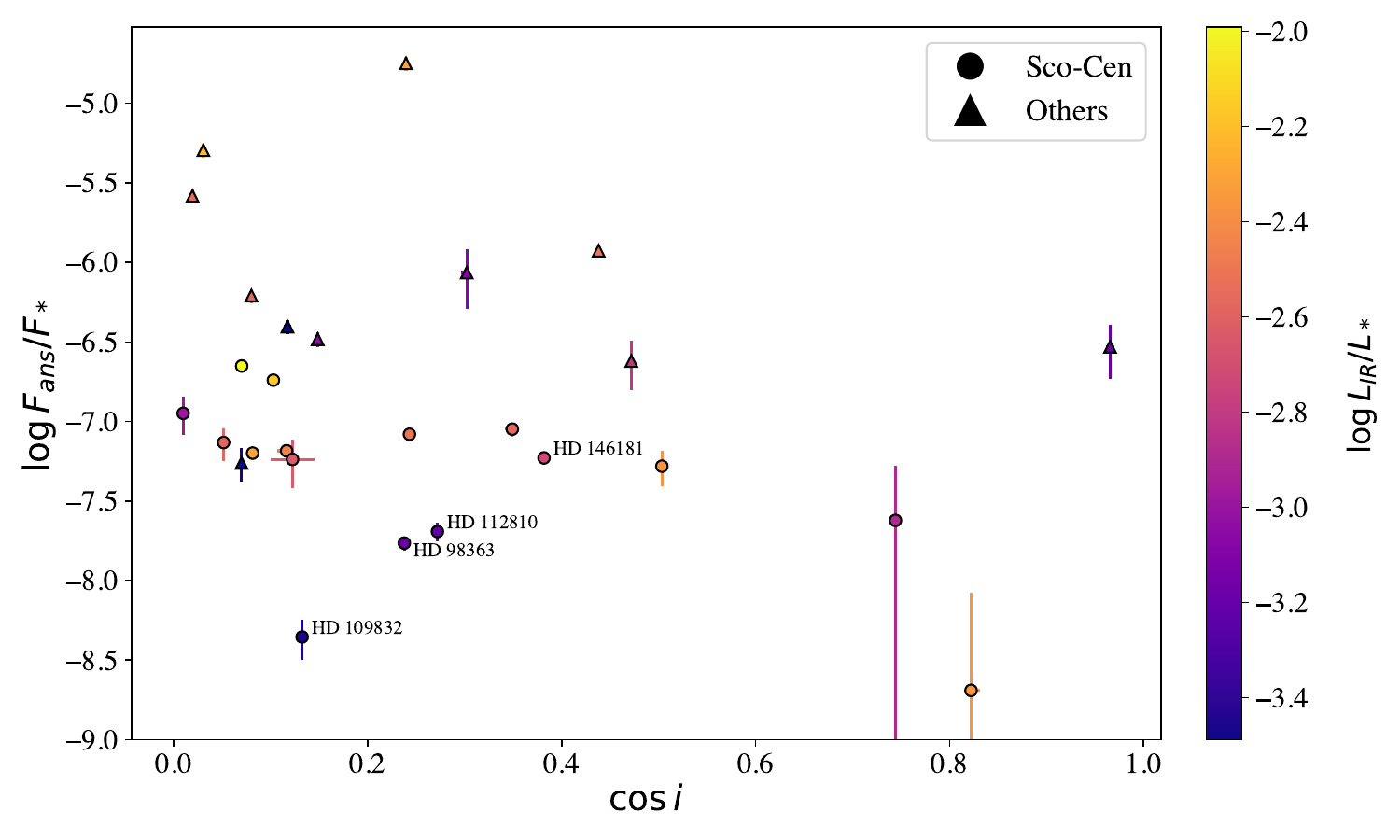}
    \caption{$F_{ans}/F_*$ as a function of system inclination and $L_{IR}/L_*$. Similar ranges in inclination are observed between both the Sco-Cen and non-Sco-Cen samples, also indicated by the K-S test p-value. Overall, this suggests that inclination and limb brightening is not a significant factor in accentuating the differences in $F_{ans}/F_*$ between Sco-Cen and non-Sco-Cen objects.}
    \label{fig:incvsansae}
\end{figure*}

Given that debris disks are preferentially forward scattering, this conclusion may be biased by individual disk inclination, as a higher inclination would increase detectability of a system. To assess the impact of inclination on $F_{ans}/F_*$, we plot $F_{ans}/F_*$ as a function of system inclination, shown in Figure \ref{fig:incvsansae}. A similar range of inclinations are present in both samples, and a K-S test of inclinations gives a p-value of 0.96, suggesting that both samples stem from similar inclination populations. The division between Sco-Cen and other systems, however, is still apparent in $F_{ans}/F_*$ despite having similar values of $L_{IR}/L_*$.

Inclination may also bias measurements of $F_{ans}/F_*$, where higher inclination systems may be subject to the effects of limb brightening. A weak negative correlation of $F_{ans}/F_*$ and $\cos i$ is observed among the sample. In Appendix \ref{sec:limb_brightening}, we generate a range of identical disk models with different inclinations to test how $F_{ans}/F_*$ may correlate with inclination. While a negative trend is observed, the slope is shallower than the slope observed in Figure \ref{fig:incvsansae}, spanning less than an order of magnitude. While the test relies on the assumption of a given scattering phase function, the measurement at the ansae should still represent the surface brightness at 90$\degr$ scattering angle. Regardless, the GPIES and DISCS samples are ultimately biased towards the highest inclination systems, likely an effect of the preferential forward scattering nature of dust grains, and an appropriate sample of low inclination disks is needed to fully understand if limb brightening is a significant biasing factor.

The difference in $F_{ans}/F_*$ between Sco-Cen systems and the other debris disks is somewhat unexpected. Measuring the flux ratio at the ansae should control against differences in system inclination and distance, and normalizing against the disk radius would eliminate any dependence on the decrease in intensity as a function of disk size. In the GPI disks sample, Sco-Cen is generally much farther away than other systems with an average distance of $\sim$118 pc compared to other GPI-imaged debris disks (average distance$\sim$60 pc). For two disks with identical surface density distributions and viewing inclinations, a greater amount of disk surface area is captured per unit solid angle for a far system compared to a more nearby system, leading to more distant disks having comparable surface brightness by factors less than an order of magnitude and a distant system having a higher $F_{ans}/F_*$ compared to a nearby system (see also Appendix \ref{sec:distance_appendix}). Nevertheless, we observe lower $F_{ans}/F_*$ by as much as 2-3 orders of magnitude in the high IR excess regime ($\log L_{IR}/L_* \sim -2$). While a positive correlation between the ansae flux ratio and the IR excess is naively expected, the offset between the non-Sco-Cen and Sco-Cen samples is surprising. The similar ranges in IR excess but distinct ranges in $F_{ans}/F_*$ between both the non-Sco-Cen and Sco-Cen samples suggest that while the range in total mass of large, $\gtrsim 5 \mu$m-sized dust grains is similar between both samples, the range in total mass of small, (sub)micron-sized dust grains dust are not. We investigate other trends with $F_{ans}/F_*$ in order to postulate a few possible explanations for this unexpected behavior. The measured $F_{ans}/F_*$ and all values used in the analysis for \S \ref{sec:obsbiases}-\ref{sec:dynamical} are shown in Table \ref{tab:sect7props}.

\subsubsection{Observational Biases} \label{sec:obsbiases}
One of the key criteria for defining the GPIES polarimetric disk survey sample in \cite{esposito2020} was previously resolved scattered light detections. Of the detected GPIES disk sample, 16 out of 24 detections satisfied this criteria and include the entirety of the non-Sco-Cen sample except for HR 7012. The non-Sco-Cen sample is therefore likely biased towards brighter scattered light debris disks, such as HD 32297, $\beta$ Pic, and AU Mic. The \cite{esposito2020} sample included many disks previously detected with HST/STIS and HST/NICMOS, and not all of these systems were ultimately detected with GPI. The detected non-Sco-Cen sample is therefore likely not representative of the non-Sco-Cen system $F_{ans}/F_*$ in general. This is further evident from the 20 non-detections (40\% detection rate) of non-Sco-Cen systems that span the same range of \Lir as the detected non-Sco-Cen sample, and \cite{esposito2020} notes that a handful of nondetected, previously-resolved systems are known to have intrinsically low scattered light brightness likely below the GPI detection limit, including for systems where \Lir $\sim10^{-3}$. While the impact of disk inclination must still be accounted for, \cite{esposito2020} was not able to detect four Sco-Cen and non-Sco-Cen systems with \Lir $>3\times10^{-4}$ and $i > 70^{\circ}$ and four systems where \Lir $\sim10^{-3}$ and $i<40^{\circ}$, even though the detected sample spans these same properties.

However, apart from preferentially higher $L_{IR}/L_*$, the DISCS sample itself is less likely to be biased (81\% detection rate, 60\% of systems observed), as the measured $F_{ans}/F_*$ probe down to the GPI detection limit in addition to not all being ``hand-picked" from knowledge of previous resolved scattered light detections. In combination with the common age and formation environment of Sco-Cen stars, the Sco-Cen imaged debris disks could be treated as benchmarks against which to compare other scattered light-imaged debris disks with spectral types earlier than mid-K. This is further exemplified by the tightness of the correlation and small standard deviations in the best-fit line parameters in $L_{IR}/L_*$ and $F_{ans}/F_*$ for Sco-Cen systems, with the only significant outlier having $\log F_{ans}/F_* \lesssim -8.5$ at $\log L_{IR}/L_* \sim -2.3$. This measurement is from HD 156623, a system with properties more akin to transition/hybrid disks rather than collision-dominated debris disks (see \citealt{lewis2024} for a detailed discussion).

While observational bias is a likely explanation for the discrepancy between non-Sco-Cen and Sco-Cen systems, it does not provide an explanation for the 1-2 orders of magnitude in difference in $F_{ans}/F_*$ between the two samples at similar $L_{IR}/L_*$, and additional factors must be investigated.

\subsubsection{Intrinsic Grain Properties} 
Differing grain compositions, grain size distributions, porosities, and other intrinsic grain properties affect the scattering efficiencies and phase functions of dust grains \citep[e.g.,][]{duchene2020,arriaga2020,milli2015}. Differences in intrinsic dust grain properties could therefore affect $F_{ans}/F_*$, although such an interpretation is difficult to confirm without constraining the properties in radiative transfer modeling. Several studies often encounter difficulties in constraining such properties, largely due to the simplicity of dust grain models \citep[e.g.,][]{duchene2020,arriaga2020}. Furthermore, from an evolutionary perspective, it is unclear why certain grain properties would differ between Sco-Cen and other systems. One system where intrinsic grain properties may explain its scattered light brightness discrepancy is HR 4796A, where its measured scattering phase function appears distinct compared to many dust scattering phase functions \citep{milli2019}.

\subsubsection{Age} 
Sco-Cen is generally younger than other observed debris disk systems, with ages ranging from 10-20 Myr compared to the mostly older GPI debris disk sample as a whole (ranging from 25--500 Myr). This age is in a range where debris disks are expected to be at their brightest \citep{wyatt2008}, but relative to their host star, the Sco-Cen sample is consistently fainter. Overall, we expect debris disks to fade in scattered light brightness over time, as radiation pressure ejects the smallest dust grains from a system. Interestingly, we see the opposite, where the non-Sco-Cen sample of debris disks that are older have higher $F_{ans}/F_*$. The two younger non-Sco-Cen debris disks (CE Ant and HR 4796A) also have higher $F_{ans}/F_*$ than Sco-Cen systems, further complicating any trends that may be seen as a function of age.

\subsubsection{Radiation Environment} \label{sec:fluxratio-radiation}
\begin{figure*}
    \centering
    \includegraphics[width=\linewidth]{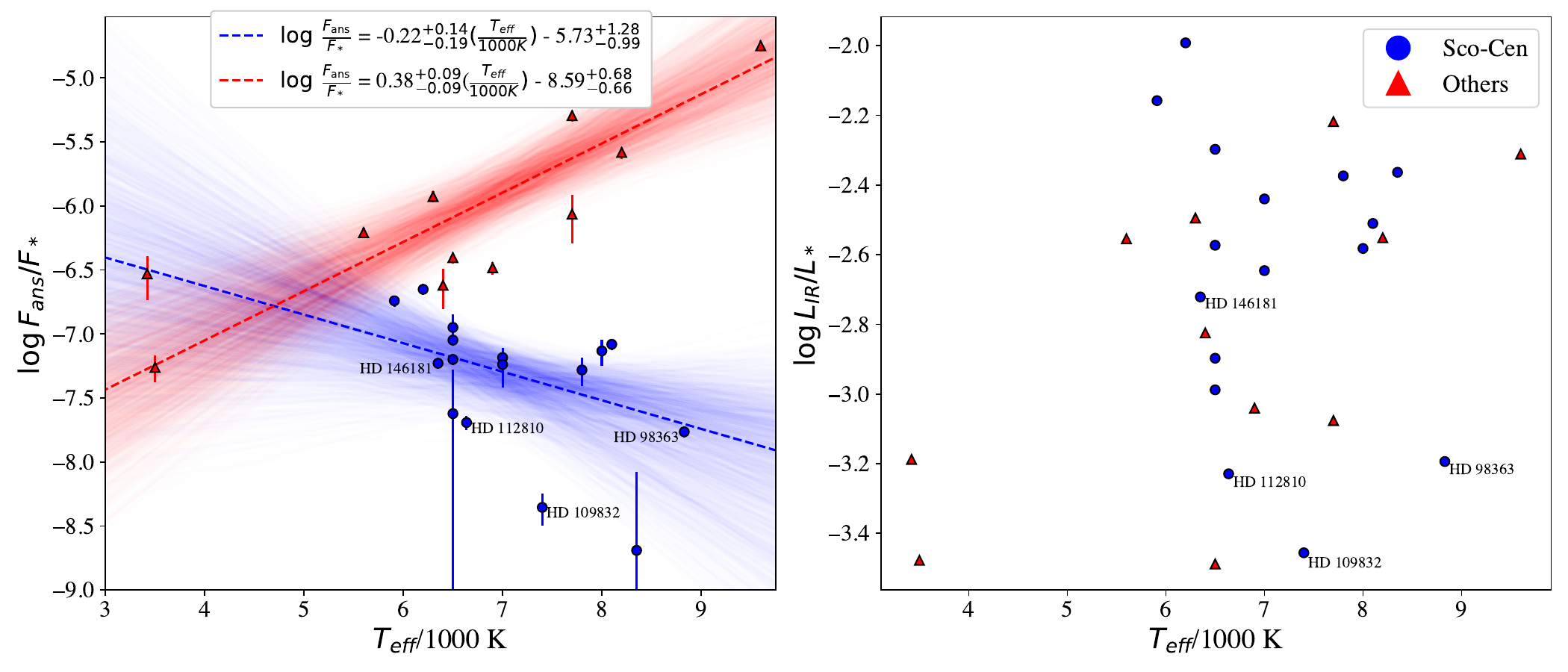}
    \caption{$F_{ans}/F_*$ (\textit{left}) and IR excess (\textit{right}) as a function of host star $T_{eff}$. The two dashed lines in the left panel represent the median likelihood line parameters with 1$\sigma$ uncertainties from separate linear bootstrapping fits (in light blue and red). No correlation appears present between $T_{eff}$ and $L_{IR}/L_*$, but a positive correlation is seen between $T_{eff}$ and $F_{ans}/F_*$ for non-Sco-Cen systems. Curiously, the division between Sco-cen and non-Sco-Cen systems is even more apparent, with a negative correlation for Sco-Cen targets compared to the rest of the GPI debris disk sample.}
    \label{fig:AnsaeVTeffVExcess}
\end{figure*}

Radiation environments have significant effects on the transport of dust grains. To identify the relationship between dust brightness in both scattered light and thermal emission, we plot both $F_{ans}/F_*$ and IR excess as a function of host star effective temperature in Figure \ref{fig:AnsaeVTeffVExcess}. As expected for medium-large grains, no correlation is seen between IR excess and host star $T_{eff}$, as the grains contributing to the IR excess are well above the blowout size and are not strongly affected by radiation pressure. When we compare Sco-Cen and non-Sco-Cen $F_{ans}/F_*$, we find a negative linear correlation in Sco-Cen systems (Pearson $r=-0.54$, p-value of 0.03), while maintaining a positive correlation in the other systems (Pearson $r=0.84$, p-value of 0.001). This is also affirmed from the linear bootstrapping fit applied to both samples separately. In the Sco-Cen systems, this could suggest that the increased luminosity around earlier-type stars is effectively blowing out the smallest dust grains, leading to a decreased scattered light brightness. The positive correlation seen in the non-Sco-Cen systems could suggest that the small dust grains contributing to scattering are more numerous relative to later-type non-Sco-Cen systems and similar-type Sco-Cen systems. In combination with the relatively older age (with the exception of HR 4796A having the second highest $F_{ans}/F_*$ in the left plot of Figure \ref{fig:AnsaeVTeffVExcess}), collisional mechanisms could be replenishing small dust grains in these systems at an increased rate compared to Sco-Cen systems. Ultimately, both the Sco-Cen and non-Sco-Cen systems lack sufficient sampling around later spectral types such as G-, K-, and M-dwarfs. Such debris disks have rarely been resolved in imaging relative to A- and F-stars \citep[e.g.,][]{Cronin-Coltsmann2023}. \cite{jang-condell2015} found that most IR excess measurements of Upper Scorpius members were consistent with T Tauri/protoplanetary disks. This is exemplified in an ALMA survey of later-type Upper Scorpius members \citep{carpenter2025}, which only identified $\sim$10\% of M-dwarfs in US as debris disk systems. In total, seven Sco-Cen G/K systems with excesses consistent with debris disks and \Lir $>2.5\times10^{-4}$ are missing from the sample due to faint primary star magnitude. The successful detection of debris disks around these systems would provide valuable missing information to confirm the trends seen in Figure \ref{fig:AnsaeVTeffVExcess}.

\subsubsection{Gas}
\begin{figure*}
    \centering
    \includegraphics[width=\linewidth]{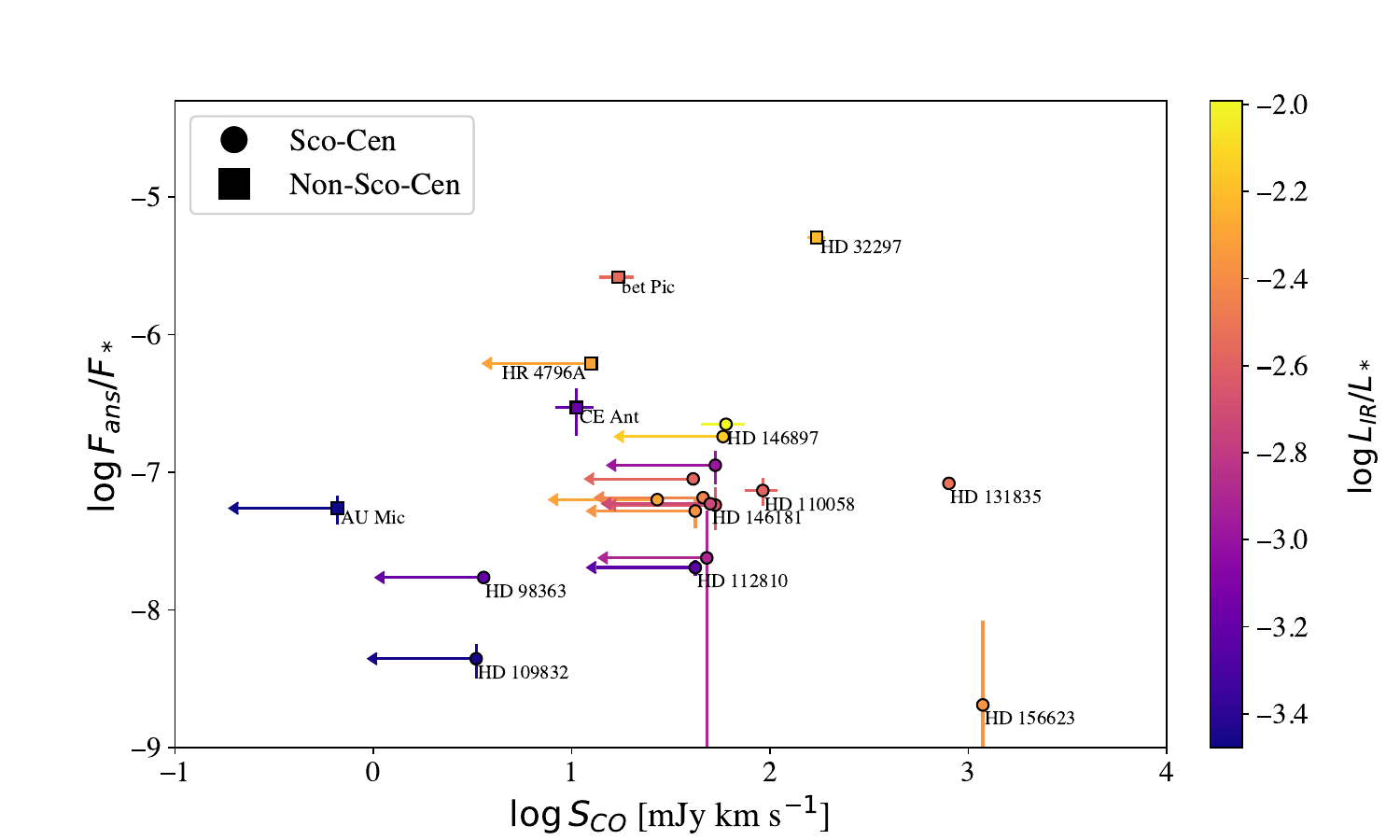}
    \caption{$F_{ans}/F_*$ as a function of $S_{CO}$ (normalized by distance). The systems highlighted in this work (HD 98363, HD 109832, HD 112810, and HD 146181) do not have confirmed detections of gas. The only systems with confirmed CO gas detections are $\beta$ Pic, CE Ant, HD 32297, HD 110058, HD 131835, HD 146897, and HD 156623, with all other measurements being 3$\sigma$ upper limits. Due to the small sample size, it is difficult to extrapolate on how $F_{ans}/F_*$ may correlate with the presence of gas. While a negative trend appears among the detected sample, the upper limits for other systems complicate this observation.}
    \label{fig:ansaeGas}
\end{figure*}

While sufficient radiation pressure can blow out the smallest dust grains of a system, the presence of gas can dynamically influence grains in different ways \citep{takeuchi2001}. \cite{takeuchi2001} has shown that the interplay between radiative and drag forces in a gas-rich debris disk can lead to extended morphologies rather than narrow rings for grains $\lesssim10\mu m$. For systems with similar levels of IR excess, this could suggest that a gas-rich disk could have a broader radial structure than a gas-poor disk, spreading the distribution of small dust grains over a larger area, decreasing the surface brightness per unit solid angle and attenuating $F_{ans}/F_*$ as a result. We investigate trends with gas by collating $^{12}CO(2\rightarrow1)$ measurements from the literature\footnote{The measurement of CO gas from CE Ant is the ($3\rightarrow2$) transition from \cite{matra2019_AUMic}. While this is not equivalent to the $^{12}CO(2\rightarrow1)$ flux, the difference is not expected to be greater than a factor of $\sim 2$ \citep{pericaud2017}.}. Sco-Cen measurements and 3$\sigma$ upper limits are sourced from \cite{lieman-sifry2016}, \cite{moor2017}, and \cite{fehr2022}. The only non-Sco-Cen systems with $F_{ans}/F_*$ measurements and gas measurements and upper limits are AU Mic \citep{daley2019}, $\beta$ Pic \citep{matra2017}, CE Ant \citep{matra2019_AUMic}, HD 32297 \citep{macgregor2018}, and HR 4796A \citep{kennedy2018}. In Figure \ref{fig:ansaeGas}, we plot $\log F_{ans}/F_*$ versus $\log S_{CO}$ in mJy km s$^{-1}$ normalized by system distance $(d^2/100pc)$. Trends are difficult to extrapolate because of the small sample of systems with significant gas detections. Among the systems with significant detections, there appears to be a tentative negative trend, but the detection upper limits for the other systems plotted weaken this interpretation. Regardless, the broad radial distribution of dust grains as a result of gas drag may be exemplified in HD 156623 \citep{lewis2024}, where the inner edge of the disk is not resolved in GPI polarimetric imaging. The system also has the largest CO flux and lowest $F_{ans}/F_*$ of the DISCS and GPIES samples.

\subsubsection{Enhanced Dynamical Activity} \label{sec:dynamical}
Dynamical environments may play a role in determining dust surface density distributions and steeper grain size distributions, where brighter systems could indicate a higher rate of collisional activity, e.g. driven by planet-disk interactions \citep{lee2016} that trigger a collisional cascade and skew the grain size distribution towards smaller grain sizes. Such processes are needed to explain the brightness of resolved scattered light debris disk systems as a way of replenishing small dust grains in an evolved system in general. If we consider Sco-Cen systems as benchmarks for scattered light brightness at a given age, then the non-Sco-Cen debris disks detected with GPI may have anomalously enhanced collisional activity compared to typical Sco-Cen debris disks. This could explain why some systems are bright enough to be resolved from ground-based imaging despite their older ages (e.g. HD 157587; 500$\pm$335 Myr). In fact, systems with the highest $F_{ans}/F_*$ measurements include HD 32297 and $\beta$ Pic, systems where planet-disk interactions likely explain observed substructures \citep[e.g.,][]{apai2015,lee2016} or may have other collisional mechanisms at play \citep[e.g. ISM sculpting,][]{debes2009}.

We attempted to identify links between our findings on scattered light brightness and likely indicators of enhanced dynamical activity by comparing $F_{ans}/F_*$, the IR excess, measured values of brightness asymmetry from \cite{crotts2024} and this work, and the grain size distribution power law index $q$ queried from various literature sources that investigated individual systems \citep{johnson2012,wahhaj2016,thilliez2017,bruzzone2018,esposito2018,lohne2020,crotts2021,norfolk2021}. Performing a two-sample K-S test on the values in brightness asymmetry gives a p-value of 0.86, suggesting that there are no distinct differences in disk brightness asymmetries between Sco-Cen and other debris disks, and no correlations were found between brightness asymmetry, $F_{ans}/F_*$, and IR excess. Similarly, no correlations were found between $F_{ans}/F_*$ and $q$ (Pearson correlation $r=-0.04$, p-value of 0.91), and a low-significance linear correlation was found between IR excess and $q$ (Pearson $r=0.10$, p-value of 0.75). Ultimately, these comparisons are limited by small sample sizes and the inconsistent manner in which $q$ was determined across multiple studies.

\cite{crotts2024} identified a tentative positive correlation between measured disk aspect ratio (defined as the measured FWHM of the disk divided by $R_0$) and \textit{q}, which could suggest a closer relationship between grain size distribution and disk morphology rather than disk brightness. Disk aspect ratio could also be related to $F_{ans}/F_*$, as differences in surface density per unit solid angle would ultimately affect the measured surface brightness. Most systems, however, have similar aspect ratios (with a few exceptions present in both Sco-Cen and non-Sco-Cen systems), and the comparison is difficult to assess as the aspect ratio does not distinguish between radial and vertical width. Ultimately, the connections between observed asymmetry, surface brightness, and dynamical activity are highly complex and can vary significantly depending on the occurrence of different collisional mechanisms.

\begin{deluxetable*}{cccccccccc}

\tablecaption{Combined GPIES and DISCS Detected Sample Properties.}


\tablehead{\colhead{Name} & \colhead{\Lir} & \colhead{$F_{ans}/F_*$} & \colhead{$i$} & \colhead{$S_{total}$} & \colhead{$S_{CO}$} &\colhead{$T_{eff}$} & \colhead{$R_d$} & \colhead{Brightness Asym.} & \colhead{$q$} \\ 
\colhead{} & \colhead{($10^{-4}$)} & \colhead{($10^{-7}$)} & \colhead{(deg)} & \colhead{(mJy)} & \colhead{(mJy km s$^{-1}$)} & \colhead{(K)} & \colhead{(au)} & \colhead{} & \colhead{} } 

\startdata
\multicolumn{10}{c}{DISCS Sample}\\\hline
HD 106906 & 50.4 & 0.63$\pm$0.06 & 85.34$\substack{+0.05\\-0.06}$ & 0.22$\pm$0.04 (1) & $<$27 (1) & 6500 & 107.98$\substack{+0.69\\-0.79}$ & 1.28$\pm$0.04 & 3.19$\substack{+0.11\\-0.20}$ (12) \\
HD 110058 & 26.2 & 0.74$\pm$0.17 & 87.06$\substack{+0.23\\-0.19}$ & 0.71$\pm$0.11 (2) & 92$\pm$17 (2) & 8000 & 59.56$\substack{+12.47\\-0.77}$ & 1.23$\pm$0.03 & ... \\
HD 111161 & 42.3 & 0.52$\pm$0.13 & 59.78$\substack{+0.08\\-0.06}$ & 0.13$\pm$0.05 (2) & $<42$ (2) & 7800 & 72.48$\substack{+0.08\\-0.09}$ & 1.04$\pm$0.07 & ... \\
HD 111520 & 10.3 & 1.13$\pm$0.31 & 89.45$\substack{+0.27\\-0.27}$ & 0.18$\pm$0.05 (2) & $<53$ (2) & 6500 & 91.42$\substack{+13.8\\-10.1}$ & 1.78$\pm$0.09 & ... \\
HD 114082 & 36.3 & 0.65$\pm$0.06 & 83.32$\substack{+0.54\\-0.20}$ & 0.43$\pm$0.05 (2) & $<46$ (2) &7000 & 28.50$\substack{+1.40\\-0.19}$ & 1.14$\pm$0.04 & $>$3.9 (13) \\
HD 115600 & 22.6 & 0.58$\pm$0.20 & 82.97$\substack{+1.32\\-1.30}$ & 0.18$\pm$0.05 (2) & $<53$ (2) &7000 & 44.02$\substack{+7.36\\-7.32}$ & 1.01$\pm$0.05 & 3.65$\pm$0.15 (14) \\
HD 117214 & 26.7 & 0.90$\pm$0.03 & 69.57$\substack{+0.46\\-0.34}$ & 0.27$\pm$0.05 (2) & $<41$ (2) &6500 & 42.77$\substack{+0.09\\-0.10}$ & 1.14$\pm$0.04 & ... \\
HD 129590 & 69.6 & 1.82$\pm$0.09 & 84.11$\substack{+0.28\\-0.26}$ & 1.46$\pm$0.15 (2) & $<58$ (2) &5910 & 45.50$\substack{+0.48\\-1.08}$ & 1.09$\pm$0.06 & ... \\
HD 131835 & 30.9 & 0.83$\pm$0.07 & 75.94$\substack{+0.23\\-0.23}$ & 2.90$\pm$0.15 (2) & 798$\pm$35 (2) &8100 & 89.62$\substack{+0.81\\-0.80}$ & 1.11$\pm$0.06 & 3.13$\pm$0.07 (15) \\
HD 143675 & 5.6 & ... & ... & $<$0.23 (3) & $<$7.8 (3) &7900 & ... & ... & ... \\
HD 145560 & 12.7 & 0.24$\pm$0.29 & 41.91$\substack{+0.49\\-0.09}$ & 1.85$\pm$0.12 (2) & $<48$ (2) &6500 & 81.23$\substack{+0.06\\-0.05}$ & 1.02$\pm$0.05 & ... \\
HD 146897 & 101.9 & 2.23$\pm$0.10 & 85.99$\substack{+0.01\\-0.01}$ & 1.30$\pm$0.12 (2) & 60$\pm$15 (2) &6200 & 51.84$\substack{+0.19\\-0.78}$ & 1.2$\pm$0.15 & ... \\
HD 156623 & 43.3 & 0.02$\pm$0.06 & 34.70$\substack{+0.46\\-0.96}$ & 0.72$\pm$0.11 (2) & 1183$\pm$37 (2) &8350 & 52.56$\substack{+0.69\\-0.25}$ & 1.05$\pm$0.02 & ... \\
HD 98363 & 6.4 & 0.17$\pm$0.01 & $76.25\substack{+0.32\\-0.47}$ & 0.11$\pm$0.03 (3) & $<3.6$ (3) &8830 & $62.43\substack{+1.42\\-0.86}$ & 1.03$\pm$0.10 & ... \\
HD 109832 & 3.5 & 0.04$\pm$0.01 & $82.28\substack{+0.30\\-0.35}$ & $<$0.11 (3) & $<$3.3 (3) &7400 & $42.55\substack{+11.10\\-0.57}$ & 1.95$\pm$0.38 & ... \\
HD 112810 & 5.9 & 0.20$\pm$0.03 & $74.23\pm0.15$ & 0.52$\pm$0.09 (2) & $<42$ (2) &6637 & $113.04\substack{+0.99\\-1.06}$ & 1.13$\pm$0.19 & ... \\
HD 146181 & 19.0 & 0.59$\pm$0.02 & $67.56\substack{+0.25\\-0.27}$ & 0.88$\pm$0.09 (2) & $<50$ (2) &6350 & $82.67\substack{+0.43\\-0.30}$ & 1.31$\pm$0.11 & ... \\\hline
\multicolumn{10}{c}{Non-Sco-Cen Sample}\\\hline
AU Mic & 3.33 & 0.55$\pm$0.13 & 86.00$\substack{+0.01\\-0.01}$ & 7.14$\substack{+0.12\\-0.25}$ (4) & $<70$ (9) &3500 & 9.91$\substack{+0.01\\-0.01}$ & 1.95$\pm$0.20 & $<$3.33 (15) \\
$\beta$ Pic & 28.1 & 26.16$\pm$2.81 & 88.90$\substack{+0.09\\-0.10}$ & 20.0$\pm$2.0 (5) & 455$\pm$91 (10) & 8200 & 27.06$\substack{+1.18\\-0.34}$ & 1.04$\pm$0.02 & 3.49$\pm$0.06 (15) \\
CE Ant & 6.49 & 2.95$\pm$1.11 & 15.10$\substack{+0.95\\-1.05}$ & 2.1$\pm$0.4 (6) & 91$\pm$20 (11) &3420 & 27.13$\substack{+0.01\\-0.01}$ & 1.01$\pm$0.02 & ... \\
HD 30447 & 9.11 & 3.29$\pm$0.37 & 81.47$\substack{+0.05\\-0.05}$ & ... & ... & 6900 & 75.43$\substack{+0.68\\-0.72}$ & 1.19$\pm$0.08 & ... \\
HD 32297 & 60.5 & 50.70$\pm$1.68 & 88.26$\substack{+0.04\\-0.04}$ & 3.04$\pm$0.21 (7) & 102$\pm$10 (7) &7700 & 105.85$\substack{+1.62\\-1.06}$ & 1.13$\pm$0.05 & 3.07$\pm$0.12 (16) \\
HD 35841 & 3.25 & 3.95$\pm$0.34 & 83.27$\substack{+0.20\\-0.24}$ & ... & ... & 6500 & 39.12$\substack{+0.36\\-0.28}$ & 1.08$\pm$0.13 & 2.90$\substack{+0.10\\-0.20}$ (17) \\
HD 61005 & 27.9 & 6.18$\pm$0.45 & 85.41$\substack{+0.06\\-0.06}$ & 4.82$\pm$0.29 (7) & ... &5600 & 50.21$\substack{+0.17\\-0.17}$ & 1.82$\pm$0.09 & 3.33$\pm$0.04 (15) \\
HD 157587 & 32.0 & 11.83$\pm$1.00 & 64.02$\substack{+0.04\\-0.02}$ & ... & ... & 6300 & 81.24$\substack{+0.05\\-0.04}$ & 1.18$\pm$0.05 & 3.73$\substack{+0.81\\-0.08}$ (18) \\
HD 191089 & 15.0 & 2.40$\pm$0.83 & 61.85$\substack{+0.09\\-0.08}$ & ... & ... & 6400 & 46.96$\substack{+0.01\\-0.03}$ & 1.08$\pm$0.12 & ... \\
HR 4796A & 48.9 & 178$\pm$2.70 & 76.15$\substack{+0.06\\-0.07}$ & 14.8$\pm$1.50 (8) & $<25$ (8) & 9600 & 77.71$\substack{+0.05\\-0.04}$ & 1.02$\pm$0.02 & 3.43$\pm$0.06 (15) \\
HR 7012 & 8.39 & 8.63$\pm$3.50 & 72.42$\substack{+0.35\\-0.16}$ & ... & ... & 7700 & 8.77$\substack{+0.08\\-0.05}$ & 1.94$\pm$0.49 & 3.95$\pm$0.10 (19) \\
\enddata

\tablecomments{\Lir and $T_{eff}$ are collated from \citet{esposito2020} and this work. $i$, $R_d$, and brightness asymmetry are collated from \citet{crotts2024} and this work. $S_{total}$ is either the 1.24 or 1.33mm integrated continuum flux, and $S_{CO}$ is the flux of the $^{12}CO(2\rightarrow1)$ transition, except for CE Ant, which is the $^{12}CO(3\rightarrow2)$ transition.} \label{tab:sect7props}

\tablerefs{1. \citet{fehr2022}, 2. \citet{lieman-sifry2016}, 3. \citet{moor2017}, 4. \citet{macgregor2013}, 5. \citet{matra2019}, 6. \citet{bayo2019}, 7. \citet{macgregor2018}, 8. \citet{kennedy2018}, 9. \citet{daley2019}, 10. \citet{matra2017}, 11. \citet{matra2019_AUMic} 12. \citet{crotts2021}, 13. \citet{wahhaj2016}, 14. \citet{thilliez2017}, 15. \citet{lohne2020}, 16. \citet{norfolk2021}, 17. \citet{esposito2018}, 18. \citet{bruzzone2018}, 19. \citet{johnson2012}}

\end{deluxetable*}

\subsubsection{Sample Limitations}
Our analysis of $F_{ans}/F_*$ is limited by the small sample of GPI-imaged, non-Sco-Cen debris disks and the incomplete information a polarized intensity surface brightness measurement provides. The non-Sco-Cen sample is small and biased towards the most exceptional examples of active debris disks, with \cite{esposito2020} not able to detect at least 20 non-Sco-Cen systems that span the range of \Lir analyzed in this work. Additionally, measurements of the total intensity surface brightness and polarization fraction profiles would provide more evidence of differing grain size distributions between systems (suggesting differing levels of collisional activity). Such an analysis was not conducted for the comparison in this work, as high SNR measurements of the polarization fraction are dependent on the total intensity images of each system, which are severely impacted by PSF subtraction post-processing \citep[e.g.,][]{milli2012} and require forward modeling \citep[e.g.,][]{hom2024}, only feasible for high SNR systems.

Despite the limited sample size, the trends we are able to identify in the Sco-Cen sample suggest that the debris disks in the OB association may be valuable benchmarks in evaluating the properties of other debris disks in scattered light, and that the small set of bright debris disks imaged from ground-based observations are exceptional examples of systems with significantly more small dust grains, which may have been created through collisional mechanisms such as planet-disk interactions.

\subsection{Sample Detection Metrics}
\begin{figure*}
    \centering
    \includegraphics[width=\linewidth]{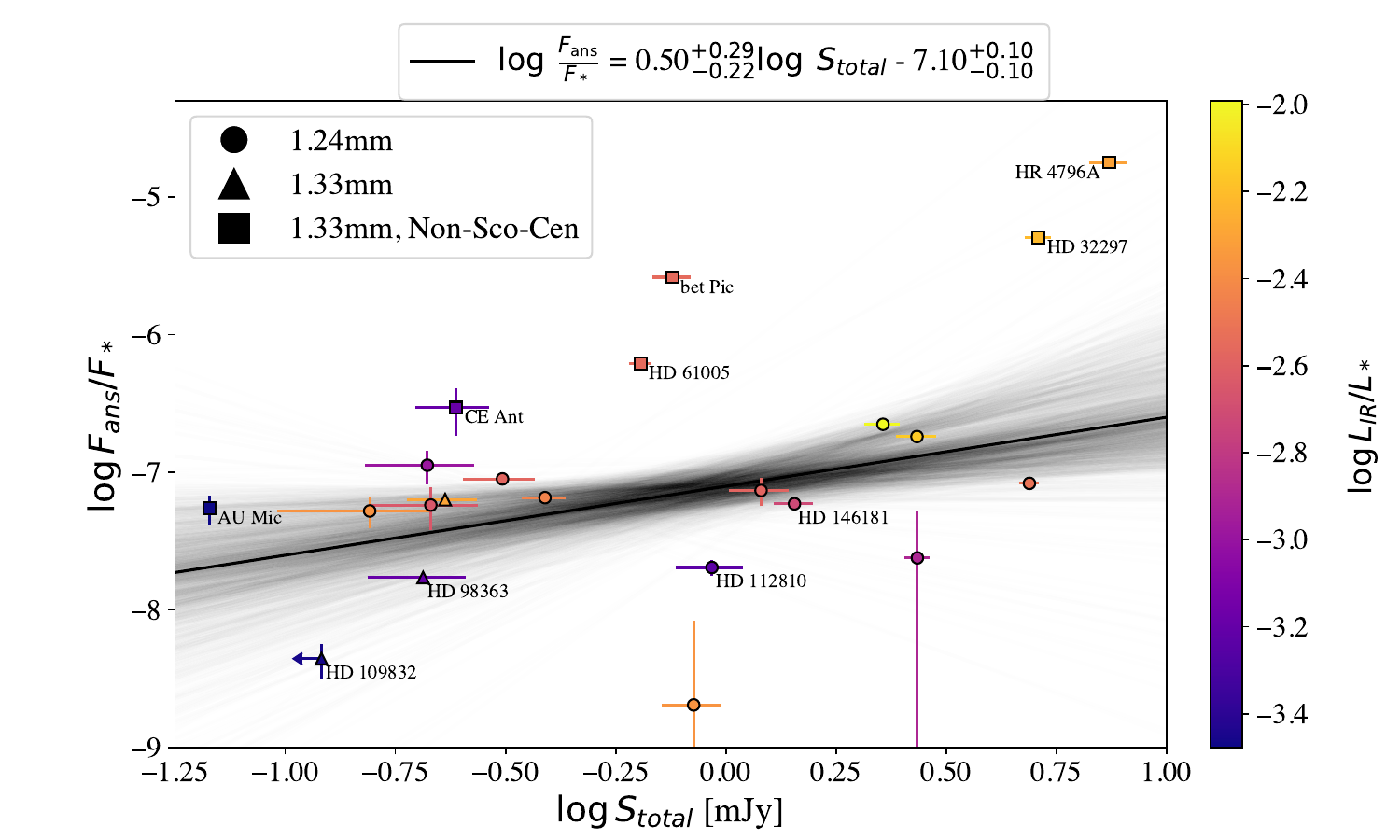}
    \caption{Ansae flux ratio measurements of Sco-Cen debris disk targets as a function of 1.24mm continuum emission from \citet{lieman-sifry2016} (or 1.33mm continuum emission in triangular points, from \citealt{moor2017} and \citealt{fehr2022}) normalized by system distance, with color indicating $log L_{IR}/L_*$. The solid line represents the median likelihood result with 1$\sigma$ uncertainties from linear bootstrap fitting (gray lines) of Sco-Cen systems. Ansae flux ratio increases slowly as a function of continuum emission at a lower rate than the IR excess, but there is large scatter in the measurements. The HD 109832 measurement is an upper limit, and is not included in the bootstrapping fit. 1.33mm continuum measurements of non-Sco-Cen systems are also plotted as squares, showing an offset similar to Figure \ref{fig:ansaeVsexcess}. \textit{Non-Sco-Cen measurement references:} 1. \citet{bayo2019}, 2. \citet{kennedy2018}, 3. \citet{macgregor2013}, 4. \citet{macgregor2018}, 5. \citet{matra2019}.}
    \label{fig:ansaeALMA}
\end{figure*}

One of the motivations of DISCS was to assess whether infrared excess as a primary detection formalism for imaging debris disks in scattered light was valid, as postulated by \cite{esposito2020}. Despite a small sample size, the high success rate of 4 out of 7 detections in the supplemental sample provides further evidence that high IR excess is a reliable predictability metric for detecting circumstellar disk systems in scattered light, even though excess measurements probe a separate emission regime. Indeed, the 2-3 non-detections had the lowest $L_{IR}/L_*$ of the sample as a whole. \cite{esposito2020} utilized a predictability metric to prioritize polarimetric debris disk observations, but found that the metric they used did not outperform relying on excess measurements alone, and even a post-survey revised metric did not outperform excess measurements by a significant margin. Therefore, prioritizing observations in order of high to low IR excess level provides a simple and robust approach to optimizing detection in scattered light debris disk surveys. Although the level of IR excess is likely an important predictor of a scattered light detection, our analysis in \S \ref{sec:scatteredlightfluxratio} suggests that it may not reliably predict scattered light brightness (with the caveat of our biased sample). A similar conclusion was reached in \cite{schneider2014}.

Several debris disks have been detected in far-IR emission, both unresolved and resolved with ALMA. Similar to the IR excess, one can imagine using ALMA mm continuum detection as a factor in prioritizing debris disk observations. Sco-Cen debris disks in particular have been measured in 1.24mm continuum and \textit{CO(2-1)} emission in \cite{lieman-sifry2016}. This study provides a uniformly measured sample in which to assess how the amount of 1.24mm flux may inform the scattered light brightness of a system. To assess this, we plot our measured $F_{ans}/F_*$ as a function of available $S_{total}$ measurements from \cite{lieman-sifry2016} normalized by system distance in Figure \ref{fig:ansaeALMA}, with colors indicating $log{L_{IR}/L_*}$. For systems that were not observed as part of \cite{lieman-sifry2016}, we instead plot ansae flux ratio as a function of 1.33mm continuum measurements from other sources \citep{moor2017,fehr2022}. To investigate if we see similar offsets between the Sco-Cen and non-Sco-Cen sample, we also plot available 1.33mm continuum measurements of non-Sco-Cen systems, collated from various literature sources. A low-significance, slowly-increasing trend (slope of $0.50\substack{+0.29\\-0.22}$) is observed among the Sco-Cen sample and contains large scatter. Similarly, as seen in Figure \ref{fig:ansaeVsexcess}, the increase in $L_{IR}/L_*$ appears to have a modest increase as both $F_{ans}/F_*$ and $S_{total}$ increase, with the linear fits in Figures \ref{fig:ansaeVsexcess} and \ref{fig:ansaeALMA} consistent within 2$\sigma$. This could suggest that mm continuum emission may be a marginally worse predictor of scattered light detection than IR excess. The sample has a wider range spanning over an order of magnitude in 1.24mm emission with similar $F_{ans}/F_*$, and the correlation is not as tight as the comparison with $L_{IR}/L_*$. Two Sco-Cen systems (HD 138813 and HD 142315) observed as part of the GPIES polarimetry debris disk survey \cite{esposito2020} were not detected despite having significant detections in \cite{lieman-sifry2016} and comparatively similar IR excess values as the rest of the detected GPIES sample. Conversely, HD 109832 and HD 143675 have been detected in scattered light but not in 1.33mm continuum emission \citep{moor2017} in addition to having relatively low \Lir, while the remaining non-detections from DISCS have the lowest IR excess levels in the sample. Similar to Figure \ref{fig:ansaeVsexcess}, there is an offset in $F_{ans}/F_*$ between non-Sco-Cen and Sco-Cen systems over a similar range in integrated mm continuum flux, although there appears to be more of an overlap in the low mm flux regime compared to the IR excess. These cases further emphasize possible gaps in linking thermal emission measurements of debris disk systems to a scattered light brightness.

The Sco-Cen sample alone is not large enough to add confidence in this assessment, and a larger sample of scattered light and ALMA-detected systems is needed that includes both non-Sco-Cen and Sco-Cen members. Other factors, such as disk inclination and scattering phase function, would also need to be controlled against, as the disk ansae are not the brightest source of scattered light flux in high inclination systems. Although ALMA detection alone does not facilitate a guaranteed scattered light detection, it is able to supplement the IR excess as a way of ranking the likelihood of a scattered light disk detection. Although the exact scattered light brightness is not easily predicted by extrapolating IR excess and mm continuum measurements, these values may be able to set a scattered light brightness floor that may lie below the GPI polarimetry detection limit, given the number of systems detected in thermal emission but not in scattered light. Additional measurements of debris disks in both scattered light and thermal emission are needed to quantify the nature of this floor and the range at which scattered light brightness can vary at a given IR excess level.

\section{Summary} \label{sec:conclusion}
We have initiated a survey of debris disks in the Scorpius-Centaurus OB association leveraging the power of polarimetric differential imaging with GPI and a selection metric based entirely on level of IR excess. In combination with the observations conducted in \cite{esposito2020}, we have observed 60\% of high-IR excess Sco-Cen systems observable with GPI down to $L_{IR}/L* = 2.5 \times 10^{-4}$ and 100\% of systems down to \Lir $\sim8\times 10^{-4}$ for spectral types earlier than mid-K. From observing 7 targets, we detect 4 debris disks in polarized intensity light for the first time. From our model characterization, we find that each new detection is noteworthy in its own right; significant morphological asymmetries are present in HD 98363 with a derived eccentricity $>0$ at a 3$\sigma$ level and HD 109832 providing tentative evidence for a warp. The HD 112810 debris disk is the faintest disk ever resolved by GPI at the $\sim50\mu$Jy level, and combined with a complementary total intensity detection in \cite{matthews2023}, measurements of the polarization fraction could shed more light on the system. Finally, the narrow and inclined ring around HD 146181 bears a tentative brightness asymmetry between its ansae and has an ideal inclination at which to measure the scattering phase function. A detection in total intensity in combination with the polarized intensity image can inform characterization efforts of the scattering properties of its dust grains. The new detections supplement an already rich sample of resolved debris disks in the Sco-Cen OB association, providing further evidence that the association sits at the perfect age in which to investigate young planetary system evolution and architectures immediately after the protoplanetary disk stage.

An investigation of morphological properties derived from resolved scattered light imaging and unresolved thermal emission measurements further supports inconsistencies between the two approaches, with the $\Gamma$ correction factor to the blackbody radius $R_{bb}$ appearing only robust for HD 146181 and under- or overestimates radii for all other systems.

Measurements of polarized intensity surface brightness at disk ansae are used to compare the debris disks in Sco-Cen to the sample of non-Sco-Cen GPI-imaged debris disks. The wide range in scattered light contrast for similar values of IR excess introduce new questions in the study of exceptionally bright systems such as $\beta$ Pictoris, HD 32297, and HR 4796A. The trends observed in scattered light contrast and other system properties such as host star $T_{eff}$ demonstrate that the Sco-Cen sample of debris disks may be critical benchmarks in assessing the characteristics and properties of debris disks as a whole. The wide range of scattered light contrast compared to thermal emission measurements suggest deeper gaps in our understanding of the connection between the smallest dust grains that scatter incident photons and the larger thermally emitting dust grains, with various collisional and transport mechanisms likely playing a more significant role in determining the scattered light brightness of a system.

The results of our survey loosely support the prediction from \cite{esposito2020} that the level of IR excess and mm continuum emission correlate with scattered light detection. However, this correlation may be weak, with significant ranges in scattered light brightness seen at a given \Lir, further suggesting deeper gaps in our understanding of the link between small and large dust grains in a debris disk.

Our analysis of Sco-Cen and its relationship to the broader population of scattered light-resolved debris disks is limited by the number of detections, particularly non-Sco-Cen targets, and our inability to resolve substructures (suggestive of collisional activity) at high fidelity. The planned upgrade to GPI, GPI2.0 \citep{chilcote2024}, will be installed at Gemini-North in 2025, allowing new DISCS observations of younger debris and transition disks in Upper Scorpius (US) in addition to older systems such as BD+45 598 \citep{hinkley2021}. Upper Centaurus Lupus (UCL) and Lower Centaurus Crux (LCC) are still accessible with VLT-SPHERE, and the planned upgrade, SPHERE+ \citep{boccaletti2020}, in combination with observational techniques like star-hopping \citep{wahhaj2021,olofsson2024} can also provide new detections and enhanced SNR to improve our disk and sample characterization efforts. DISCS II will leverage these new capabilities to further expand the sample of resolved, scattered light Sco-Cen systems, particularly for later spectral types that GPI was not sensitive to. To help bridge the gap between the scattered light and thermal emission properties of disk systems, new observations at high spatial resolution may also provide additional insight, with programs like the ARKS ALMA Large Program (Marino et al. In prep, 2022.1.00338.L) providing unprecedented spatially resolved mapping of debris disk substructures, and upcoming instrumentation on ELTs providing similarly spectacular spatial resolution at scattered light wavelengths.

\section*{Acknowledgements}
The authors would like to acknowledge the anonymous referee for their comprehensive and constructive review. The authors would also like to acknowledge Drs. Ewan Douglas and Steve Ertel for useful discussions supporting this work. This work is based on observations obtained at the Gemini Observatory, which is operated by the Association of Universities for Research in Astronomy, Inc., under a cooperative agreement with the NSF on behalf of the Gemini partnership: the National Science Foundation (United States), the National Research Council (Canada), CONICYT (Chile), Ministerio de Ciencia, Tecnolog\'ia e Innovaci\'on Productiva (Argentina), and Minist\'erio da Ci\^encia, Tecnologia e Inova\c c\~ao (Brazil). This work has made use of data from the European Space Agency (ESA) mission {\it Gaia} (\url{https://www.cosmos.esa.int/gaia}), processed by the {\it Gaia} Data Processing and Analysis Consortium (DPAC, \url{https://www.cosmos.esa.int/web/gaia/dpac/consortium}). Funding for the DPAC has been provided by national institutions, in particular the institutions participating in the {\it Gaia} Multilateral Agreement. Based on observations collected at the European Organisation for Astronomical Research in the Southern Hemisphere under ESO programmes 1100.C-0481(F) and 097.C-0330(A). This research has made use of the SIMBAD and VizieR databases, operated at CDS, Strasbourg, France.

This material is based upon High Performance Computing (HPC) resources supported by the University of Arizona TRIF, UITS, and Research, Innovation, and Impact (RII) and maintained by the UArizona Research Technologies department.

The raw data files for all observational sequences are available through the Gemini archive\footnote{\url{https://archive.gemini.edu/searchform}}. The reduced data files, modeling data files, and other materials underlying this article will be shared on reasonable request to the corresponding author. The Python scripts used in this article to create some figures and perform some calculations are available on Github\footnote{\url{https://github.com/jrhom1/DISCS_Paper1}}. Python scripts for performing spine-fitting analyses are described in \cite{crotts2024}.

\software{Gemini Planet Imager Data Pipeline (\citealt{perrin2014,perrin2016}, \url{http://ascl.net/1411.018}), \texttt{pyKLIP} (\citealt{wang2015}, \url{http://ascl.net/1506.001}), numpy, scipy, Astropy \citep{astropy2018}, matplotlib \citep{matplotlib2007, matplotlib_v2.0.2}, iPython \citep{ipython2007}, emcee (\citealt{foreman-mackey2013}, \url{http://ascl.net/1303.002}), corner (\citealt{foreman-mackey2017}, \url{http://ascl.net/1702.002}), \texttt{MCFOST} (\citealt{pinte2006}), \texttt{pymcfost}, \texttt{uncertainties}, (\citealt{lebigot2010uncertainties}).}

\facilities{Gemini:South, ESO:VLT}

\appendix

\section{Vetting Candidate Companions} \label{sec:vettingcompanions}
Point sources are easily detected in our total intensity reductions of HD 108904, HD 113556, and HD 119718. The companions are relatively bright, but the probability of them being associated with our target stars is low because of their low galactic latitudes. Regardless, both HD 108904 and HD 113556 have archival high resolution observations, and therefore the astrometry of the companions can be compared against the absolute astrometry of the target stars by following methods described in both \cite{nielsen2013} and \cite{nielsen2017}.

HD 108904 has three detected candidate companions in our images, a bright candidate with an airy ring at $PA\sim139\degr$/$\sim1\farcs28$ separation and fainter companions at $PA\sim64\degr$/$\sim0\farcs78$ and $PA\sim200\degr$/$\sim0\farcs97$ separations. The system was observed in two other epochs (2018 and 2019) as a part of the GPIES campaign in spectroscopic mode. The brightest candidate was also seen in archival Gemini/NICI imaging \citep{janson2013}. The astrometric analysis gives a high likelihood of the bright candidate being a background object. While the fainter candidates were not seen in NICI imaging, the offsets in position between the 2018 and 2019 GPI epochs are identical between the fainter candidates and the bright candidate, suggesting that the fainter candidates are also background objects.

\begin{figure*}
    \centering
    \includegraphics[width=0.9\textwidth]{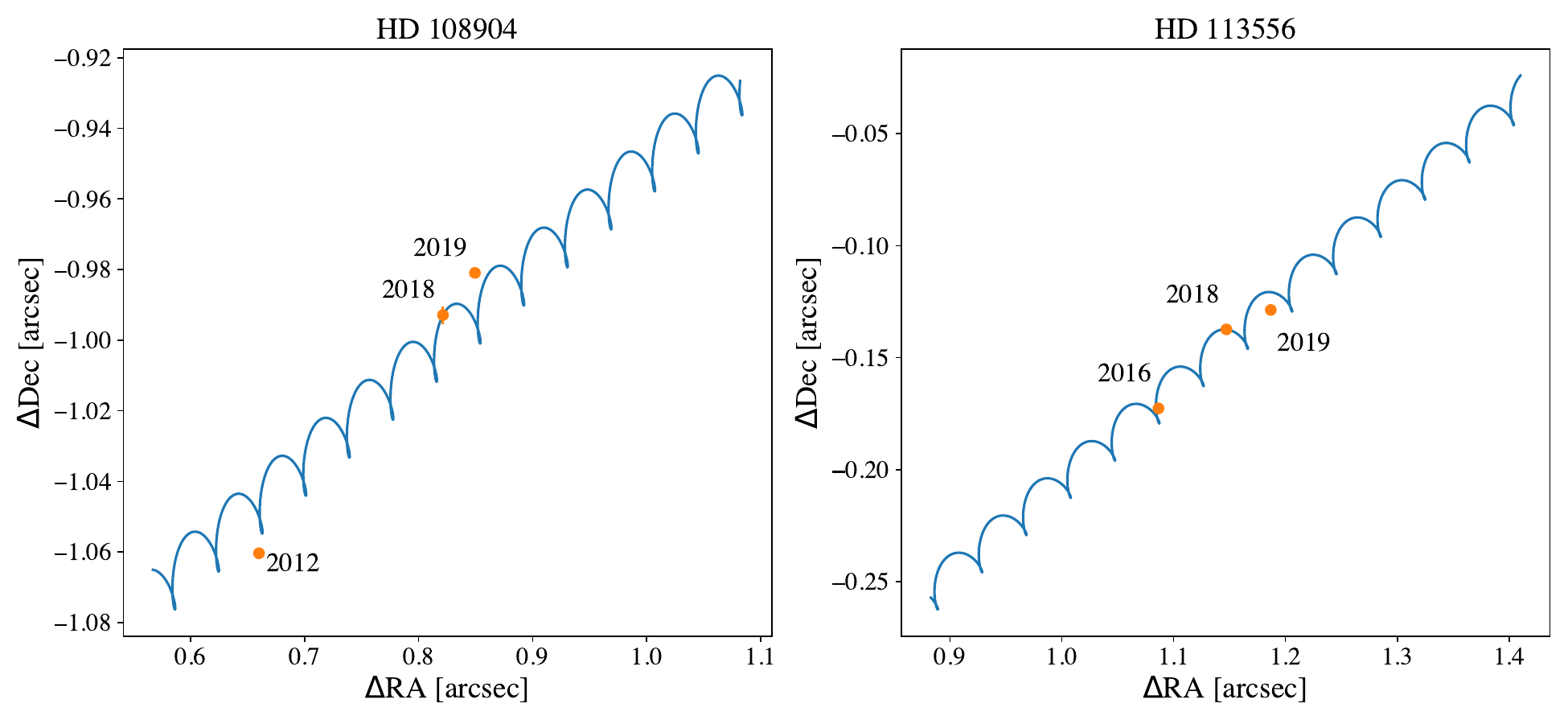}
    \caption{Background tracks calculated for both HD 108904 (left) and HD 113556 (right). The $\Delta$RA and $\Delta$Dec offsets for both candidate companions are consistent with being background objects.}
    \label{fig:astrometry}
\end{figure*}

HD 113556 has one candidate companion at $PA\sim96\degr$ and separation of $\sim1\farcs2$. 
The system has two additional epochs of archival observations (see Table \ref{tab:SPHERE_obs}), both from VLT-SPHERE/IRDIS in 2016 (with the \textit{B\_H} filter) as a part of the SPHERE/SHINE program (PI: Beuzit, 1100.C-0481(F)) and from SPHERE/IRDIFS in 2018 (PI: Olofsson, 097.C-0330(A)) with the \textit{K1K2} filter combination for IRDIS imaging and the \textit{YJH} disperser for IFS imaging. The public SPHERE IRDIS and IFS datasets for HD 113556 were queried and downloaded from the ESO archive. The query includes all relevant calibration files including dark frames, flat fields, and sky frames (for \textit{K}-band calibration). Pre-processing utilizes \texttt{vlt-sphere} \citep{vigan2020}, a Python wrapper that interfaces with ESO data reduction tools \citep{ESO2014} and performs dark correction, bad pixel correction, flat field calibration, and geometric distortion correction. Sky subtraction was also performed for IRDIS-\textit{K1K2} datacubes, and the pipeline also determines a wavelength solution for IFS datacubes from calibration sequences of an arc lamp. For post-processing, we utilize both ADI and spectral differential imaging \citep[SDI;][]{vigan2010} with \texttt{pyKLIP} subdividing our image into 10 annuli and selecting 50 KL modes. From the reduced \textit{B\_H} and \textit{K1K2} IRDIS images, our candidate is clearly visible. Two other bright exterior candidates are also visible in the HD 113556 image, outside of the FOV of GPI. From calculating the expected background object track, we find that the $\Delta$RA and $\Delta$Dec offset values for the candidate are consistent with being a background object in all three epochs, as shown in Figure \ref{fig:astrometry}.

HD 119718 has one bright candidate located at $PA\sim210\degr$ and $\sim1\farcs29$ separation. We measure a flux corresponding to $H \sim$ 22.5. At the distance of HD 119718, $M_H \sim$ 17.2. This roughly corresponds to a late-T dwarf companion assuming that the candidate is bound to the system. There are no additional archival high resolution observations of the system, and additional epochs would be needed to confirm the nature of this candidate. We chose not to pursue additional observations, as the likelihood of this candidate being a binary companion is low due to the low galactic latitude of the star. 
\begin{deluxetable}{ccccccc}
\tablecaption{Summary of SPHERE IFS and IRDIS observations.}\label{tab:SPHERE_obs}  

\tablenum{6}
\tablehead{\colhead{Name} & \colhead{Date} & \colhead{Mode} & \colhead{Filter} & \colhead{$N$} & \colhead{$t_{exp}$} & \colhead{$\Delta \theta$} \\ 
\colhead{} & \colhead{(UT)} & \colhead{} & \colhead{} & \colhead{} & \colhead{(s)} & \colhead{(deg)}} 
\startdata
HD 113556  & 2016 April 4 & IRDIS & \textit{B\_H} &32 & 64 & 15$\degr$ \\
      & 2018 April 17 & IRDIFS & \textit{K1K2} & 16 & 96 & 30$\degr$ \\
\enddata
\tablecomments{$N$ refers to the number of exposures.}
  
\end{deluxetable}

\section{Disk Characterization Corner Plots}
\label{sec:cornerplots}
In Figures 13.1-13.4, we show the 5-dimensional corner plots from our spine-fitting analysis after excluding a certain number of iterations as burn-in.







\begin{figure}
\figurenum{13.1}
\plotone{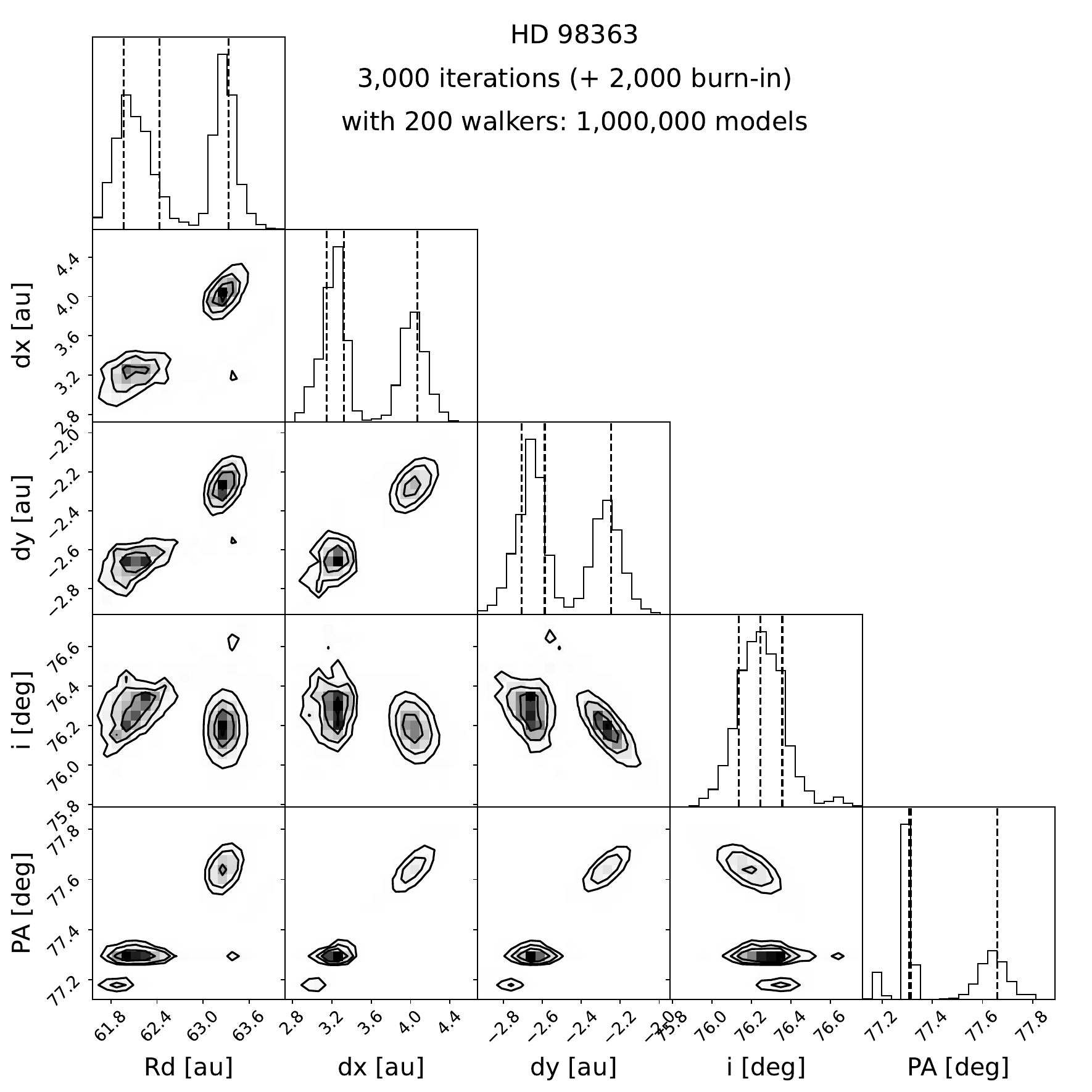}
\caption{Posterior distribution functions excluding burn-in iterations for the spine-fitting analysis of HD 98363. The complete figure set (4 images) is available in the online journal.}
\end{figure}

\begin{figure}
\figurenum{13.2}
\plotone{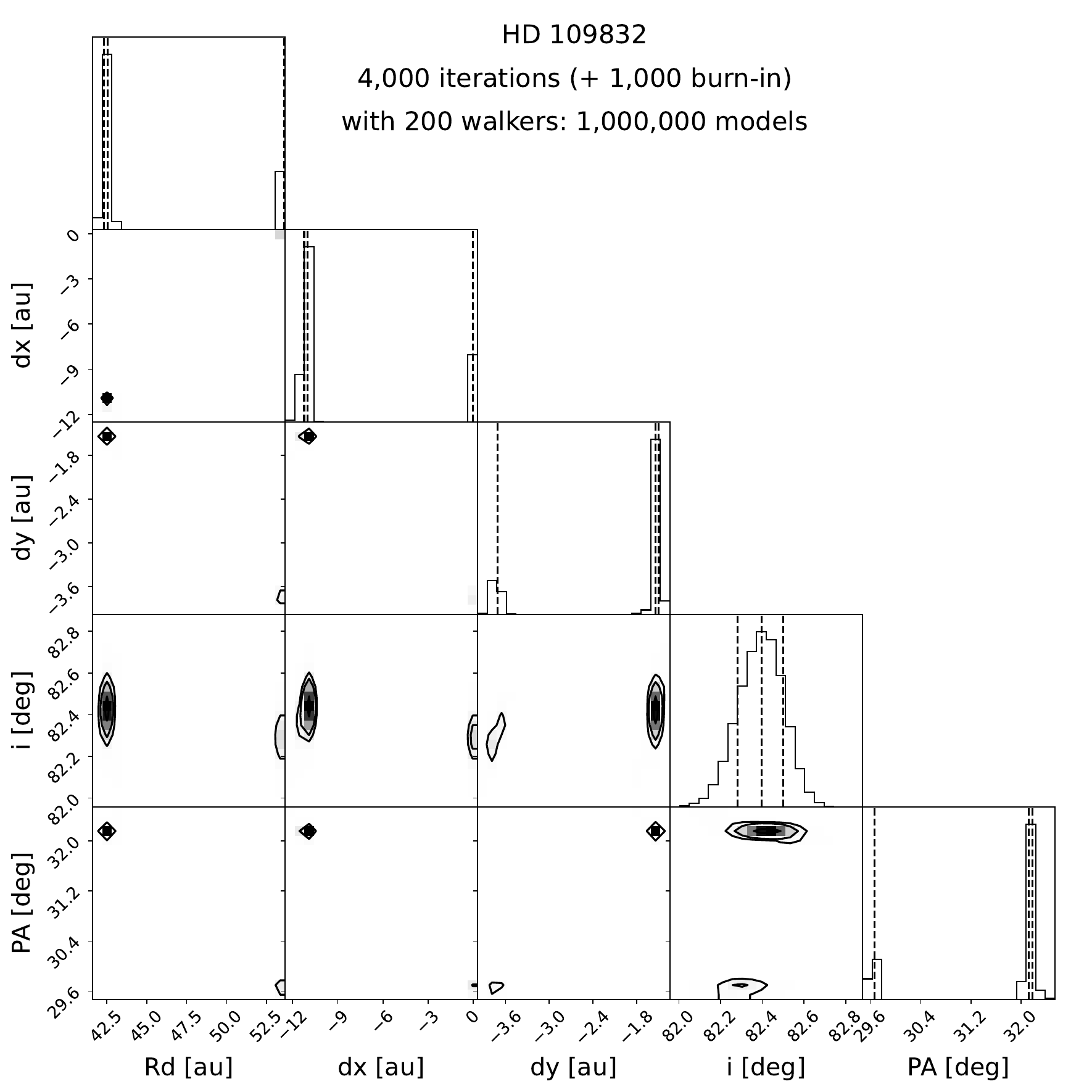}
\caption{Posterior distribution functions excluding burn-in iterations for the spine-fitting analysis of HD 109832.}
\end{figure}

\begin{figure}
\figurenum{13.3}
\plotone{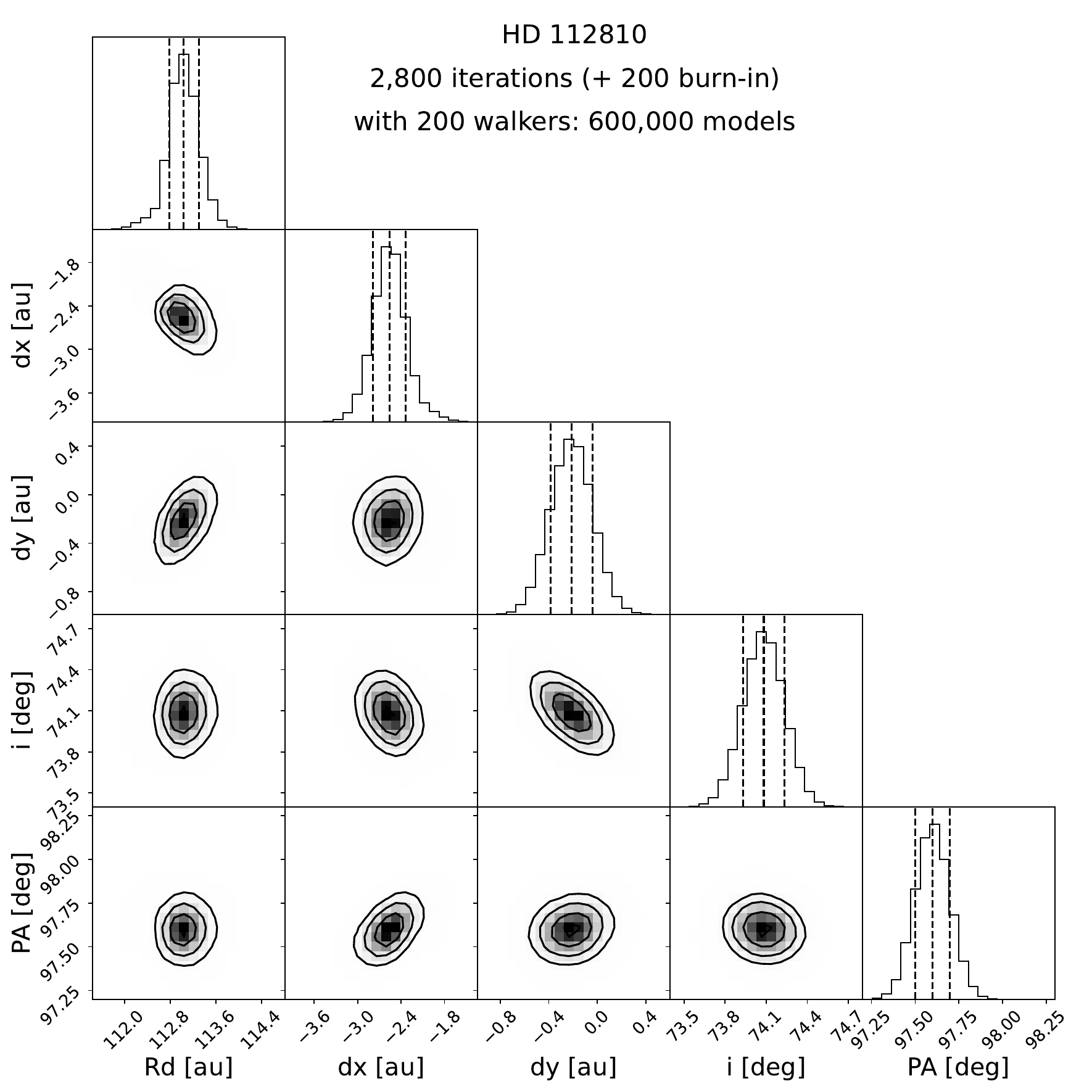}
\caption{Posterior distribution functions excluding burn-in iterations for the spine-fitting analysis of HD 112810.}
\end{figure}

\begin{figure}
\figurenum{13.4}
\plotone{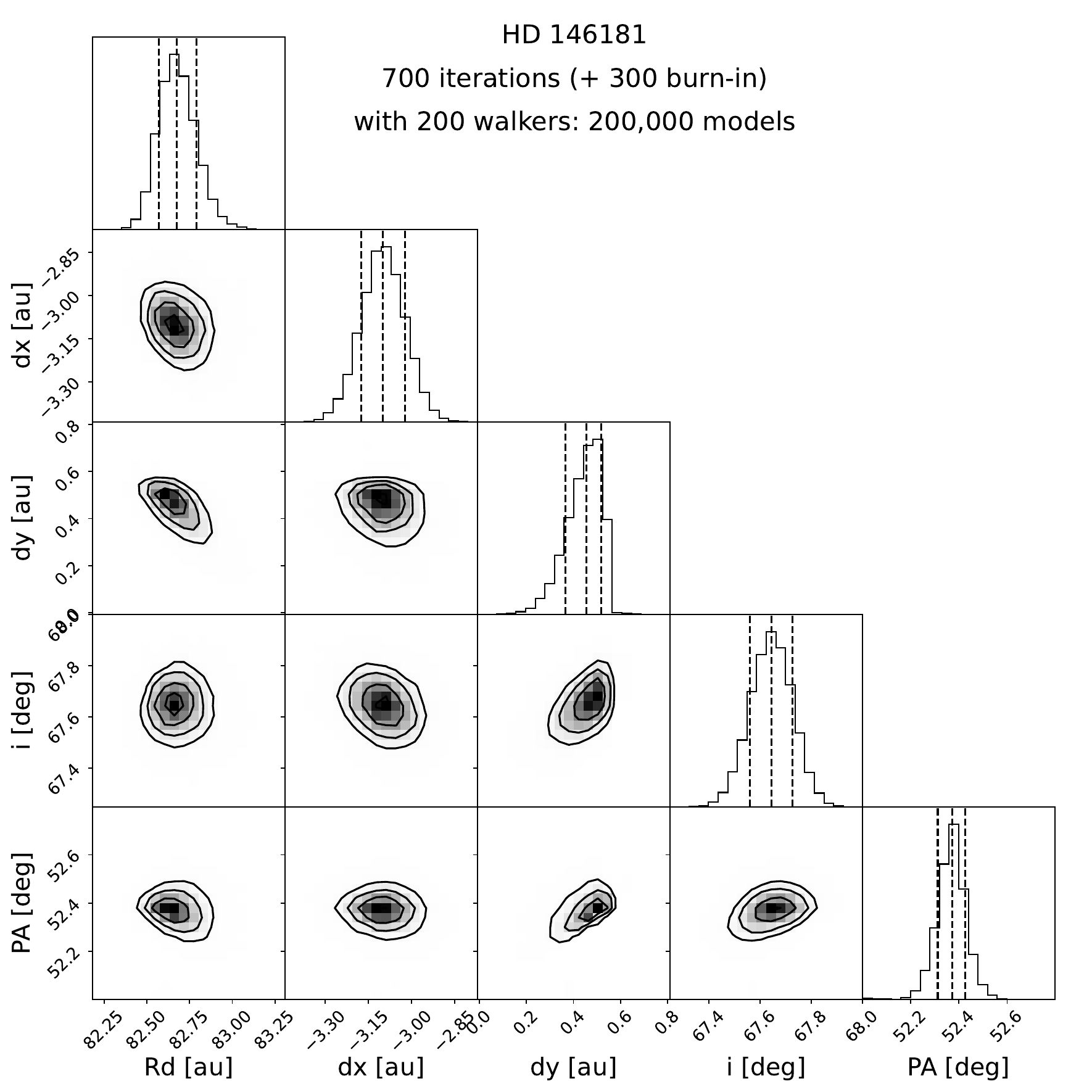}
\caption{Posterior distribution functions excluding burn-in iterations for the spine-fitting analysis of HD 146181.}
\end{figure}

\section{Impact of Distance on Ansae Flux Ratio for Identical Systems} \label{sec:distance_appendix}
\begin{figure}
    \centering
    \figurenum{14}
    \includegraphics[width=0.75\linewidth]{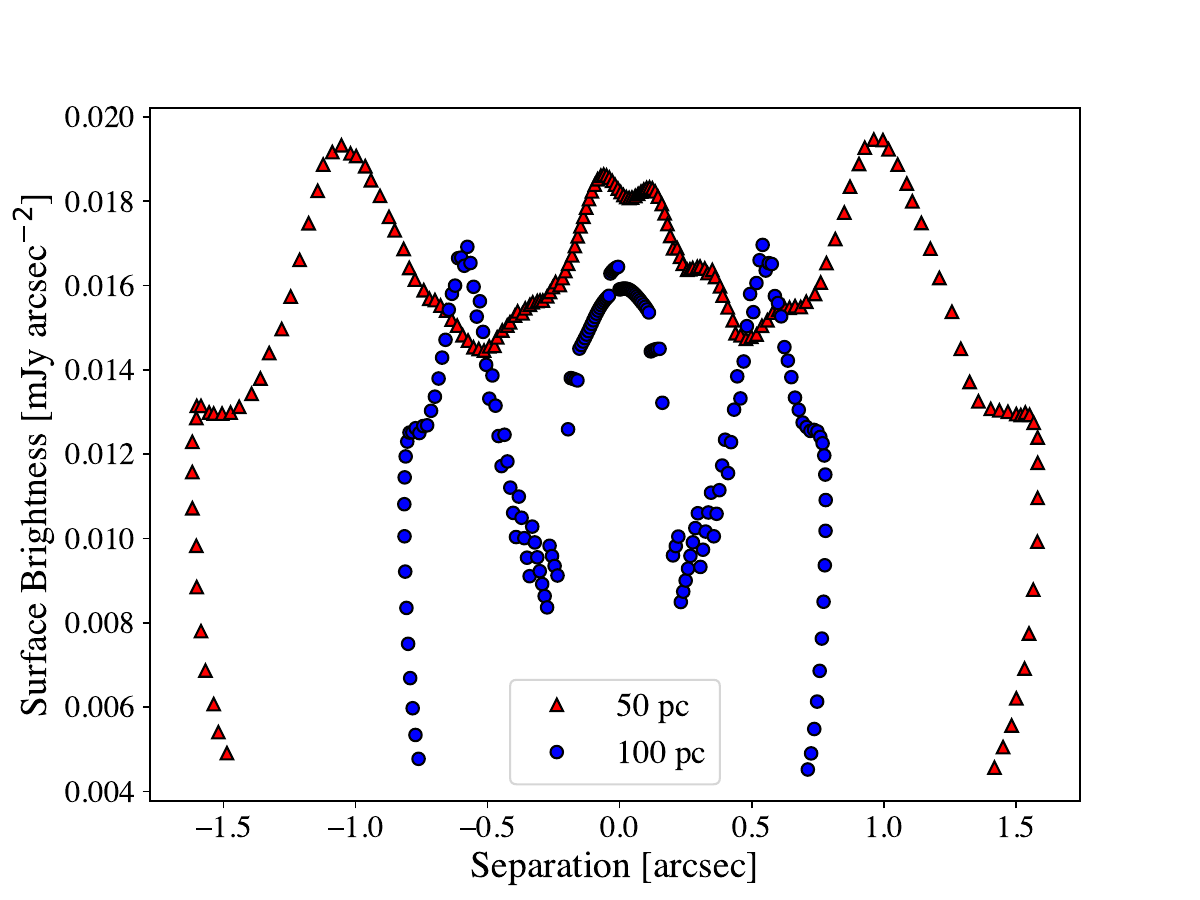}
    \caption{Comparison of surface brightness profiles between two identical convolved disk models at distances of 50 and 100 pc. The measurement of surface brightness around the fitted ansae of both models is comparable, with the 100 pc disk ansae fainter by a factor of $\sim$1.12. As a result, $F_{ans}/F_*$ for the 100 pc model is higher by a factor of $\sim$3.58 compared to the 50 pc model.}
    \label{fig:closefarcompare}
\end{figure}

For two resolved, identical debris disks located at different distances from the observer, we expect the measurement of $F_{ans}/F_*$ to be larger for the more distant system compared to the closer system, as more disk surface area is captured within a pixel. To assert this, we create two identical debris disks models using the radiative transfer code \texttt{MCFOST} \citep{pinte2006}. The model assumes a ring-like geometry following the surface density distribution described in \cite{augereau2001} with $i = 70\degr$, $PA = 90\degr$, $R_{\rm in} = 50.0$ au, $R_C = 80.0$ au, an aspect ratio of 0.05, $\alpha_{\rm in} = 5.0$, $\alpha_{\rm out} = -2.5$, and host star of spectral type F6V. The only parameter different between the two models is the system distance: 50 pc for the nearby system and 100 pc for the distant system. We convolve each model with a Gaussian of FWHM $=$ 3.8 pixels to match the size of a GPI PSF FWHM. We then fit for the spine and measure the surface brightness profiles for both convolved models in the same manner as the spine-fitting analysis described in \S \ref{sec:spinemodel_setup}. We show the measured surface brightness profiles in Figure \ref{fig:closefarcompare} only differ by a factor of $\sim$1.12, and find that the measured $F_{ans}/F_*$ of the 100 pc model to be higher than the 50 pc model by a factor of $\sim$3.58. This is contrary to the measurements of $F_{ans}/F_*$ between the Sco-Cen and non-Sco-Cen debris disks, suggesting that the two samples likely have distinct surface density distributions.

\section{Impact of Limb Brightening on Ansae Flux Ratio for Identical Systems} \label{sec:limb_brightening}
\begin{figure}
    \centering
    \figurenum{15}
    \includegraphics[width=0.75\linewidth]{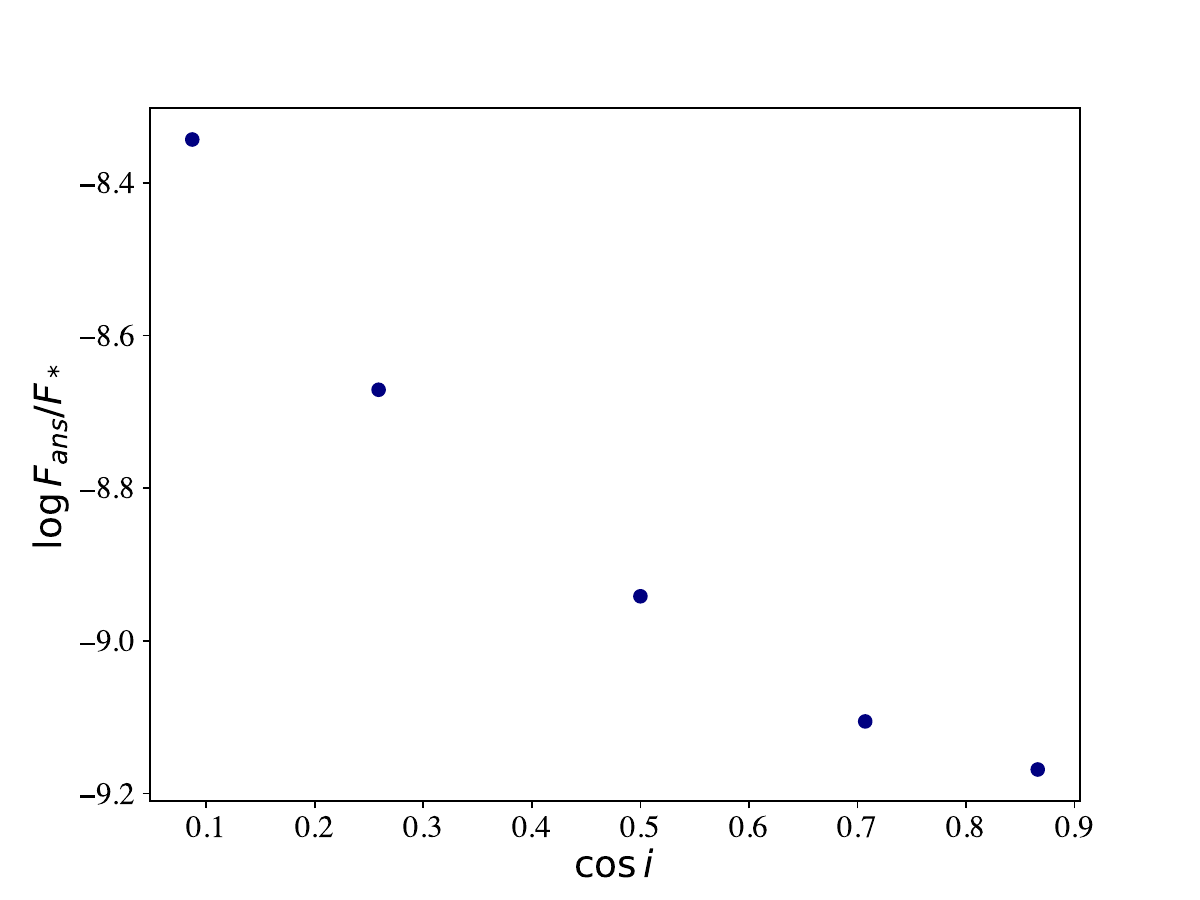}
    \caption{$F_{ans}/F_*$ as a function of $\cos i$ for identical systems at different inclinations. The slope of the trend is negative, but spans less than an order of magnitude in difference. By comparison, the GPIES and DISCS samples combined span several orders of magnitude across the same space in inclination.}
    \label{fig:limb_brightening_test}
\end{figure}

While ansae flux ratio measurements are designed to avoid measurement bias of scattered light surface brightenesses of different inclination systems, the effects of limb brightening could bias $F_{ans}/F_*$ upward with increasing inclination due to increasing dust column density. To test the impact of limb brightening, we construct identical radiative transfer disk models with \texttt{MCFOST}, using the same model parameters as described in Appendix \S \ref{sec:distance_appendix} for a system at 100 pc, and vary the inclination from 30$\degr$ to 85$\degr$. We measure $F_{ans}/F_*$ for each system, shown in Figure \ref{fig:limb_brightening_test}. Similar to Figure \ref{fig:incvsansae}, there is a negative trend between $F_{ans}/F_*$ and $\cos i$, but the slope of this trend appears shallower in this simple test case, with the difference between the highest and lowest inclination systems spanning less than an order of magnitude. By comparison, the combined GPIES and DISCS samples span a few orders of magnitude in the same range of inclination.

\bibliographystyle{aasjournal}
\bibliography{export-bibtex}

\newcommand{\noop}[1]{}
\begin{thebibliography}{}
\expandafter\ifx\csname natexlab\endcsname\relax\def\natexlab#1{#1}\fi
\providecommand{\url}[1]{\href{#1}{#1}}
\providecommand{\dodoi}[1]{}
\providecommand{\doarXiv}[1]{\href{https://arxiv.org/abs/#1}{\nolinkurl{https://arxiv.org/abs/#1}}}

\bibitem[{{Amara} \& {Quanz}(2012)}]{amara2012}
{Amara}, A., \& {Quanz}, S.~P. 2012, \href{http://dx.doi.org/10.1111/j.1365-2966.2012.21918.x}{\color{magenta}\mnras}, \href{https://ui.adsabs.harvard.edu/abs/2012MNRAS.427..948A}{\color{blue}427}, \href{https://ui.adsabs.harvard.edu/abs/2012MNRAS.427..948A}{\color{blue}948}

\bibitem[{{Apai} {et~al.}(2015){Apai}, {Schneider}, {Grady}, {Wyatt}, {Lagrange}, {Kuchner}, {Stark}, \& {Lubow}}]{apai2015}
{Apai}, D., {Schneider}, G., {Grady}, C.~A., {et~al.} 2015, \href{http://dx.doi.org/10.1088/0004-637X/800/2/136}{\color{magenta}\apj}, \href{https://ui.adsabs.harvard.edu/abs/2015ApJ...800..136A}{\color{blue}800}, \href{https://ui.adsabs.harvard.edu/abs/2015ApJ...800..136A}{\color{blue}136}

\bibitem[{{Arriaga} {et~al.}(2020){Arriaga}, {Fitzgerald}, {Duch{\^e}ne}, {Kalas}, {Millar-Blanchaer}, {Perrin}, {Chen}, {Mazoyer}, {Ammons}, {Bailey}, {Barman}, {Bulger}, {Chilcote}, {Cotten}, {De Rosa}, {Doyon}, {Esposito}, {Follette}, {Gerard}, {Goodsell}, {Graham}, {Greenbaum}, {Hibon}, {Hom}, {Hung}, {Ingraham}, {Konopacky}, {Macintosh}, {Maire}, {Marchis}, {Marley}, {Marois}, {Metchev}, {Nielsen}, {Oppenheimer}, {Palmer}, {Patience}, {Poyneer}, {Pueyo}, {Rajan}, {Rameau}, {Rantakyr{\"o}}, {Ruffio}, {Savransky}, {Schneider}, {Sivaramakrishnan}, {Song}, {Soummer}, {Thomas}, {Wang}, {Ward-Duong}, \& {Wolff}}]{arriaga2020}
{Arriaga}, P., {Fitzgerald}, M.~P., {Duch{\^e}ne}, G., {et~al.} 2020, \href{http://dx.doi.org/10.3847/1538-3881/ab91b1}{\color{magenta}\aj}, \href{https://ui.adsabs.harvard.edu/abs/2020AJ....160...79A}{\color{blue}160}, \href{https://ui.adsabs.harvard.edu/abs/2020AJ....160...79A}{\color{blue}79}

\bibitem[{{Astropy Collaboration} {et~al.}(2018){Astropy Collaboration}, {Price-Whelan}, {Sip{\H{o}}cz}, {G{\"u}nther}, {Lim}, {Crawford}, {Conseil}, {Shupe}, {Craig}, {Dencheva}, {Ginsburg}, {Vand erPlas}, {Bradley}, {P{\'e}rez-Su{\'a}rez}, {de Val-Borro}, {Aldcroft}, {Cruz}, {Robitaille}, {Tollerud}, {Ardelean}, {Babej}, {Bach}, {Bachetti}, {Bakanov}, {Bamford}, {Barentsen}, {Barmby}, {Baumbach}, {Berry}, {Biscani}, {Boquien}, {Bostroem}, {Bouma}, {Brammer}, {Bray}, {Breytenbach}, {Buddelmeijer}, {Burke}, {Calderone}, {Cano Rodr{\'\i}guez}, {Cara}, {Cardoso}, {Cheedella}, {Copin}, {Corrales}, {Crichton}, {D'Avella}, {Deil}, {Depagne}, {Dietrich}, {Donath}, {Droettboom}, {Earl}, {Erben}, {Fabbro}, {Ferreira}, {Finethy}, {Fox}, {Garrison}, {Gibbons}, {Goldstein}, {Gommers}, {Greco}, {Greenfield}, {Groener}, {Grollier}, {Hagen}, {Hirst}, {Homeier}, {Horton}, {Hosseinzadeh}, {Hu}, {Hunkeler}, {Ivezi{\'c}}, {Jain}, {Jenness}, {Kanarek}, {Kendrew}, {Kern}, {Kerzendorf}, {Khvalko}, {King}, {Kirkby}, {Kulkarni},
  {Kumar}, {Lee}, {Lenz}, {Littlefair}, {Ma}, {Macleod}, {Mastropietro}, {McCully}, {Montagnac}, {Morris}, {Mueller}, {Mumford}, {Muna}, {Murphy}, {Nelson}, {Nguyen}, {Ninan}, {N{\"o}the}, {Ogaz}, {Oh}, {Parejko}, {Parley}, {Pascual}, {Patil}, {Patil}, {Plunkett}, {Prochaska}, {Rastogi}, {Reddy Janga}, {Sabater}, {Sakurikar}, {Seifert}, {Sherbert}, {Sherwood-Taylor}, {Shih}, {Sick}, {Silbiger}, {Singanamalla}, {Singer}, {Sladen}, {Sooley}, {Sornarajah}, {Streicher}, {Teuben}, {Thomas}, {Tremblay}, {Turner}, {Terr{\'o}n}, {van Kerkwijk}, {de la Vega}, {Watkins}, {Weaver}, {Whitmore}, {Woillez}, {Zabalza}, \& {Astropy Contributors}}]{astropy2018}
{Astropy Collaboration}, {Price-Whelan}, A.~M., {Sip{\H{o}}cz}, B.~M., {et~al.} 2018, \href{http://dx.doi.org/10.3847/1538-3881/aabc4f}{\color{magenta}\aj}, \href{https://ui.adsabs.harvard.edu/abs/2018AJ....156..123A}{\color{blue}156}, \href{https://ui.adsabs.harvard.edu/abs/2018AJ....156..123A}{\color{blue}123}

\bibitem[{{Augereau} {et~al.}(2001){Augereau}, {Lagrange}, {Mouillet}, \& {M{\'e}nard}}]{augereau2001}
{Augereau}, J.~C., {Lagrange}, A.~M., {Mouillet}, D., \& {M{\'e}nard}, F. 2001, \href{http://dx.doi.org/10.1051/0004-6361:20000328}{\color{magenta}\aap}, \href{https://ui.adsabs.harvard.edu/abs/2001A&A...365...78A}{\color{blue}365}, \href{https://ui.adsabs.harvard.edu/abs/2001A&A...365...78A}{\color{blue}78}

\bibitem[{{Aumann} {et~al.}(1984){Aumann}, {Gillett}, {Beichman}, {de Jong}, {Houck}, {Low}, {Neugebauer}, {Walker}, \& {Wesselius}}]{aumann1984}
{Aumann}, H.~H., {Gillett}, F.~C., {Beichman}, C.~A., {et~al.} 1984, \href{http://dx.doi.org/10.1086/184214}{\color{magenta}\apjl}, \href{https://ui.adsabs.harvard.edu/abs/1984ApJ...278L..23A}{\color{blue}278}, \href{https://ui.adsabs.harvard.edu/abs/1984ApJ...278L..23A}{\color{blue}L23}

\bibitem[{{Backman} \& {Paresce}(1993)}]{backman1993}
{Backman}, D.~E., \& {Paresce}, F. 1993, \href{https://ui.adsabs.harvard.edu/abs/1993prpl.conf.1253B}{\color{blue}1253}

\bibitem[{{Bailey} {et~al.}(2014){Bailey}, {Meshkat}, {Reiter}, {Morzinski}, {Males}, {Su}, {Hinz}, {Kenworthy}, {Stark}, {Mamajek}, {Briguglio}, {Close}, {Follette}, {Puglisi}, {Rodigas}, {Weinberger}, \& {Xompero}}]{bailey2014}
{Bailey}, V., {Meshkat}, T., {Reiter}, M., {et~al.} 2014, \href{http://dx.doi.org/10.1088/2041-8205/780/1/L4}{\color{magenta}\apjl}, \href{https://ui.adsabs.harvard.edu/abs/2014ApJ...780L...4B}{\color{blue}780}, \href{https://ui.adsabs.harvard.edu/abs/2014ApJ...780L...4B}{\color{blue}L4}

\bibitem[{{Bayo} {et~al.}(2019){Bayo}, {Olofsson}, {Matr{\`a}}, {Beam{\'\i}n}, {Gallardo}, {de Gregorio-Monsalvo}, {Booth}, {Zamora}, {Iglesias}, {Henning}, {Schreiber}, \& {C{\'a}ceres}}]{bayo2019}
{Bayo}, A., {Olofsson}, J., {Matr{\`a}}, L., {et~al.} 2019, \href{http://dx.doi.org/10.1093/mnras/stz1133}{\color{magenta}\mnras}, \href{https://ui.adsabs.harvard.edu/abs/2019MNRAS.486.5552B}{\color{blue}486}, \href{https://ui.adsabs.harvard.edu/abs/2019MNRAS.486.5552B}{\color{blue}5552}

\bibitem[{{Benisty} {et~al.}(2017){Benisty}, {Stolker}, {Pohl}, {de Boer}, {Lesur}, {Dominik}, {Dullemond}, {Langlois}, {Min}, {Wagner}, {Henning}, {Juhasz}, {Pinilla}, {Facchini}, {Apai}, {van Boekel}, {Garufi}, {Ginski}, {M{\'e}nard}, {Pinte}, {Quanz}, {Zurlo}, {Boccaletti}, {Bonnefoy}, {Beuzit}, {Chauvin}, {Cudel}, {Desidera}, {Feldt}, {Fontanive}, {Gratton}, {Kasper}, {Lagrange}, {LeCoroller}, {Mouillet}, {Mesa}, {Sissa}, {Vigan}, {Antichi}, {Buey}, {Fusco}, {Gisler}, {Llored}, {Magnard}, {Moeller-Nilsson}, {Pragt}, {Roelfsema}, {Sauvage}, \& {Wildi}}]{benisty2017}
{Benisty}, M., {Stolker}, T., {Pohl}, A., {et~al.} 2017, \href{http://dx.doi.org/10.1051/0004-6361/201629798}{\color{magenta}\aap}, \href{https://ui.adsabs.harvard.edu/abs/2017A&A...597A..42B}{\color{blue}597}, \href{https://ui.adsabs.harvard.edu/abs/2017A&A...597A..42B}{\color{blue}A42}

\bibitem[{{Beuzit} {et~al.}(2019){Beuzit}, {Vigan}, {Mouillet}, {Dohlen}, {Gratton}, {Boccaletti}, {Sauvage}, {Schmid}, {Langlois}, {Petit}, {Baruffolo}, {Feldt}, {Milli}, {Wahhaj}, {Abe}, {Anselmi}, {Antichi}, {Barette}, {Baudrand}, {Baudoz}, {Bazzon}, {Bernardi}, {Blanchard}, {Brast}, {Bruno}, {Buey}, {Carbillet}, {Carle}, {Cascone}, {Chapron}, {Charton}, {Chauvin}, {Claudi}, {Costille}, {De Caprio}, {de Boer}, {Delboulb{\'e}}, {Desidera}, {Dominik}, {Downing}, {Dupuis}, {Fabron}, {Fantinel}, {Farisato}, {Feautrier}, {Fedrigo}, {Fusco}, {Gigan}, {Ginski}, {Girard}, {Giro}, {Gisler}, {Gluck}, {Gry}, {Henning}, {Hubin}, {Hugot}, {Incorvaia}, {Jaquet}, {Kasper}, {Lagadec}, {Lagrange}, {Le Coroller}, {Le Mignant}, {Le Ruyet}, {Lessio}, {Lizon}, {Llored}, {Lundin}, {Madec}, {Magnard}, {Marteaud}, {Martinez}, {Maurel}, {M{\'e}nard}, {Mesa}, {M{\"o}ller-Nilsson}, {Moulin}, {Moutou}, {Orign{\'e}}, {Parisot}, {Pavlov}, {Perret}, {Pragt}, {Puget}, {Rabou}, {Ramos}, {Reess}, {Rigal}, {Rochat}, {Roelfsema}, {Rousset},
  {Roux}, {Saisse}, {Salasnich}, {Santambrogio}, {Scuderi}, {Segransan}, {Sevin}, {Siebenmorgen}, {Soenke}, {Stadler}, {Suarez}, {Tiph{\`e}ne}, {Turatto}, {Udry}, {Vakili}, {Waters}, {Weber}, {Wildi}, {Zins}, \& {Zurlo}}]{beuzit2019}
{Beuzit}, J.~L., {Vigan}, A., {Mouillet}, D., {et~al.} 2019, \href{http://dx.doi.org/10.1051/0004-6361/201935251}{\color{magenta}\aap}, \href{https://ui.adsabs.harvard.edu/abs/2019A&A...631A.155B}{\color{blue}631}, \href{https://ui.adsabs.harvard.edu/abs/2019A&A...631A.155B}{\color{blue}A155}

\bibitem[{{Blaauw}(1946)}]{blaauw1946}
{Blaauw}, A. 1946PGro, \href{https://ui.adsabs.harvard.edu/abs/1946PGro...52....1B}{\color{blue}52}, \href{https://ui.adsabs.harvard.edu/abs/1946PGro...52....1B}{\color{blue}1}

\bibitem[{{Boccaletti} {et~al.}(2020){Boccaletti}, {Chauvin}, {Mouillet}, {Absil}, {Allard}, {Antoniucci}, {Augereau}, {Barge}, {Baruffolo}, {Baudino}, {Baudoz}, {Beaulieu}, {Benisty}, {Beuzit}, {Bianco}, {Biller}, {Bonavita}, {Bonnefoy}, {Bos}, {Bouret}, {Brandner}, {Buchschache}, {Carry}, {Cantalloube}, {Cascone}, {Carlotti}, {Charnay}, {Chiavassa}, {Choquet}, {Clenet}, {Crida}, {De Boer}, {De Caprio}, {Desidera}, {Desert}, {Delisle}, {Delorme}, {Dohlen}, {Doelman}, {Dominik}, {Orazi}, {Dougados}, {Doute}, {Fedele}, {Feldt}, {Ferreira}, {Fontanive}, {Fusco}, {Galicher}, {Garufi}, {Gendron}, {Ghedina}, {Ginski}, {Gonzalez}, {Gratadour}, {Gratton}, {Guillot}, {Haffert}, {Hagelberg}, {Henning}, {Huby}, {Janson}, {Kamp}, {Keller}, {Kenworthy}, {Kervella}, {Kral}, {Kuhn}, {Lagadec}, {Laibe}, {Langlois}, {Lagrange}, {Launhardt}, {Leboulleux}, {Le Coroller}, {Li Causi}, {Loupias}, {Maire}, {Marleau}, {Martinache}, {Martinez}, {Mary}, {Mattioli}, {Mazoyer}, {Meheut}, {Menard}, {Mesa}, {Meunier}, {Miguel}, {Milli},
  {Min}, {Molliere}, {Mordasini}, {Moretto}, {Mugnier}, {Muro Arena}, {Nardetto}, {Diaye}, {Nesvadba}, {Pedichini}, {Pinilla}, {Por}, {Potier}, {Quanz}, {Rameau}, {Roelfsema}, {Rouan}, {Rigliaco}, {Salasnich}, {Samland}, {Sauvage}, {Schmid}, {Segransan}, {Snellen}, {Snik}, {Soulez}, {Stadler}, {Stam}, {Tallon}, {Thebault}, {Thiebaut}, {Tschudi}, {Udry}, {van Holstein}, {Vernazza}, {Vidal}, {Vigan}, {Waters}, {Wildi}, {Willson}, {Zanutta}, {Zavagno}, \& {Zurlo}}]{boccaletti2020}
{Boccaletti}, A., {Chauvin}, G., {Mouillet}, D., {et~al.} 2020, \href{https://arxiv.org/abs/2003.05714}{\color{magenta}arXiv}, \href{https://ui.adsabs.harvard.edu/abs/2020arXiv200305714B}{\color{blue}arXiv:2003.05714}

\bibitem[{{Bohn} {et~al.}(2019){Bohn}, {Kenworthy}, {Ginski}, {Benisty}, {de Boer}, {Keller}, {Mamajek}, {Meshkat}, {Muro-Arena}, {Pecaut}, {Snik}, {Wolff}, \& {Reggiani}}]{bohn2019}
{Bohn}, A.~J., {Kenworthy}, M.~A., {Ginski}, C., {et~al.} 2019, \href{http://dx.doi.org/10.1051/0004-6361/201834523}{\color{magenta}\aap}, \href{https://ui.adsabs.harvard.edu/abs/2019A&A...624A..87B}{\color{blue}624}, \href{https://ui.adsabs.harvard.edu/abs/2019A&A...624A..87B}{\color{blue}A87}

\bibitem[{{Bonnefoy} {et~al.}(2017){Bonnefoy}, {Milli}, {M{\'e}nard}, {Vigan}, {Lagrange}, {Delorme}, {Boccaletti}, {Lazzoni}, {Galicher}, {Desidera}, {Chauvin}, {Augereau}, {Mouillet}, {Pinte}, {van der Plas}, {Gratton}, {Beust}, \& {Beuzit}}]{bonnefoy2017}
{Bonnefoy}, M., {Milli}, J., {M{\'e}nard}, F., {et~al.} 2017, \href{http://dx.doi.org/10.1051/0004-6361/201628929}{\color{magenta}\aap}, \href{https://ui.adsabs.harvard.edu/abs/2017A&A...597L...7B}{\color{blue}597}, \href{https://ui.adsabs.harvard.edu/abs/2017A&A...597L...7B}{\color{blue}L7}

\bibitem[{{Bonnefoy} {et~al.}(2021){Bonnefoy}, {Milli}, {Menard}, {Delorme}, {Chomez}, {Bonavita}, {Lagrange}, {Vigan}, {Augereau}, {Beuzit}, {Biller}, {Boccaletti}, {Chauvin}, {Desidera}, {Faramaz}, {Galicher}, {Gratton}, {Hinkley}, {Lazzoni}, {Matthews}, {Mesa}, {Mordasini}, {Mouillet}, {Olofsson}, \& {Pinte}}]{bonnefoy2021}
{Bonnefoy}, M., {Milli}, J., {Menard}, F., {et~al.} 2021, \href{http://dx.doi.org/10.1051/0004-6361/202141848}{\color{magenta}\aap}, \href{https://ui.adsabs.harvard.edu/abs/2021A&A...655A..62B}{\color{blue}655}, \href{https://ui.adsabs.harvard.edu/abs/2021A&A...655A..62B}{\color{blue}A62}

\bibitem[{{Booth} {et~al.}(2013){Booth}, {Kennedy}, {Sibthorpe}, {Matthews}, {Wyatt}, {Duch{\^e}ne}, {Kavelaars}, {Rodriguez}, {Greaves}, {Koning}, {Vican}, {Rieke}, {Su}, {Moro-Mart{\'\i}n}, \& {Kalas}}]{booth2013}
{Booth}, M., {Kennedy}, G., {Sibthorpe}, B., {et~al.} 2013, \href{http://dx.doi.org/10.1093/mnras/sts117}{\color{magenta}\mnras}, \href{https://ui.adsabs.harvard.edu/abs/2013MNRAS.428.1263B}{\color{blue}428}, \href{https://ui.adsabs.harvard.edu/abs/2013MNRAS.428.1263B}{\color{blue}1263}

\bibitem[{{Bruzzone}(2018)}]{bruzzone2018}
{Bruzzone}, J.~S. 2018, PhD thesis, University of Western Ontario, Canada

\bibitem[{{Cameron}(1997)}]{cameron1997}
{Cameron}, A.~G.~W. 1997, \href{http://dx.doi.org/10.1006/icar.1996.5642}{\color{magenta}\icarus}, \href{https://ui.adsabs.harvard.edu/abs/1997Icar..126..126C}{\color{blue}126}, \href{https://ui.adsabs.harvard.edu/abs/1997Icar..126..126C}{\color{blue}126}

\bibitem[{{Carpenter} {et~al.}(2025){Carpenter}, {Esplin}, {Luhman}, {Mamajek}, \& {Andrews}}]{carpenter2025}
{Carpenter}, J.~M., {Esplin}, T.~L., {Luhman}, K.~L., {et~al.} 2025, \href{http://dx.doi.org/10.3847/1538-4357/ad8ebc}{\color{magenta}\apj}, \href{https://ui.adsabs.harvard.edu/abs/2025ApJ...978..117C}{\color{blue}978}, \href{https://ui.adsabs.harvard.edu/abs/2025ApJ...978..117C}{\color{blue}117}

\bibitem[{{Chauvin} {et~al.}(2017){Chauvin}, {Desidera}, {Lagrange}, {Vigan}, {Gratton}, {Langlois}, {Bonnefoy}, {Beuzit}, {Feldt}, {Mouillet}, {Meyer}, {Cheetham}, {Biller}, {Boccaletti}, {D'Orazi}, {Galicher}, {Hagelberg}, {Maire}, {Mesa}, {Olofsson}, {Samland}, {Schmidt}, {Sissa}, {Bonavita}, {Charnay}, {Cudel}, {Daemgen}, {Delorme}, {Janin-Potiron}, {Janson}, {Keppler}, {Le Coroller}, {Ligi}, {Marleau}, {Messina}, {Molli{\`e}re}, {Mordasini}, {M{\"u}ller}, {Peretti}, {Perrot}, {Rodet}, {Rouan}, {Zurlo}, {Dominik}, {Henning}, {Menard}, {Schmid}, {Turatto}, {Udry}, {Vakili}, {Abe}, {Antichi}, {Baruffolo}, {Baudoz}, {Baudrand}, {Blanchard}, {Bazzon}, {Buey}, {Carbillet}, {Carle}, {Charton}, {Cascone}, {Claudi}, {Costille}, {Deboulbe}, {De Caprio}, {Dohlen}, {Fantinel}, {Feautrier}, {Fusco}, {Gigan}, {Giro}, {Gisler}, {Gluck}, {Hubin}, {Hugot}, {Jaquet}, {Kasper}, {Madec}, {Magnard}, {Martinez}, {Maurel}, {Le Mignant}, {M{\"o}ller-Nilsson}, {Llored}, {Moulin}, {Orign{\'e}}, {Pavlov}, {Perret}, {Petit},
  {Pragt}, {Puget}, {Rabou}, {Ramos}, {Rigal}, {Rochat}, {Roelfsema}, {Rousset}, {Roux}, {Salasnich}, {Sauvage}, {Sevin}, {Soenke}, {Stadler}, {Suarez}, {Weber}, {Wildi}, {Antoniucci}, {Augereau}, {Baudino}, {Brandner}, {Engler}, {Girard}, {Gry}, {Kral}, {Kopytova}, {Lagadec}, {Milli}, {Moutou}, {Schlieder}, {Szul{\'a}gyi}, {Thalmann}, \& {Wahhaj}}]{chauvin2017}
{Chauvin}, G., {Desidera}, S., {Lagrange}, A.~M., {et~al.} 2017, \href{http://dx.doi.org/10.1051/0004-6361/201731152}{\color{magenta}\aap}, \href{https://ui.adsabs.harvard.edu/abs/2017A&A...605L...9C}{\color{blue}605}, \href{https://ui.adsabs.harvard.edu/abs/2017A&A...605L...9C}{\color{blue}L9}

\bibitem[{{Chen} {et~al.}(2014){Chen}, {Mittal}, {Kuchner}, {Forrest}, {Lisse}, {Manoj}, {Sargent}, \& {Watson}}]{chen2014}
{Chen}, C.~H., {Mittal}, T., {Kuchner}, M., {et~al.} 2014, \href{http://dx.doi.org/10.1088/0067-0049/211/2/25}{\color{magenta}\apjs}, \href{https://ui.adsabs.harvard.edu/abs/2014ApJS..211...25C}{\color{blue}211}, \href{https://ui.adsabs.harvard.edu/abs/2014ApJS..211...25C}{\color{blue}25}

\bibitem[{{Chilcote} {et~al.}(2024){Chilcote}, {Konopacky}, {Hamper}, {Macintosh}, {Marois}, {Savransky}, {Soummer}, {V{\'e}ran}, {Agapito}, {Aleman}, {Bonaglia}, {Boucher}, {Burke}, {Chambouleyron}, {De Rosa}, {Do {\'O}}, {Dunn}, {Engstrom}, {Esposito}, {Filion}, {Fitzsimmons}, {Galvan}, {Kerley}, {Landry}, {Lardi{\`e}re}, {Levinstein}, {Limbach}, {Maire}, {Matzner}, {Mocnik}, {Nickson}, {Nielsen}, {Nguyen}, {Nguyen}, {Perera}, {Peng}, {Perrin}, {Por}, {Pueyo}, {Quiroz}, {Rantakyr{\"o}}, {Sands}, {Seifahrt}, {Singh}, \& {Wolf}}]{chilcote2024}
{Chilcote}, J., {Konopacky}, Q., {Hamper}, R., {et~al.} 2024, \href{http://dx.doi.org/10.1117/12.3020642}{\color{magenta}Proc.~SPIE}, \href{https://ui.adsabs.harvard.edu/abs/2024SPIE13096E..99C}{\color{blue}13096}, \href{https://ui.adsabs.harvard.edu/abs/2024SPIE13096E..99C}{\color{blue}1309699}

\bibitem[{{Cotten} \& {Song}(2016)}]{cotten2016}
{Cotten}, T.~H., \& {Song}, I. 2016, \href{http://dx.doi.org/10.3847/0067-0049/225/1/15}{\color{magenta}\apjs}, \href{https://ui.adsabs.harvard.edu/abs/2016ApJS..225...15C}{\color{blue}225}, \href{https://ui.adsabs.harvard.edu/abs/2016ApJS..225...15C}{\color{blue}15}

\bibitem[{{Cronin-Coltsmann} {et~al.}(2023){Cronin-Coltsmann}, {Kennedy}, {Kral}, {Lestrade}, {Marino}, {Matr{\`a}}, \& {Wyatt}}]{Cronin-Coltsmann2023}
{Cronin-Coltsmann}, P.~F., {Kennedy}, G.~M., {Kral}, Q., {et~al.} 2023, \href{http://dx.doi.org/10.1093/mnras/stad3083}{\color{magenta}\mnras}, \href{https://ui.adsabs.harvard.edu/abs/2023MNRAS.526.5401C}{\color{blue}526}, \href{https://ui.adsabs.harvard.edu/abs/2023MNRAS.526.5401C}{\color{blue}5401}

\bibitem[{{Crotts} {et~al.}(2021){Crotts}, {Matthews}, {Esposito}, {Duch{\^e}ne}, {Kalas}, {Chen}, {Arriaga}, {Millar-Blanchaer}, {Debes}, {Draper}, {Fitzgerald}, {Hom}, {MacGregor}, {Mazoyer}, {Patience}, {Rice}, {Weinberger}, {Wilner}, \& {Wolff}}]{crotts2021}
{Crotts}, K.~A., {Matthews}, B.~C., {Esposito}, T.~M., {et~al.} 2021, \href{http://dx.doi.org/10.3847/1538-4357/abff5c}{\color{magenta}\apj}, \href{https://ui.adsabs.harvard.edu/abs/2021ApJ...915...58C}{\color{blue}915}, \href{https://ui.adsabs.harvard.edu/abs/2021ApJ...915...58C}{\color{blue}58}

\bibitem[{{Crotts} {et~al.}(2022){Crotts}, {Draper}, {Matthews}, {Duch{\^e}ne}, {Esposito}, {Wilner}, {Mazoyer}, {Padgett}, {Kalas}, \& {Stapelfeldt}}]{crotts2022}
{Crotts}, K.~A., {Draper}, Z.~H., {Matthews}, B.~C., {et~al.} 2022, \href{http://dx.doi.org/10.3847/1538-4357/ac6c86}{\color{magenta}\apj}, \href{https://ui.adsabs.harvard.edu/abs/2022ApJ...932...23C}{\color{blue}932}, \href{https://ui.adsabs.harvard.edu/abs/2022ApJ...932...23C}{\color{blue}23}

\bibitem[{{Crotts} {et~al.}(2024){Crotts}, {Matthews}, {Duch{\^e}ne}, {Esposito}, {Dong}, {Hom}, {Oppenheimer}, {Rice}, {Wolff}, {Chen}, {Do {\'O}}, {Kalas}, {Lewis}, {Weinberger}, {Wilner}, {Ammons}, {Arriaga}, {De Rosa}, {Debes}, {Fitzgerald}, {Gonzales}, {Hines}, {Hinkley}, {Hughes}, {Kolokolova}, {Lee}, {L{\'o}pez}, {Macintosh}, {Mazoyer}, {Metchev}, {Millar-Blanchaer}, {Nielsen}, {Patience}, {Perrin}, {Pueyo}, {Rantakyr{\"o}}, {Ren}, {Schneider}, {Soummer}, \& {Stark}}]{crotts2024}
{Crotts}, K.~A., {Matthews}, B.~C., {Duch{\^e}ne}, G., {et~al.} 2024, \href{http://dx.doi.org/10.3847/1538-4357/ad0e69}{\color{magenta}\apj}, \href{https://ui.adsabs.harvard.edu/abs/2024ApJ...961..245C}{\color{blue}961}, \href{https://ui.adsabs.harvard.edu/abs/2024ApJ...961..245C}{\color{blue}245}

\bibitem[{{Currie} {et~al.}(2015){Currie}, {Cloutier}, {Brittain}, {Grady}, {Burrows}, {Muto}, {Kenyon}, \& {Kuchner}}]{currie2015a}
{Currie}, T., {Cloutier}, R., {Brittain}, S., {et~al.} 2015, \href{http://dx.doi.org/10.1088/2041-8205/814/2/L27}{\color{magenta}\apjl}, \href{https://ui.adsabs.harvard.edu/abs/2015ApJ...814L..27C}{\color{blue}814}, \href{https://ui.adsabs.harvard.edu/abs/2015ApJ...814L..27C}{\color{blue}L27}

\bibitem[{{Daley} {et~al.}(2019){Daley}, {Hughes}, {Carter}, {Flaherty}, {Lambros}, {Pan}, {Schlichting}, {Chiang}, {Wyatt}, {Wilner}, {Andrews}, \& {Carpenter}}]{daley2019}
{Daley}, C., {Hughes}, A.~M., {Carter}, E.~S., {et~al.} 2019, \href{http://dx.doi.org/10.3847/1538-4357/ab1074}{\color{magenta}\apj}, \href{https://ui.adsabs.harvard.edu/abs/2019ApJ...875...87D}{\color{blue}875}, \href{https://ui.adsabs.harvard.edu/abs/2019ApJ...875...87D}{\color{blue}87}

\bibitem[{{de Zeeuw} {et~al.}(1999){de Zeeuw}, {Hoogerwerf}, {de Bruijne}, {Brown}, \& {Blaauw}}]{dezeeuw1999}
{de Zeeuw}, P.~T., {Hoogerwerf}, R., {de Bruijne}, J.~H.~J., {et~al.} 1999, \href{http://dx.doi.org/10.1086/300682}{\color{magenta}\aj}, \href{https://ui.adsabs.harvard.edu/abs/1999AJ....117..354D}{\color{blue}117}, \href{https://ui.adsabs.harvard.edu/abs/1999AJ....117..354D}{\color{blue}354}

\bibitem[{{Debes} {et~al.}(2009){Debes}, {Weinberger}, \& {Kuchner}}]{debes2009}
{Debes}, J.~H., {Weinberger}, A.~J., \& {Kuchner}, M.~J. 2009, \href{http://dx.doi.org/10.1088/0004-637X/702/1/318}{\color{magenta}\apj}, \href{https://ui.adsabs.harvard.edu/abs/2009ApJ...702..318D}{\color{blue}702}, \href{https://ui.adsabs.harvard.edu/abs/2009ApJ...702..318D}{\color{blue}318}

\bibitem[{{Draper} {et~al.}(2016){Draper}, {Duch{\^e}ne}, {Millar-Blanchaer}, {Matthews}, {Wang}, {Kalas}, {Graham}, {Padgett}, {Ammons}, {Bulger}, {Chen}, {Chilcote}, {Doyon}, {Fitzgerald}, {Follette}, {Gerard}, {Greenbaum}, {Hibon}, {Hinkley}, {Macintosh}, {Ingraham}, {Lafreni{\`e}re}, {Marchis}, {Marois}, {Nielsen}, {Oppenheimer}, {Patel}, {Patience}, {Perrin}, {Pueyo}, {Rajan}, {Rameau}, {Sivaramakrishnan}, {Vega}, {Ward-Duong}, \& {Wolff}}]{draper2016}
{Draper}, Z.~H., {Duch{\^e}ne}, G., {Millar-Blanchaer}, M.~A., {et~al.} 2016, \href{http://dx.doi.org/10.3847/0004-637X/826/2/147}{\color{magenta}\apj}, \href{https://ui.adsabs.harvard.edu/abs/2016ApJ...826..147D}{\color{blue}826}, \href{https://ui.adsabs.harvard.edu/abs/2016ApJ...826..147D}{\color{blue}147}

\bibitem[{Droettboom {et~al.}(2017)Droettboom, Caswell, Hunter, Firing, Nielsen, Varoquaux, Root, Elson, Dale, Lee, de~Andrade, Sepp{\"a}nen, McDougall, May, Lee, Straw, Stansby, Hobson, Yu, Ma, Gohlke, Silvester, Moad, Schulz, Vincent, W{\"u}rtz, Ariza, Cimarron, Hisch, \& Kniazev}]{matplotlib_v2.0.2}
Droettboom, M., Caswell, T.~A., Hunter, J., {et~al.} 2017, matplotlib/matplotlib v2.0.2

\bibitem[{{Duch{\^e}ne} {et~al.}(2020){Duch{\^e}ne}, {Rice}, {Hom}, {Zalesky}, {Esposito}, {Millar-Blanchaer}, {Ren}, {Kalas}, {Fitzgerald}, {Arriaga}, {Bruzzone}, {Bulger}, {Chen}, {Chiang}, {Cotten}, {Czekala}, {De Rosa}, {Dong}, {Draper}, {Follette}, {Graham}, {Hung}, {Lopez}, {Macintosh}, {Matthews}, {Mazoyer}, {Metchev}, {Patience}, {Perrin}, {Rameau}, {Song}, {Stahl}, {Wang}, {Wolff}, {Zuckerman}, {Ammons}, {Bailey}, {Barman}, {Chilcote}, {Doyon}, {Gerard}, {Goodsell}, {Greenbaum}, {Hibon}, {Ingraham}, {Konopacky}, {Maire}, {Marchis}, {Marley}, {Marois}, {Nielsen}, {Oppenheimer}, {Palmer}, {Poyneer}, {Pueyo}, {Rajan}, {Rantakyr{\"o}}, {Ruffio}, {Savransky}, {Schneider}, {Sivaramakrishnan}, {Soummer}, {Thomas}, \& {Ward-Duong}}]{duchene2020}
{Duch{\^e}ne}, G., {Rice}, M., {Hom}, J., {et~al.} 2020, \href{http://dx.doi.org/10.3847/1538-3881/ab8881}{\color{magenta}\aj}, \href{https://ui.adsabs.harvard.edu/abs/2020AJ....159..251D}{\color{blue}159}, \href{https://ui.adsabs.harvard.edu/abs/2020AJ....159..251D}{\color{blue}251}

\bibitem[{{ESO CPL Development Team}(2014)}]{ESO2014}
{ESO CPL Development Team}. 2014, {CPL: Common Pipeline Library}, Astrophysics Source Code Library, record ascl:1402.010, Astrophysics Source Code Library, record ascl:1402.010

\bibitem[{{Esposito} {et~al.}(2018){Esposito}, {Duch{\^e}ne}, {Kalas}, {Rice}, {Choquet}, {Ren}, {Perrin}, {Chen}, {Arriaga}, {Chiang}, {Nielsen}, {Graham}, {Wang}, {De Rosa}, {Follette}, {Ammons}, {Ansdell}, {Bailey}, {Barman}, {Sebasti{\'a}n Bruzzone}, {Bulger}, {Chilcote}, {Cotten}, {Doyon}, {Fitzgerald}, {Goodsell}, {Greenbaum}, {Hibon}, {Hung}, {Ingraham}, {Konopacky}, {Larkin}, {Macintosh}, {Maire}, {Marchis}, {Marois}, {Mazoyer}, {Metchev}, {Millar-Blanchaer}, {Oppenheimer}, {Palmer}, {Patience}, {Poyneer}, {Pueyo}, {Rajan}, {Rameau}, {Rantakyr{\"o}}, {Ryan}, {Savransky}, {Schneider}, {Sivaramakrishnan}, {Song}, {Soummer}, {Thomas}, {Wallace}, {Ward-Duong}, {Wiktorowicz}, \& {Wolff}}]{esposito2018}
{Esposito}, T.~M., {Duch{\^e}ne}, G., {Kalas}, P., {et~al.} 2018, \href{http://dx.doi.org/10.3847/1538-3881/aacbc9}{\color{magenta}\aj}, \href{https://ui.adsabs.harvard.edu/abs/2018AJ....156...47E}{\color{blue}156}, \href{https://ui.adsabs.harvard.edu/abs/2018AJ....156...47E}{\color{blue}47}

\bibitem[{{Esposito} {et~al.}(2020){Esposito}, {Kalas}, {Fitzgerald}, {Millar-Blanchaer}, {Duch{\^e}ne}, {Patience}, {Hom}, {Perrin}, {De Rosa}, {Chiang}, {Czekala}, {Macintosh}, {Graham}, {Ansdell}, {Arriaga}, {Bruzzone}, {Bulger}, {Chen}, {Cotten}, {Dong}, {Draper}, {Follette}, {Hung}, {Lopez}, {Matthews}, {Mazoyer}, {Metchev}, {Rameau}, {Ren}, {Rice}, {Song}, {Stahl}, {Wang}, {Wolff}, {Zuckerman}, {Ammons}, {Bailey}, {Barman}, {Chilcote}, {Doyon}, {Gerard}, {Goodsell}, {Greenbaum}, {Hibon}, {Hinkley}, {Ingraham}, {Konopacky}, {Maire}, {Marchis}, {Marley}, {Marois}, {Nielsen}, {Oppenheimer}, {Palmer}, {Poyneer}, {Pueyo}, {Rajan}, {Rantakyr{\"o}}, {Ruffio}, {Savransky}, {Schneider}, {Sivaramakrishnan}, {Soummer}, {Thomas}, \& {Ward-Duong}}]{esposito2020}
{Esposito}, T.~M., {Kalas}, P., {Fitzgerald}, M.~P., {et~al.} 2020, \href{http://dx.doi.org/10.3847/1538-3881/ab9199}{\color{magenta}\aj}, \href{https://ui.adsabs.harvard.edu/abs/2020AJ....160...24E}{\color{blue}160}, \href{https://ui.adsabs.harvard.edu/abs/2020AJ....160...24E}{\color{blue}24}

\bibitem[{{Fehr} {et~al.}(2022){Fehr}, {Hughes}, {Dawson}, {Marino}, {Ackelsberg}, {Kittling}, {Flaherty}, {Nesvold}, {Carpenter}, {Andrews}, {Matthews}, {Crotts}, \& {Kalas}}]{fehr2022}
{Fehr}, A.~J., {Hughes}, A.~M., {Dawson}, R.~I., {et~al.} 2022, \href{http://dx.doi.org/10.3847/1538-4357/ac9235}{\color{magenta}\apj}, \href{https://ui.adsabs.harvard.edu/abs/2022ApJ...939...56F}{\color{blue}939}, \href{https://ui.adsabs.harvard.edu/abs/2022ApJ...939...56F}{\color{blue}56}

\bibitem[{{Foreman-Mackey}(2017)}]{foreman-mackey2017}
{Foreman-Mackey}, D. 2017, {corner.py: Corner plots}

\bibitem[{{Foreman-Mackey} {et~al.}(2013){Foreman-Mackey}, {Hogg}, {Lang}, \& {Goodman}}]{foreman-mackey2013}
{Foreman-Mackey}, D., {Hogg}, D.~W., {Lang}, D., \& {Goodman}, J. 2013, \href{http://dx.doi.org/10.1086/670067}{\color{magenta}\pasp}, \href{https://ui.adsabs.harvard.edu/abs/2013PASP..125..306F}{\color{blue}125}, \href{https://ui.adsabs.harvard.edu/abs/2013PASP..125..306F}{\color{blue}306}

\bibitem[{{Gaia Collaboration}(2020)}]{gaia2020}
{Gaia Collaboration}. 2020, {VizieR Online Data Catalog: Gaia EDR3 (Gaia Collaboration, 2020)}, VizieR On-line Data Catalog: I/350. Originally published in: 2021A\&A...649A...1G; doi:10.5270/esa-1ug, VizieR On-line Data Catalog: I/350. Originally published in: 2021A\&A...649A...1G; doi:10.5270/esa-1ug

\bibitem[{{Goldman} {et~al.}(2018){Goldman}, {R{\"o}ser}, {Schilbach}, {Mo{\'o}r}, \& {Henning}}]{goldman2018}
{Goldman}, B., {R{\"o}ser}, S., {Schilbach}, E., {et~al.} 2018, \href{http://dx.doi.org/10.3847/1538-4357/aae64c}{\color{magenta}\apj}, \href{https://ui.adsabs.harvard.edu/abs/2018ApJ...868...32G}{\color{blue}868}, \href{https://ui.adsabs.harvard.edu/abs/2018ApJ...868...32G}{\color{blue}32}

\bibitem[{{Hinkley} {et~al.}(2021){Hinkley}, {Matthews}, {Lefevre}, {Lestrade}, {Kennedy}, {Mawet}, {Stapelfeldt}, {Ray}, {Mamajek}, {Bowler}, {Wilner}, {Williams}, {Ansdell}, {Wyatt}, {Lau}, {Phillips}, {Fernandez}, {Gagn{\'e}}, {Bubb}, {Sutlieff}, {Wilson}, {Matthews}, {Ngo}, {Piskorz}, {Crepp}, {Gonzalez}, {Mann}, \& {Mace}}]{hinkley2021}
{Hinkley}, S., {Matthews}, E.~C., {Lefevre}, C., {et~al.} 2021, \href{http://dx.doi.org/10.3847/1538-4357/abec6e}{\color{magenta}\apj}, \href{https://ui.adsabs.harvard.edu/abs/2021ApJ...912..115H}{\color{blue}912}, \href{https://ui.adsabs.harvard.edu/abs/2021ApJ...912..115H}{\color{blue}115}

\bibitem[{{Holland} {et~al.}(1998){Holland}, {Greaves}, {Zuckerman}, {Webb}, {McCarthy}, {Coulson}, {Walther}, {Dent}, {Gear}, \& {Robson}}]{holland1998}
{Holland}, W.~S., {Greaves}, J.~S., {Zuckerman}, B., {et~al.} 1998, \href{http://dx.doi.org/10.1038/33874}{\color{magenta}\nat}, \href{https://ui.adsabs.harvard.edu/abs/1998Natur.392..788H}{\color{blue}392}, \href{https://ui.adsabs.harvard.edu/abs/1998Natur.392..788H}{\color{blue}788}

\bibitem[{{Hom} {et~al.}(2020){Hom}, {Patience}, {Esposito}, {Duch{\^e}ne}, {Worthen}, {Kalas}, {Jang-Condell}, {Saboi}, {Arriaga}, {Mazoyer}, {Wolff}, {Millar-Blanchaer}, {Fitzgerald}, {Perrin}, {Chen}, {Macintosh}, {Matthews}, {Wang}, {Graham}, {Marchis}, {Ammons}, {Bailey}, {Barman}, {Bulger}, {Chilcote}, {Cotten}, {De Rosa}, {Doyon}, {Follette}, {Goodsell}, {Greenbaum}, {Hibon}, {Ingraham}, {Konopacky}, {Larkin}, {Maire}, {Marley}, {Marois}, {Matthews}, {Metchev}, {Nielsen}, {Oppenheimer}, {Palmer}, {Poyneer}, {Pueyo}, {Rajan}, {Rameau}, {Rantakyr{\"o}}, {Ren}, {Savransky}, {Schneider}, {Sivaramakrishnan}, {Song}, {Soummer}, {Tallis}, {Thomas}, {Wallace}, {Ward-Duong}, {Wiktorowicz}, \& {Zuckerman}}]{hom2020}
{Hom}, J., {Patience}, J., {Esposito}, T.~M., {et~al.} 2020, \href{http://dx.doi.org/10.3847/1538-3881/ab5af2}{\color{magenta}\aj}, \href{https://ui.adsabs.harvard.edu/abs/2020AJ....159...31H}{\color{blue}159}, \href{https://ui.adsabs.harvard.edu/abs/2020AJ....159...31H}{\color{blue}31}

\bibitem[{{Hom} {et~al.}(2024){Hom}, {Patience}, {Chen}, {Duch{\^e}ne}, {Mazoyer}, {Millar-Blanchaer}, {Esposito}, {Kalas}, {Crotts}, {Gonzales}, {Kolokolova}, {Lewis}, {Matthews}, {Rice}, {Weinberger}, {Wilner}, {Wolff}, {Bruzzone}, {Choquet}, {Debes}, {De Rosa}, {Donaldson}, {Draper}, {Fitzgerald}, {Hines}, {Hinkley}, {Hughes}, {L{\'o}pez}, {Marchis}, {Metchev}, {Moro-Martin}, {Nesvold}, {Nielsen}, {Oppenheimer}, {Padgett}, {Perrin}, {Pueyo}, {Rantakyr{\"o}}, {Ren}, {Schneider}, {Soummer}, {Song}, \& {Stark}}]{hom2024}
{Hom}, J., {Patience}, J., {Chen}, C.~H., {et~al.} 2024, \href{http://dx.doi.org/10.1093/mnras/stae368}{\color{magenta}\mnras}, \href{https://ui.adsabs.harvard.edu/abs/2024MNRAS.528.6959H}{\color{blue}528}, \href{https://ui.adsabs.harvard.edu/abs/2024MNRAS.528.6959H}{\color{blue}6959}

\bibitem[{{Houk}(1978)}]{houk1978}
{Houk}, N. 1978, {Michigan catalogue of two-dimensional spectral types for the HD stars}

\bibitem[{{Houk} \& {Cowley}(1975)}]{houk1975}
{Houk}, N., \& {Cowley}, A.~P. 1975, {University of Michigan Catalogue of two-dimensional spectral types for the HD stars. Volume I. Declinations -90\_ to -53\_{\textflorin}0.}

\bibitem[{{Hughes} {et~al.}(2018){Hughes}, {Duch{\^e}ne}, \& {Matthews}}]{hughes2018}
{Hughes}, A.~M., {Duch{\^e}ne}, G., \& {Matthews}, B.~C. 2018, \href{http://dx.doi.org/10.1146/annurev-astro-081817-052035}{\color{magenta}\araa}, \href{https://ui.adsabs.harvard.edu/abs/2018ARA&A..56..541H}{\color{blue}56}, \href{https://ui.adsabs.harvard.edu/abs/2018ARA&A..56..541H}{\color{blue}541}

\bibitem[{{Hung} {et~al.}(2015){Hung}, {Duch{\^e}ne}, {Arriaga}, {Fitzgerald}, {Maire}, {Marois}, {Millar-Blanchaer}, {Bruzzone}, {Rajan}, {Pueyo}, {Kalas}, {De Rosa}, {Graham}, {Konopacky}, {Wolff}, {Ammons}, {Chen}, {Chilcote}, {Draper}, {Esposito}, {Gerard}, {Goodsell}, {Greenbaum}, {Hibon}, {Hinkley}, {Macintosh}, {Marchis}, {Metchev}, {Nielsen}, {Oppenheimer}, {Patience}, {Perrin}, {Rantakyr{\"o}}, {Sivaramakrishnan}, {Wang}, {Ward-Duong}, \& {Wiktorowicz}}]{hung2015}
{Hung}, L.-W., {Duch{\^e}ne}, G., {Arriaga}, P., {et~al.} 2015, \href{http://dx.doi.org/10.1088/2041-8205/815/1/L14}{\color{magenta}\apjl}, \href{https://ui.adsabs.harvard.edu/abs/2015ApJ...815L..14H}{\color{blue}815}, \href{https://ui.adsabs.harvard.edu/abs/2015ApJ...815L..14H}{\color{blue}L14}

\bibitem[{{Hung} {et~al.}(2016){Hung}, {Bruzzone}, {Millar-Blanchaer}, {Wang}, {Arriaga}, {Metchev}, {Fitzgerald}, {Sivaramakrishnan}, \& {Perrin}}]{hung2016}
{Hung}, L.-W., {Bruzzone}, S., {Millar-Blanchaer}, M.~A., {et~al.} 2016, \href{http://dx.doi.org/10.1117/12.2233665}{\color{magenta}Proc.~SPIE}, \href{https://ui.adsabs.harvard.edu/abs/2016SPIE.9908E..3AH}{\color{blue}9908}, \href{https://ui.adsabs.harvard.edu/abs/2016SPIE.9908E..3AH}{\color{blue}99083A}

\bibitem[{{Hunter}(2007)}]{matplotlib2007}
{Hunter}, J.~D. 2007, \href{http://dx.doi.org/10.1109/MCSE.2007.55}{\color{magenta}CSE}, \href{https://ui.adsabs.harvard.edu/abs/2007CSE.....9...90H}{\color{blue}9}, \href{https://ui.adsabs.harvard.edu/abs/2007CSE.....9...90H}{\color{blue}90}

\bibitem[{{Jang-Condell} {et~al.}(2015){Jang-Condell}, {Chen}, {Mittal}, {Manoj}, {Watson}, {Lisse}, {Nesvold}, \& {Kuchner}}]{jang-condell2015}
{Jang-Condell}, H., {Chen}, C.~H., {Mittal}, T., {et~al.} 2015, \href{http://dx.doi.org/10.1088/0004-637X/808/2/167}{\color{magenta}\apj}, \href{https://ui.adsabs.harvard.edu/abs/2015ApJ...808..167J}{\color{blue}808}, \href{https://ui.adsabs.harvard.edu/abs/2015ApJ...808..167J}{\color{blue}167}

\bibitem[{{Janson} {et~al.}(2013){Janson}, {Lafreni{\`e}re}, {Jayawardhana}, {Bonavita}, {Girard}, {Brandeker}, \& {Gizis}}]{janson2013}
{Janson}, M., {Lafreni{\`e}re}, D., {Jayawardhana}, R., {et~al.} 2013, \href{http://dx.doi.org/10.1088/0004-637X/773/2/170}{\color{magenta}\apj}, \href{https://ui.adsabs.harvard.edu/abs/2013ApJ...773..170J}{\color{blue}773}, \href{https://ui.adsabs.harvard.edu/abs/2013ApJ...773..170J}{\color{blue}170}

\bibitem[{{Janson} {et~al.}(2016){Janson}, {Thalmann}, {Boccaletti}, {Maire}, {Zurlo}, {Marzari}, {Meyer}, {Carson}, {Augereau}, {Garufi}, {Henning}, {Desidera}, {Asensio-Torres}, \& {Pohl}}]{janson2016}
{Janson}, M., {Thalmann}, C., {Boccaletti}, A., {et~al.} 2016, \href{http://dx.doi.org/10.3847/2041-8205/816/1/L1}{\color{magenta}\apjl}, \href{https://ui.adsabs.harvard.edu/abs/2016ApJ...816L...1J}{\color{blue}816}, \href{https://ui.adsabs.harvard.edu/abs/2016ApJ...816L...1J}{\color{blue}L1}

\bibitem[{{Johnson} {et~al.}(2012){Johnson}, {Lisse}, {Chen}, {Melosh}, {Wyatt}, {Thebault}, {Henning}, {Gaidos}, {Elkins-Tanton}, {Bridges}, \& {Morlok}}]{johnson2012}
{Johnson}, B.~C., {Lisse}, C.~M., {Chen}, C.~H., {et~al.} 2012, \href{http://dx.doi.org/10.1088/0004-637X/761/1/45}{\color{magenta}\apj}, \href{https://ui.adsabs.harvard.edu/abs/2012ApJ...761...45J}{\color{blue}761}, \href{https://ui.adsabs.harvard.edu/abs/2012ApJ...761...45J}{\color{blue}45}

\bibitem[{{Kalas} {et~al.}(2015){Kalas}, {Rajan}, {Wang}, {Millar-Blanchaer}, {Duchene}, {Chen}, {Fitzgerald}, {Dong}, {Graham}, {Patience}, {Macintosh}, {Murray-Clay}, {Matthews}, {Rameau}, {Marois}, {Chilcote}, {De Rosa}, {Doyon}, {Draper}, {Lawler}, {Ammons}, {Arriaga}, {Bulger}, {Cotten}, {Follette}, {Goodsell}, {Greenbaum}, {Hibon}, {Hinkley}, {Hung}, {Ingraham}, {Konapacky}, {Lafreniere}, {Larkin}, {Long}, {Maire}, {Marchis}, {Metchev}, {Morzinski}, {Nielsen}, {Oppenheimer}, {Perrin}, {Pueyo}, {Rantakyr{\"o}}, {Ruffio}, {Saddlemyer}, {Savransky}, {Schneider}, {Sivaramakrishnan}, {Soummer}, {Song}, {Thomas}, {Vasisht}, {Ward-Duong}, {Wiktorowicz}, \& {Wolff}}]{kalas2015}
{Kalas}, P.~G., {Rajan}, A., {Wang}, J.~J., {et~al.} 2015, \href{http://dx.doi.org/10.1088/0004-637X/814/1/32}{\color{magenta}\apj}, \href{https://ui.adsabs.harvard.edu/abs/2015ApJ...814...32K}{\color{blue}814}, \href{https://ui.adsabs.harvard.edu/abs/2015ApJ...814...32K}{\color{blue}32}

\bibitem[{{Kennedy} {et~al.}(2018){Kennedy}, {Marino}, {Matr{\`a}}, {Pani{\'c}}, {Wilner}, {Wyatt}, \& {Yelverton}}]{kennedy2018}
{Kennedy}, G.~M., {Marino}, S., {Matr{\`a}}, L., {et~al.} 2018, \href{http://dx.doi.org/10.1093/mnras/sty135}{\color{magenta}\mnras}, \href{https://ui.adsabs.harvard.edu/abs/2018MNRAS.475.4924K}{\color{blue}475}, \href{https://ui.adsabs.harvard.edu/abs/2018MNRAS.475.4924K}{\color{blue}4924}

\bibitem[{{Krivov} {et~al.}(2007){Krivov}, {Queck}, {L{\"o}hne}, \& {Srem{\v{c}}evi{\'c}}}]{krivov2007}
{Krivov}, A.~V., {Queck}, M., {L{\"o}hne}, T., \& {Srem{\v{c}}evi{\'c}}, M. 2007, \href{http://dx.doi.org/10.1051/0004-6361:20065584}{\color{magenta}\aap}, \href{https://ui.adsabs.harvard.edu/abs/2007A&A...462..199K}{\color{blue}462}, \href{https://ui.adsabs.harvard.edu/abs/2007A&A...462..199K}{\color{blue}199}

\bibitem[{{Lafreni{\`e}re} {et~al.}(2007){Lafreni{\`e}re}, {Marois}, {Doyon}, {Nadeau}, \& {Artigau}}]{lafreniere2007}
{Lafreni{\`e}re}, D., {Marois}, C., {Doyon}, R., {et~al.} 2007, \href{http://dx.doi.org/10.1086/513180}{\color{magenta}\apj}, \href{https://ui.adsabs.harvard.edu/abs/2007ApJ...660..770L}{\color{blue}660}, \href{https://ui.adsabs.harvard.edu/abs/2007ApJ...660..770L}{\color{blue}770}

\bibitem[{{Lagrange} {et~al.}(2009){Lagrange}, {Gratadour}, {Chauvin}, {Fusco}, {Ehrenreich}, {Mouillet}, {Rousset}, {Rouan}, {Allard}, {Gendron}, {Charton}, {Mugnier}, {Rabou}, {Montri}, \& {Lacombe}}]{lagrange2009}
{Lagrange}, A.~M., {Gratadour}, D., {Chauvin}, G., {et~al.} 2009, \href{http://dx.doi.org/10.1051/0004-6361:200811325}{\color{magenta}\aap}, \href{https://ui.adsabs.harvard.edu/abs/2009A&A...493L..21L}{\color{blue}493}, \href{https://ui.adsabs.harvard.edu/abs/2009A&A...493L..21L}{\color{blue}L21}

\bibitem[{{Lagrange} {et~al.}(2016){Lagrange}, {Langlois}, {Gratton}, {Maire}, {Milli}, {Olofsson}, {Vigan}, {Bailey}, {Mesa}, {Chauvin}, {Boccaletti}, {Galicher}, {Girard}, {Bonnefoy}, {Samland}, {Menard}, {Henning}, {Kenworthy}, {Thalmann}, {Beust}, {Beuzit}, {Brandner}, {Buenzli}, {Cheetham}, {Janson}, {le Coroller}, {Lannier}, {Mouillet}, {Peretti}, {Perrot}, {Salter}, {Sissa}, {Wahhaj}, {Abe}, {Desidera}, {Feldt}, {Madec}, {Perret}, {Petit}, {Rabou}, {Soenke}, \& {Weber}}]{lagrange2016}
{Lagrange}, A.~M., {Langlois}, M., {Gratton}, R., {et~al.} 2016, \href{http://dx.doi.org/10.1051/0004-6361/201527264}{\color{magenta}\aap}, \href{https://ui.adsabs.harvard.edu/abs/2016A&A...586L...8L}{\color{blue}586}, \href{https://ui.adsabs.harvard.edu/abs/2016A&A...586L...8L}{\color{blue}L8}

\bibitem[{Lebigot(2010)}]{lebigot2010uncertainties}
Lebigot, E.~O. 2010, URL http://pythonhosted. org/uncertainties

\bibitem[{{Lee} \& {Chiang}(2016)}]{lee2016}
{Lee}, E.~J., \& {Chiang}, E. 2016, \href{http://dx.doi.org/10.3847/0004-637X/827/2/125}{\color{magenta}\apj}, \href{https://ui.adsabs.harvard.edu/abs/2016ApJ...827..125L}{\color{blue}827}, \href{https://ui.adsabs.harvard.edu/abs/2016ApJ...827..125L}{\color{blue}125}

\bibitem[{{Lewis} {et~al.}(2024){Lewis}, {Fitzgerald}, {Esposito}, {Arriaga}, {L{\'o}pez}, {Crotts}, {Duch{\^e}ne}, {Follette}, {Hom}, {Kalas}, {Matthews}, {Millar-Blanchaer}, {Wilner}, {Mazoyer}, \& {Macintosh}}]{lewis2024}
{Lewis}, B.~L., {Fitzgerald}, M.~P., {Esposito}, T.~M., {et~al.} 2024, \href{http://dx.doi.org/10.3847/1538-3881/ad67e1}{\color{magenta}\aj}, \href{https://ui.adsabs.harvard.edu/abs/2024AJ....168..142L}{\color{blue}168}, \href{https://ui.adsabs.harvard.edu/abs/2024AJ....168..142L}{\color{blue}142}

\bibitem[{{Lieman-Sifry} {et~al.}(2016){Lieman-Sifry}, {Hughes}, {Carpenter}, {Gorti}, {Hales}, \& {Flaherty}}]{lieman-sifry2016}
{Lieman-Sifry}, J., {Hughes}, A.~M., {Carpenter}, J.~M., {et~al.} 2016, \href{http://dx.doi.org/10.3847/0004-637X/828/1/25}{\color{magenta}\apj}, \href{https://ui.adsabs.harvard.edu/abs/2016ApJ...828...25L}{\color{blue}828}, \href{https://ui.adsabs.harvard.edu/abs/2016ApJ...828...25L}{\color{blue}25}

\bibitem[{{L{\"o}hne}(2020)}]{lohne2020}
{L{\"o}hne}, T. 2020, \href{http://dx.doi.org/10.1051/0004-6361/202037858}{\color{magenta}\aap}, \href{https://ui.adsabs.harvard.edu/abs/2020A&A...641A..75L}{\color{blue}641}, \href{https://ui.adsabs.harvard.edu/abs/2020A&A...641A..75L}{\color{blue}A75}

\bibitem[{{L{\'o}pez} {et~al.}(In Prep){L{\'o}pez}, {Fitzgerald}, {Esposito}, {Arriaga}, {Lewis}, {Duch{\^e}ne}, {Millar-Blanchaer}, {Wang}, {Chen}, {Benac}, {Savransky}, \& {Kalas}}]{lopez2023}
{L{\'o}pez}, R.~A., {Fitzgerald}, M.~P., {Esposito}, T.~M., {et~al.} In Prep

\bibitem[{{MacGregor} {et~al.}(2013){MacGregor}, {Wilner}, {Rosenfeld}, {Andrews}, {Matthews}, {Hughes}, {Booth}, {Chiang}, {Graham}, {Kalas}, {Kennedy}, \& {Sibthorpe}}]{macgregor2013}
{MacGregor}, M.~A., {Wilner}, D.~J., {Rosenfeld}, K.~A., {et~al.} 2013, \href{http://dx.doi.org/10.1088/2041-8205/762/2/L21}{\color{magenta}\apjl}, \href{https://ui.adsabs.harvard.edu/abs/2013ApJ...762L..21M}{\color{blue}762}, \href{https://ui.adsabs.harvard.edu/abs/2013ApJ...762L..21M}{\color{blue}L21}

\bibitem[{{MacGregor} {et~al.}(2018){MacGregor}, {Weinberger}, {Hughes}, {Wilner}, {Currie}, {Debes}, {Donaldson}, {Redfield}, {Roberge}, \& {Schneider}}]{macgregor2018}
{MacGregor}, M.~A., {Weinberger}, A.~J., {Hughes}, A.~M., {et~al.} 2018, \href{http://dx.doi.org/10.3847/1538-4357/aaec71}{\color{magenta}\apj}, \href{https://ui.adsabs.harvard.edu/abs/2018ApJ...869...75M}{\color{blue}869}, \href{https://ui.adsabs.harvard.edu/abs/2018ApJ...869...75M}{\color{blue}75}

\bibitem[{{Macintosh} {et~al.}(2014){Macintosh}, {Graham}, {Ingraham}, {Konopacky}, {Marois}, {Perrin}, {Poyneer}, {Bauman}, {Barman}, {Burrows}, {Cardwell}, {Chilcote}, {De Rosa}, {Dillon}, {Doyon}, {Dunn}, {Erikson}, {Fitzgerald}, {Gavel}, {Goodsell}, {Hartung}, {Hibon}, {Kalas}, {Larkin}, {Maire}, {Marchis}, {Marley}, {McBride}, {Millar-Blanchaer}, {Morzinski}, {Norton}, {Oppenheimer}, {Palmer}, {Patience}, {Pueyo}, {Rantakyro}, {Sadakuni}, {Saddlemyer}, {Savransky}, {Serio}, {Soummer}, {Sivaramakrishnan}, {Song}, {Thomas}, {Wallace}, {Wiktorowicz}, \& {Wolff}}]{macintosh2014}
{Macintosh}, B., {Graham}, J.~R., {Ingraham}, P., {et~al.} 2014, \href{http://dx.doi.org/10.1073/pnas.1304215111}{\color{magenta}PNAS}, \href{https://ui.adsabs.harvard.edu/abs/2014PNAS..11112661M}{\color{blue}111}, \href{https://ui.adsabs.harvard.edu/abs/2014PNAS..11112661M}{\color{blue}12661}

\bibitem[{{Marois} {et~al.}(2006){Marois}, {Lafreni{\`e}re}, {Doyon}, {Macintosh}, \& {Nadeau}}]{marois2006}
{Marois}, C., {Lafreni{\`e}re}, D., {Doyon}, R., {et~al.} 2006, \href{http://dx.doi.org/10.1086/500401}{\color{magenta}\apj}, \href{https://ui.adsabs.harvard.edu/abs/2006ApJ...641..556M}{\color{blue}641}, \href{https://ui.adsabs.harvard.edu/abs/2006ApJ...641..556M}{\color{blue}556}

\bibitem[{{Matr{\`a}} {et~al.}(2019{\natexlab{a}}){Matr{\`a}}, {{\"O}berg}, {Wilner}, {Olofsson}, \& {Bayo}}]{matra2019_AUMic}
{Matr{\`a}}, L., {{\"O}berg}, K.~I., {Wilner}, D.~J., {et~al.} 2019{\natexlab{a}}, \href{http://dx.doi.org/10.3847/1538-3881/aaff5b}{\color{magenta}\aj}, \href{https://ui.adsabs.harvard.edu/abs/2019AJ....157..117M}{\color{blue}157}, \href{https://ui.adsabs.harvard.edu/abs/2019AJ....157..117M}{\color{blue}117}

\bibitem[{{Matr{\`a}} {et~al.}(2019{\natexlab{b}}){Matr{\`a}}, {Wyatt}, {Wilner}, {Dent}, {Marino}, {Kennedy}, \& {Milli}}]{matra2019}
{Matr{\`a}}, L., {Wyatt}, M.~C., {Wilner}, D.~J., {et~al.} 2019{\natexlab{b}}, \href{http://dx.doi.org/10.3847/1538-3881/ab06c0}{\color{magenta}\aj}, \href{https://ui.adsabs.harvard.edu/abs/2019AJ....157..135M}{\color{blue}157}, \href{https://ui.adsabs.harvard.edu/abs/2019AJ....157..135M}{\color{blue}135}

\bibitem[{{Matr{\`a}} {et~al.}(2017){Matr{\`a}}, {Dent}, {Wyatt}, {Kral}, {Wilner}, {Pani{\'c}}, {Hughes}, {de Gregorio-Monsalvo}, {Hales}, {Augereau}, {Greaves}, \& {Roberge}}]{matra2017}
{Matr{\`a}}, L., {Dent}, W.~R.~F., {Wyatt}, M.~C., {et~al.} 2017, \href{http://dx.doi.org/10.1093/mnras/stw2415}{\color{magenta}\mnras}, \href{https://ui.adsabs.harvard.edu/abs/2017MNRAS.464.1415M}{\color{blue}464}, \href{https://ui.adsabs.harvard.edu/abs/2017MNRAS.464.1415M}{\color{blue}1415}

\bibitem[{{Matthews} {et~al.}(2023){Matthews}, {Bonnefoy}, {Xie}, {Desgrange}, {Desidera}, {Delorme}, {Milli}, {Olofsson}, {Barbato}, {Ceva}, {Augereau}, {Biller}, {Chen}, {Faramaz-Gorka}, {Galicher}, {Hinkley}, {Lagrange}, {M{\'e}nard}, {Pinte}, \& {Stapelfeldt}}]{matthews2023}
{Matthews}, E.~C., {Bonnefoy}, M., {Xie}, C., {et~al.} 2023, \href{http://dx.doi.org/10.1051/0004-6361/202347335}{\color{magenta}\aap}, \href{https://ui.adsabs.harvard.edu/abs/2023A&A...679A..58M}{\color{blue}679}, \href{https://ui.adsabs.harvard.edu/abs/2023A&A...679A..58M}{\color{blue}A58}

\bibitem[{{Mazoyer} {et~al.}(2020){Mazoyer}, {Arriaga}, {Hom}, {Millar-Blanchaer}, {Chen}, {Wang}, {Duch{\^e}ne}, {Patience}, \& {Pueyo}}]{mazoyer2020}
{Mazoyer}, J., {Arriaga}, P., {Hom}, J., {et~al.} 2020, \href{http://dx.doi.org/10.1117/12.2560091}{\color{magenta}Proc.~SPIE}, \href{https://ui.adsabs.harvard.edu/abs/2020SPIE11447E..59M}{\color{blue}11447}, \href{https://ui.adsabs.harvard.edu/abs/2020SPIE11447E..59M}{\color{blue}1144759}

\bibitem[{{Milli} {et~al.}(2012){Milli}, {Mouillet}, {Lagrange}, {Boccaletti}, {Mawet}, {Chauvin}, \& {Bonnefoy}}]{milli2012}
{Milli}, J., {Mouillet}, D., {Lagrange}, A.~M., {et~al.} 2012, \href{http://dx.doi.org/10.1051/0004-6361/201219687}{\color{magenta}\aap}, \href{https://ui.adsabs.harvard.edu/abs/2012A&A...545A.111M}{\color{blue}545}, \href{https://ui.adsabs.harvard.edu/abs/2012A&A...545A.111M}{\color{blue}A111}

\bibitem[{{Milli} {et~al.}(2015){Milli}, {Mawet}, {Pinte}, {Lagrange}, {Mouillet}, {Girard}, {Augereau}, {De Boer}, {Pueyo}, \& {Choquet}}]{milli2015}
{Milli}, J., {Mawet}, D., {Pinte}, C., {et~al.} 2015, \href{http://dx.doi.org/10.1051/0004-6361/201423950}{\color{magenta}\aap}, \href{https://ui.adsabs.harvard.edu/abs/2015A&A...577A..57M}{\color{blue}577}, \href{https://ui.adsabs.harvard.edu/abs/2015A&A...577A..57M}{\color{blue}A57}

\bibitem[{{Milli} {et~al.}(2017){Milli}, {Vigan}, {Mouillet}, {Lagrange}, {Augereau}, {Pinte}, {Mawet}, {Schmid}, {Boccaletti}, {Matr{\`a}}, {Kral}, {Ertel}, {Chauvin}, {Bazzon}, {M{\'e}nard}, {Beuzit}, {Thalmann}, {Dominik}, {Feldt}, {Henning}, {Min}, {Girard}, {Galicher}, {Bonnefoy}, {Fusco}, {de Boer}, {Janson}, {Maire}, {Mesa}, {Schlieder}, \& {SPHERE Consortium}}]{milli2017}
{Milli}, J., {Vigan}, A., {Mouillet}, D., {et~al.} 2017, \href{http://dx.doi.org/10.1051/0004-6361/201527838}{\color{magenta}\aap}, \href{https://ui.adsabs.harvard.edu/abs/2017A&A...599A.108M}{\color{blue}599}, \href{https://ui.adsabs.harvard.edu/abs/2017A&A...599A.108M}{\color{blue}A108}

\bibitem[{{Milli} {et~al.}(2019){Milli}, {Engler}, {Schmid}, {Olofsson}, {M{\'e}nard}, {Kral}, {Boccaletti}, {Th{\'e}bault}, {Choquet}, {Mouillet}, {Lagrange}, {Augereau}, {Pinte}, {Chauvin}, {Dominik}, {Perrot}, {Zurlo}, {Henning}, {Beuzit}, {Avenhaus}, {Bazzon}, {Moulin}, {Llored}, {Moeller-Nilsson}, {Roelfsema}, \& {Pragt}}]{milli2019}
{Milli}, J., {Engler}, N., {Schmid}, H.~M., {et~al.} 2019, \href{http://dx.doi.org/10.1051/0004-6361/201935363}{\color{magenta}\aap}, \href{https://ui.adsabs.harvard.edu/abs/2019A&A...626A..54M}{\color{blue}626}, \href{https://ui.adsabs.harvard.edu/abs/2019A&A...626A..54M}{\color{blue}A54}

\bibitem[{{Mo{\'o}r} {et~al.}(2017){Mo{\'o}r}, {Cur{\'e}}, {K{\'o}sp{\'a}l}, {{\'A}brah{\'a}m}, {Csengeri}, {Eiroa}, {Gunawan}, {Henning}, {Hughes}, {Juh{\'a}sz}, {Pawellek}, \& {Wyatt}}]{moor2017}
{Mo{\'o}r}, A., {Cur{\'e}}, M., {K{\'o}sp{\'a}l}, {\'A}., {et~al.} 2017, \href{http://dx.doi.org/10.3847/1538-4357/aa8e4e}{\color{magenta}\apj}, \href{https://ui.adsabs.harvard.edu/abs/2017ApJ...849..123M}{\color{blue}849}, \href{https://ui.adsabs.harvard.edu/abs/2017ApJ...849..123M}{\color{blue}123}

\bibitem[{{Morales} {et~al.}(2016){Morales}, {Bryden}, {Werner}, \& {Stapelfeldt}}]{morales2016}
{Morales}, F.~Y., {Bryden}, G., {Werner}, M.~W., \& {Stapelfeldt}, K.~R. 2016, \href{http://dx.doi.org/10.3847/0004-637X/831/1/97}{\color{magenta}\apj}, \href{https://ui.adsabs.harvard.edu/abs/2016ApJ...831...97M}{\color{blue}831}, \href{https://ui.adsabs.harvard.edu/abs/2016ApJ...831...97M}{\color{blue}97}

\bibitem[{{Nesvold} {et~al.}(2017){Nesvold}, {Naoz}, \& {Fitzgerald}}]{nesvold2017}
{Nesvold}, E.~R., {Naoz}, S., \& {Fitzgerald}, M.~P. 2017, \href{http://dx.doi.org/10.3847/2041-8213/aa61a7}{\color{magenta}\apjl}, \href{https://ui.adsabs.harvard.edu/abs/2017ApJ...837L...6N}{\color{blue}837}, \href{https://ui.adsabs.harvard.edu/abs/2017ApJ...837L...6N}{\color{blue}L6}

\bibitem[{{Nielsen} {et~al.}(2013){Nielsen}, {Liu}, {Wahhaj}, {Biller}, {Hayward}, {Close}, {Males}, {Skemer}, {Chun}, {Ftaclas}, {Alencar}, {Artymowicz}, {Boss}, {Clarke}, {de Gouveia Dal Pino}, {Gregorio-Hetem}, {Hartung}, {Ida}, {Kuchner}, {Lin}, {Reid}, {Shkolnik}, {Tecza}, {Thatte}, \& {Toomey}}]{nielsen2013}
{Nielsen}, E.~L., {Liu}, M.~C., {Wahhaj}, Z., {et~al.} 2013, \href{http://dx.doi.org/10.1088/0004-637X/776/1/4}{\color{magenta}\apj}, \href{https://ui.adsabs.harvard.edu/abs/2013ApJ...776....4N}{\color{blue}776}, \href{https://ui.adsabs.harvard.edu/abs/2013ApJ...776....4N}{\color{blue}4}

\bibitem[{{Nielsen} {et~al.}(2017){Nielsen}, {Rosa}, {Rameau}, {Wang}, {Esposito}, {Millar-Blanchaer}, {Marois}, {Vigan}, {Ammons}, {Artigau}, {Bailey}, {Blunt}, {Bulger}, {Chilcote}, {Cotten}, {Doyon}, {Duch{\^e}ne}, {Fabrycky}, {Fitzgerald}, {Follette}, {Gerard}, {Goodsell}, {Graham}, {Greenbaum}, {Hibon}, {Hinkley}, {Hung}, {Ingraham}, {Jensen-Clem}, {Kalas}, {Konopacky}, {Larkin}, {Macintosh}, {Maire}, {Marchis}, {Metchev}, {Morzinski}, {Murray-Clay}, {Oppenheimer}, {Palmer}, {Patience}, {Perrin}, {Poyneer}, {Pueyo}, {Rafikov}, {Rajan}, {Rantakyr{\"o}}, {Ruffio}, {Savransky}, {Schneider}, {Sivaramakrishnan}, {Song}, {Soummer}, {Thomas}, {Wallace}, {Ward-Duong}, {Wiktorowicz}, \& {Wolff}}]{nielsen2017}
{Nielsen}, E.~L., {Rosa}, R. J.~D., {Rameau}, J., {et~al.} 2017, \href{http://dx.doi.org/10.3847/1538-3881/aa8a69}{\color{magenta}\aj}, \href{https://ui.adsabs.harvard.edu/abs/2017AJ....154..218N}{\color{blue}154}, \href{https://ui.adsabs.harvard.edu/abs/2017AJ....154..218N}{\color{blue}218}

\bibitem[{{Norfolk} {et~al.}(2021){Norfolk}, {Maddison}, {Marshall}, {Kennedy}, {Duch{\^e}ne}, {Wilner}, {Pinte}, {Mo{\'o}r}, {Matthews}, {{\'A}brah{\'a}m}, {K{\'o}sp{\'a}l}, \& {van der Marel}}]{norfolk2021}
{Norfolk}, B.~J., {Maddison}, S.~T., {Marshall}, J.~P., {et~al.} 2021, \href{http://dx.doi.org/10.1093/mnras/stab1901}{\color{magenta}\mnras}, \href{https://ui.adsabs.harvard.edu/abs/2021MNRAS.507.3139N}{\color{blue}507}, \href{https://ui.adsabs.harvard.edu/abs/2021MNRAS.507.3139N}{\color{blue}3139}

\bibitem[{{Olofsson} {et~al.}(2024){Olofsson}, {Th{\'e}bault}, {Bayo}, {Henning}, \& {Milli}}]{olofsson2024}
{Olofsson}, J., {Th{\'e}bault}, P., {Bayo}, A., {et~al.} 2024, \href{http://dx.doi.org/10.1051/0004-6361/202450100}{\color{magenta}\aap}, \href{https://ui.adsabs.harvard.edu/abs/2024A&A...688A..42O}{\color{blue}688}, \href{https://ui.adsabs.harvard.edu/abs/2024A&A...688A..42O}{\color{blue}A42}

\bibitem[{{Pawellek} \& {Krivov}(2015)}]{pawellek2015}
{Pawellek}, N., \& {Krivov}, A.~V. 2015, \href{http://dx.doi.org/10.1093/mnras/stv2142}{\color{magenta}\mnras}, \href{https://ui.adsabs.harvard.edu/abs/2015MNRAS.454.3207P}{\color{blue}454}, \href{https://ui.adsabs.harvard.edu/abs/2015MNRAS.454.3207P}{\color{blue}3207}

\bibitem[{{Pawellek} {et~al.}(2024){Pawellek}, {Mo{\'o}r}, {Kirchschlager}, {Milli}, {K{\'o}sp{\'a}l}, {{\'A}brah{\'a}m}, {Marino}, {Wyatt}, {Rebollido}, {Hughes}, {Cantalloube}, \& {Henning}}]{pawellek2024}
{Pawellek}, N., {Mo{\'o}r}, A., {Kirchschlager}, F., {et~al.} 2024, \href{http://dx.doi.org/10.1093/mnras/stad3455}{\color{magenta}\mnras}, \href{https://ui.adsabs.harvard.edu/abs/2024MNRAS.527.3559P}{\color{blue}527}, \href{https://ui.adsabs.harvard.edu/abs/2024MNRAS.527.3559P}{\color{blue}3559}

\bibitem[{{Pecaut} \& {Mamajek}(2013)}]{pecaut2013}
{Pecaut}, M.~J., \& {Mamajek}, E.~E. 2013, \href{http://dx.doi.org/10.1088/0067-0049/208/1/9}{\color{magenta}\apjs}, \href{https://ui.adsabs.harvard.edu/abs/2013ApJS..208....9P}{\color{blue}208}, \href{https://ui.adsabs.harvard.edu/abs/2013ApJS..208....9P}{\color{blue}9}

\bibitem[{{Pecaut} \& {Mamajek}(2016)}]{pecaut2016}
---. 2016, \href{http://dx.doi.org/10.1093/mnras/stw1300}{\color{magenta}\mnras}, \href{https://ui.adsabs.harvard.edu/abs/2016MNRAS.461..794P}{\color{blue}461}, \href{https://ui.adsabs.harvard.edu/abs/2016MNRAS.461..794P}{\color{blue}794}

\bibitem[{{Perez} \& {Granger}(2007)}]{ipython2007}
{Perez}, F., \& {Granger}, B.~E. 2007, \href{http://dx.doi.org/10.1109/MCSE.2007.53}{\color{magenta}CSE}, \href{https://ui.adsabs.harvard.edu/abs/2007CSE.....9c..21P}{\color{blue}9}, \href{https://ui.adsabs.harvard.edu/abs/2007CSE.....9c..21P}{\color{blue}21}

\bibitem[{{P{\'e}ricaud} {et~al.}(2017){P{\'e}ricaud}, {Di Folco}, {Dutrey}, {Guilloteau}, \& {Pi{\'e}tu}}]{pericaud2017}
{P{\'e}ricaud}, J., {Di Folco}, E., {Dutrey}, A., {et~al.} 2017, \href{http://dx.doi.org/10.1051/0004-6361/201629371}{\color{magenta}\aap}, \href{https://ui.adsabs.harvard.edu/abs/2017A&A...600A..62P}{\color{blue}600}, \href{https://ui.adsabs.harvard.edu/abs/2017A&A...600A..62P}{\color{blue}A62}

\bibitem[{{Perrin} {et~al.}(2014){Perrin}, {Maire}, {Ingraham}, {Savransky}, {Millar-Blanchaer}, {Wolff}, {Ruffio}, {Wang}, {Draper}, {Sadakuni}, {Marois}, {Rajan}, {Fitzgerald}, {Macintosh}, {Graham}, {Doyon}, {Larkin}, {Chilcote}, {Goodsell}, {Palmer}, {Labrie}, {Beaulieu}, {De Rosa}, {Greenbaum}, {Hartung}, {Hibon}, {Konopacky}, {Lafreniere}, {Lavigne}, {Marchis}, {Patience}, {Pueyo}, {Rantakyr{\"o}}, {Soummer}, {Sivaramakrishnan}, {Thomas}, {Ward-Duong}, \& {Wiktorowicz}}]{perrin2014}
{Perrin}, M.~D., {Maire}, J., {Ingraham}, P., {et~al.} 2014, \href{http://dx.doi.org/10.1117/12.2055246}{\color{magenta}Proc.~SPIE}, \href{https://ui.adsabs.harvard.edu/abs/2014SPIE.9147E..3JP}{\color{blue}9147}, \href{https://ui.adsabs.harvard.edu/abs/2014SPIE.9147E..3JP}{\color{blue}91473J}

\bibitem[{{Perrin} {et~al.}(2015){Perrin}, {Duchene}, {Millar-Blanchaer}, {Fitzgerald}, {Graham}, {Wiktorowicz}, {Kalas}, {Macintosh}, {Bauman}, {Cardwell}, {Chilcote}, {De Rosa}, {Dillon}, {Doyon}, {Dunn}, {Erikson}, {Gavel}, {Goodsell}, {Hartung}, {Hibon}, {Ingraham}, {Kerley}, {Konapacky}, {Larkin}, {Maire}, {Marchis}, {Marois}, {Mittal}, {Morzinski}, {Oppenheimer}, {Palmer}, {Patience}, {Poyneer}, {Pueyo}, {Rantakyr{\"o}}, {Sadakuni}, {Saddlemyer}, {Savransky}, {Soummer}, {Sivaramakrishnan}, {Song}, {Thomas}, {Wallace}, {Wang}, \& {Wolff}}]{perrin2015}
{Perrin}, M.~D., {Duchene}, G., {Millar-Blanchaer}, M., {et~al.} 2015, \href{http://dx.doi.org/10.1088/0004-637X/799/2/182}{\color{magenta}\apj}, \href{https://ui.adsabs.harvard.edu/abs/2015ApJ...799..182P}{\color{blue}799}, \href{https://ui.adsabs.harvard.edu/abs/2015ApJ...799..182P}{\color{blue}182}

\bibitem[{{Perrin} {et~al.}(2016){Perrin}, {Ingraham}, {Follette}, {Maire}, {Wang}, {Savransky}, {Arriaga}, {Bailey}, {Bruzzone}, {Chilcote}, {De Rosa}, {Draper}, {Fitzgerald}, {Greenbaum}, {Hung}, {Konopacky}, {Macintosh}, {Marchis}, {Marois}, {Millar-Blanchaer}, {Nielsen}, {Rajan}, {Rameau}, {Rantakyro}, {Ruffio}, {Ward-Duong}, {Wolff}, \& {Zalesky}}]{perrin2016}
{Perrin}, M.~D., {Ingraham}, P., {Follette}, K.~B., {et~al.} 2016, \href{http://dx.doi.org/10.1117/12.2233197}{\color{magenta}Proc.~SPIE}, \href{https://ui.adsabs.harvard.edu/abs/2016SPIE.9908E..37P}{\color{blue}9908}, \href{https://ui.adsabs.harvard.edu/abs/2016SPIE.9908E..37P}{\color{blue}990837}

\bibitem[{{Perrot} {et~al.}(2023){Perrot}, {Olofsson}, {Kral}, {Th{\'e}bault}, {Montesinos}, {Kennedy}, {Bayo}, {Iglesias}, {van Holstein}, \& {Pinte}}]{perrot2023}
{Perrot}, C., {Olofsson}, J., {Kral}, Q., {et~al.} 2023, \href{http://dx.doi.org/10.1051/0004-6361/202244694}{\color{magenta}\aap}, \href{https://ui.adsabs.harvard.edu/abs/2023A&A...673A..39P}{\color{blue}673}, \href{https://ui.adsabs.harvard.edu/abs/2023A&A...673A..39P}{\color{blue}A39}

\bibitem[{{Pinte} {et~al.}(2006){Pinte}, {M{\'e}nard}, {Duch{\^e}ne}, \& {Bastien}}]{pinte2006}
{Pinte}, C., {M{\'e}nard}, F., {Duch{\^e}ne}, G., \& {Bastien}, P. 2006, \href{http://dx.doi.org/10.1051/0004-6361:20053275}{\color{magenta}\aap}, \href{https://ui.adsabs.harvard.edu/abs/2006A&A...459..797P}{\color{blue}459}, \href{https://ui.adsabs.harvard.edu/abs/2006A&A...459..797P}{\color{blue}797}

\bibitem[{{Preibisch} \& {Mamajek}(2008)}]{preibisch2008}
{Preibisch}, T., \& {Mamajek}, E. 2008, {The Nearest OB Association: Scorpius-Centaurus (Sco OB2)}, ed. B.~{Reipurth}, Vol.~5, \href{https://ui.adsabs.harvard.edu/abs/2008hsf2.book..235P}{\color{blue}235}

\bibitem[{{Rameau} {et~al.}(2013){Rameau}, {Chauvin}, {Lagrange}, {Meshkat}, {Boccaletti}, {Quanz}, {Currie}, {Mawet}, {Girard}, {Bonnefoy}, \& {Kenworthy}}]{rameau2013}
{Rameau}, J., {Chauvin}, G., {Lagrange}, A.~M., {et~al.} 2013, \href{http://dx.doi.org/10.1088/2041-8205/779/2/L26}{\color{magenta}\apjl}, \href{https://ui.adsabs.harvard.edu/abs/2013ApJ...779L..26R}{\color{blue}779}, \href{https://ui.adsabs.harvard.edu/abs/2013ApJ...779L..26R}{\color{blue}L26}

\bibitem[{{Rodriguez} \& {Zuckerman}(2012)}]{rodriguez2012}
{Rodriguez}, D.~R., \& {Zuckerman}, B. 2012, \href{http://dx.doi.org/10.1088/0004-637X/745/2/147}{\color{magenta}\apj}, \href{https://ui.adsabs.harvard.edu/abs/2012ApJ...745..147R}{\color{blue}745}, \href{https://ui.adsabs.harvard.edu/abs/2012ApJ...745..147R}{\color{blue}147}

\bibitem[{{Ruffio} {et~al.}(2017){Ruffio}, {Macintosh}, {Wang}, {Pueyo}, {Nielsen}, {De Rosa}, {Czekala}, {Marley}, {Arriaga}, {Bailey}, {Barman}, {Bulger}, {Chilcote}, {Cotten}, {Doyon}, {Duch{\^e}ne}, {Fitzgerald}, {Follette}, {Gerard}, {Goodsell}, {Graham}, {Greenbaum}, {Hibon}, {Hung}, {Ingraham}, {Kalas}, {Konopacky}, {Larkin}, {Maire}, {Marchis}, {Marois}, {Metchev}, {Millar-Blanchaer}, {Morzinski}, {Oppenheimer}, {Palmer}, {Patience}, {Perrin}, {Poyneer}, {Rajan}, {Rameau}, {Rantakyr{\"o}}, {Savransky}, {Schneider}, {Sivaramakrishnan}, {Song}, {Soummer}, {Thomas}, {Wallace}, {Ward-Duong}, {Wiktorowicz}, \& {Wolff}}]{ruffio2017}
{Ruffio}, J.-B., {Macintosh}, B., {Wang}, J.~J., {et~al.} 2017, \href{http://dx.doi.org/10.3847/1538-4357/aa72dd}{\color{magenta}\apj}, \href{https://ui.adsabs.harvard.edu/abs/2017ApJ...842...14R}{\color{blue}842}, \href{https://ui.adsabs.harvard.edu/abs/2017ApJ...842...14R}{\color{blue}14}

\bibitem[{{Schmid} {et~al.}(2006){Schmid}, {Joos}, \& {Tschan}}]{schmid2006}
{Schmid}, H.~M., {Joos}, F., \& {Tschan}, D. 2006, \href{http://dx.doi.org/10.1051/0004-6361:20053273}{\color{magenta}\aap}, \href{https://ui.adsabs.harvard.edu/abs/2006A&A...452..657S}{\color{blue}452}, \href{https://ui.adsabs.harvard.edu/abs/2006A&A...452..657S}{\color{blue}657}

\bibitem[{{Schneider} {et~al.}(1999){Schneider}, {Smith}, {Becklin}, {Koerner}, {Meier}, {Hines}, {Lowrance}, {Terrile}, {Thompson}, \& {Rieke}}]{schneider1999}
{Schneider}, G., {Smith}, B.~A., {Becklin}, E.~E., {et~al.} 1999, \href{http://dx.doi.org/10.1086/311921}{\color{magenta}\apjl}, \href{https://ui.adsabs.harvard.edu/abs/1999ApJ...513L.127S}{\color{blue}513}, \href{https://ui.adsabs.harvard.edu/abs/1999ApJ...513L.127S}{\color{blue}L127}

\bibitem[{{Schneider} {et~al.}(2014){Schneider}, {Grady}, {Hines}, {Stark}, {Debes}, {Carson}, {Kuchner}, {Perrin}, {Weinberger}, {Wisniewski}, {Silverstone}, {Jang-Condell}, {Henning}, {Woodgate}, {Serabyn}, {Moro-Martin}, {Tamura}, {Hinz}, \& {Rodigas}}]{schneider2014}
{Schneider}, G., {Grady}, C.~A., {Hines}, D.~C., {et~al.} 2014, \href{http://dx.doi.org/10.1088/0004-6256/148/4/59}{\color{magenta}\aj}, \href{https://ui.adsabs.harvard.edu/abs/2014AJ....148...59S}{\color{blue}148}, \href{https://ui.adsabs.harvard.edu/abs/2014AJ....148...59S}{\color{blue}59}

\bibitem[{{Smith} \& {Terrile}(1984)}]{smith1984}
{Smith}, B.~A., \& {Terrile}, R.~J. 1984, \href{http://dx.doi.org/10.1126/science.226.4681.1421}{\color{magenta}Science}, \href{https://ui.adsabs.harvard.edu/abs/1984Sci...226.1421S}{\color{blue}226}, \href{https://ui.adsabs.harvard.edu/abs/1984Sci...226.1421S}{\color{blue}1421}

\bibitem[{{Soummer} {et~al.}(2012){Soummer}, {Pueyo}, \& {Larkin}}]{soummer2012}
{Soummer}, R., {Pueyo}, L., \& {Larkin}, J. 2012, \href{http://dx.doi.org/10.1088/2041-8205/755/2/L28}{\color{magenta}\apjl}, \href{https://ui.adsabs.harvard.edu/abs/2012ApJ...755L..28S}{\color{blue}755}, \href{https://ui.adsabs.harvard.edu/abs/2012ApJ...755L..28S}{\color{blue}L28}

\bibitem[{{Stasevic} {et~al.}(2023){Stasevic}, {Milli}, {Mazoyer}, {Lagrange}, {Bonnefoy}, {Faramaz-Gorka}, {M{\'e}nard}, {Boccaletti}, {Choquet}, {Shuai}, {Olofsson}, {Chomez}, {Ren}, {Rubini}, {Desgrange}, {Gratton}, {Chauvin}, {Vigan}, \& {Matthews}}]{stasevic2023}
{Stasevic}, S., {Milli}, J., {Mazoyer}, J., {et~al.} 2023, \href{http://dx.doi.org/10.1051/0004-6361/202346720}{\color{magenta}\aap}, \href{https://ui.adsabs.harvard.edu/abs/2023A&A...678A...8S}{\color{blue}678}, \href{https://ui.adsabs.harvard.edu/abs/2023A&A...678A...8S}{\color{blue}A8}

\bibitem[{{Stuber} {et~al.}(2023){Stuber}, {L{\"o}hne}, \& {Wolf}}]{stuber2023}
{Stuber}, T.~A., {L{\"o}hne}, T., \& {Wolf}, S. 2023, \href{http://dx.doi.org/10.1051/0004-6361/202243240}{\color{magenta}\aap}, \href{https://ui.adsabs.harvard.edu/abs/2023A&A...669A...3S}{\color{blue}669}, \href{https://ui.adsabs.harvard.edu/abs/2023A&A...669A...3S}{\color{blue}A3}

\bibitem[{{Su} {et~al.}(2017){Su}, {MacGregor}, {Booth}, {Wilner}, {Flaherty}, {Hughes}, {Phillips}, {Malhotra}, {Hales}, {Morrison}, {Ertel}, {Matthews}, {Dent}, \& {Casassus}}]{su2017}
{Su}, K. Y.~L., {MacGregor}, M.~A., {Booth}, M., {et~al.} 2017, \href{http://dx.doi.org/10.3847/1538-3881/aa906b}{\color{magenta}\aj}, \href{https://ui.adsabs.harvard.edu/abs/2017AJ....154..225S}{\color{blue}154}, \href{https://ui.adsabs.harvard.edu/abs/2017AJ....154..225S}{\color{blue}225}

\bibitem[{{Takeuchi} \& {Artymowicz}(2001)}]{takeuchi2001}
{Takeuchi}, T., \& {Artymowicz}, P. 2001, \href{http://dx.doi.org/10.1086/322252}{\color{magenta}\apj}, \href{https://ui.adsabs.harvard.edu/abs/2001ApJ...557..990T}{\color{blue}557}, \href{https://ui.adsabs.harvard.edu/abs/2001ApJ...557..990T}{\color{blue}990}

\bibitem[{{Thilliez} \& {Maddison}(2017)}]{thilliez2017}
{Thilliez}, E., \& {Maddison}, S.~T. 2017, \href{http://dx.doi.org/10.1093/mnras/stw2427}{\color{magenta}\mnras}, \href{https://ui.adsabs.harvard.edu/abs/2017MNRAS.464.1434T}{\color{blue}464}, \href{https://ui.adsabs.harvard.edu/abs/2017MNRAS.464.1434T}{\color{blue}1434}

\bibitem[{{Vigan}(2020)}]{vigan2020}
{Vigan}, A. 2020, {vlt-sphere: Automatic VLT/SPHERE data reduction and analysis}, Astrophysics Source Code Library, record ascl:2009.002, Astrophysics Source Code Library, record ascl:2009.002

\bibitem[{{Vigan} {et~al.}(2010){Vigan}, {Moutou}, {Langlois}, {Allard}, {Boccaletti}, {Carbillet}, {Mouillet}, \& {Smith}}]{vigan2010}
{Vigan}, A., {Moutou}, C., {Langlois}, M., {et~al.} 2010, \href{http://dx.doi.org/10.1111/j.1365-2966.2010.16916.x}{\color{magenta}\mnras}, \href{https://ui.adsabs.harvard.edu/abs/2010MNRAS.407...71V}{\color{blue}407}, \href{https://ui.adsabs.harvard.edu/abs/2010MNRAS.407...71V}{\color{blue}71}

\bibitem[{{Wahhaj} {et~al.}(2016){Wahhaj}, {Milli}, {Kennedy}, {Ertel}, {Matr{\`a}}, {Boccaletti}, {del Burgo}, {Wyatt}, {Pinte}, {Lagrange}, {Absil}, {Choquet}, {G{\'o}mez Gonz{\'a}lez}, {Kobayashi}, {Mawet}, {Mouillet}, {Pueyo}, {Dent}, {Augereau}, \& {Girard}}]{wahhaj2016}
{Wahhaj}, Z., {Milli}, J., {Kennedy}, G., {et~al.} 2016, \href{http://dx.doi.org/10.1051/0004-6361/201629769}{\color{magenta}\aap}, \href{https://ui.adsabs.harvard.edu/abs/2016A&A...596L...4W}{\color{blue}596}, \href{https://ui.adsabs.harvard.edu/abs/2016A&A...596L...4W}{\color{blue}L4}

\bibitem[{{Wahhaj} {et~al.}(2021){Wahhaj}, {Milli}, {Romero}, {Cieza}, {Zurlo}, {Vigan}, {Pe{\~n}a}, {Valdes}, {Cantalloube}, {Girard}, \& {Pantoja}}]{wahhaj2021}
{Wahhaj}, Z., {Milli}, J., {Romero}, C., {et~al.} 2021, \href{http://dx.doi.org/10.1051/0004-6361/202038794}{\color{magenta}\aap}, \href{https://ui.adsabs.harvard.edu/abs/2021A&A...648A..26W}{\color{blue}648}, \href{https://ui.adsabs.harvard.edu/abs/2021A&A...648A..26W}{\color{blue}A26}

\bibitem[{{Wang} {et~al.}(2015){Wang}, {Ruffio}, {De Rosa}, {Aguilar}, {Wolff}, \& {Pueyo}}]{wang2015}
{Wang}, J.~J., {Ruffio}, J.-B., {De Rosa}, R.~J., {et~al.} 2015, {pyKLIP: PSF Subtraction for Exoplanets and Disks}

\bibitem[{{Wang} {et~al.}(2018){Wang}, {Perrin}, {Savransky}, {Arriaga}, {Chilcote}, {De Rosa}, {Millar-Blanchaer}, {Marois}, {Rameau}, {Wolff}, {Shapiro}, {Ruffio}, {Maire}, {Marchis}, {Graham}, {Macintosh}, {Ammons}, {Bailey}, {Barman}, {Bruzzone}, {Bulger}, {Cotten}, {Doyon}, {Duch{\^e}ne}, {Fitzgerald}, {Follette}, {Goodsell}, {Greenbaum}, {Hibon}, {Hung}, {Ingraham}, {Kalas}, {Konopacky}, {Larkin}, {Marley}, {Metchev}, {Nielsen}, {Oppenheimer}, {Palmer}, {Patience}, {Poyneer}, {Pueyo}, {Rajan}, {Rantakyr{\"o}}, {Schneider}, {Sivaramakrishnan}, {Song}, {Soummer}, {Thomas}, {Wallace}, {Ward-Duong}, \& {Wiktorowicz}}]{wang2018}
{Wang}, J.~J., {Perrin}, M.~D., {Savransky}, D., {et~al.} 2018, \href{http://dx.doi.org/10.1117/1.JATIS.4.1.018002}{\color{magenta}JATIS}, \href{http://adsabs.harvard.edu/abs/2018JATIS...4a8002W}{\color{blue}4}, \href{http://adsabs.harvard.edu/abs/2018JATIS...4a8002W}{\color{blue}018002}

\bibitem[{{Wyatt}(2008)}]{wyatt2008}
{Wyatt}, M.~C. 2008, \href{http://dx.doi.org/10.1146/annurev.astro.45.051806.110525}{\color{magenta}\araa}, \href{https://ui.adsabs.harvard.edu/abs/2008ARA&A..46..339W}{\color{blue}46}, \href{https://ui.adsabs.harvard.edu/abs/2008ARA&A..46..339W}{\color{blue}339}

\end{thebibliography}

\end{document}